\documentclass[10pt,aps,prd,twocolumn,notitlepage,showpacs,superscriptaddress]{revtex4-1}
\usepackage{amsthm,amsmath,amsfonts,amssymb,verbatim,color}
\usepackage{graphicx}
\usepackage{bm}
\usepackage{epsfig,slashed}
\usepackage{amssymb}
\usepackage{amsfonts}
\usepackage{tikz}
\usepackage{xparse}
\usepackage{ifthen}
\usepackage{lipsum}
\usepackage[colorlinks=true,citecolor=blue,
linkcolor=blue,urlcolor=blue]{hyperref}
\usepackage[caption=false]{subfig}
\preprint{\mbox{}\hfill DESY 18-107\\\mbox{}\hfill HU-EP-18/20}

\begin{document}

\newcommand{\tr}{\mathop{\mathrm{tr}}}
\newcommand{\bsigma}{\boldsymbol{\sigma}}
\newcommand{\re}{\mathop{\mathrm{Re}}}
\newcommand{\im}{\mathop{\mathrm{Im}}}
\renewcommand{\b}[1]{{\boldsymbol{#1}}}
\newcommand{\diag}{\mathrm{diag}}
\newcommand{\sign}{\mathrm{sign}}
\newcommand{\sgn}{\mathop{\mathrm{sgn}}}

\newcommand{\cl}{\mathrm{cl}}
\newcommand{\mb}{\bm}
\newcommand{\ua}{\uparrow}
\newcommand{\da}{\downarrow}
\newcommand{\ra}{\rightarrow}
\newcommand{\la}{\leftarrow}
\newcommand{\mc}{\mathcal}
\newcommand{\bs}{\boldsymbol}
\newcommand{\lra}{\leftrightarrow}
\newcommand{\nn}{\nonumber}
\newcommand{\half}{{\textstyle{\frac{1}{2}}}}
\newcommand{\mf}{\mathfrak}
\newcommand{\MF}{\text{MF}}
\newcommand{\IR}{\text{IR}}
\newcommand{\UV}{\text{UV}}
\newcommand{\sech}{\mathrm{sech}}

\newcommand*{\vcenteredhbox}[1]{\begingroup
\setbox0=\hbox{#1}\parbox{\wd0}{\box0}\endgroup}
\newcommand{\picscalefactor}{0.5}

\title{Critical behavior of the QED$_3$-Gross-Neveu-Yukawa model at four loops}

\author{Nikolai Zerf}
\affiliation{Institut f\"ur Physik, Humboldt-Universit\"at zu Berlin, Newtonstra{\ss}e 15, D-12489 Berlin, Germany}

\author{Peter Marquard}
\affiliation{Deutsches Elektronen Synchrotron (DESY), Platanenallee 6, Zeuthen, Germany}

\author{Rufus Boyack}
\affiliation{Department of Physics, University of Alberta, Edmonton, Alberta T6G 2E1, Canada}
\affiliation{Theoretical Physics Institute, University of Alberta, Edmonton, Alberta T6G 2E1, Canada}

\author{Joseph Maciejko}
\affiliation{Department of Physics, University of Alberta, Edmonton, Alberta T6G 2E1, Canada}
\affiliation{Theoretical Physics Institute, University of Alberta, Edmonton, Alberta T6G 2E1, Canada}
\affiliation{Canadian Institute for Advanced Research, Toronto, Ontario M5G 1Z8, Canada}

\date\today

\begin{abstract}
We study the universal critical properties of the QED$_3$-Gross-Neveu-Yukawa model with $N$ flavors of four-component Dirac fermions coupled to a real scalar order parameter at four-loop order in the $\epsilon$ expansion below four dimensions. For $N=1$, the model is conjectured to be infrared dual to the $SU(2)$-symmetric noncompact $\mathbb{C}$P$^1$ model, which describes the deconfined quantum critical point of the N\'eel-valence-bond-solid transition of spin-1/2 quantum antiferromagnets on the two-dimensional square lattice. For $N=2$, the model describes a quantum phase transition between an algebraic spin liquid and a chiral spin liquid in the spin-1/2 kagom\'e antiferromagnet. For general $N$ we determine the order parameter anomalous dimension, the correlation length exponent, the stability critical exponent, as well as the scaling dimensions of $SU(N)$ singlet and adjoint fermion mass bilinears at the critical point. We use Pad\'e approximants to obtain estimates of critical properties in 2+1 dimensions.
\end{abstract}

\maketitle

\section{Introduction}
\label{sec:intro}

The study of critical phenomena represents one of the cornerstones of modern condensed matter physics, and the systematic understanding of such phenomena by renormalization group (RG) methods is widely acknowledged as one of the great triumphs of theoretical physics in the twentieth century~\cite{fisher1974}. The best-known examples of critical phenomena are phase transitions in systems with an $n$-component vector order parameter, such as Ising ($n=1$), XY ($n=2$), or Heisenberg ($n=3$) magnets, which are typically described by the Wilson-Fisher fixed point of the bosonic $O(n)$ vector model~\cite{wilson1972}. Critical exponents for this model have been determined in successive refinements over the years, culminating in the recent tour-de-force calculation of critical exponents at six-loop order~\cite{batkovich2016,kompaniets2017,schnetz2018}. Combined with Pad\'e or Borel resummation techniques, the $\epsilon$ expansion below the upper critical dimension of four is known to yield precise values for the critical exponents~\cite{guida1998}.

While the $O(n)$ vector model provides a satisfactory description of phase transitions obeying the Landau-Ginzburg-Wilson (LGW) paradigm, that is, transitions at which bosonic order parameter fluctuations are the only relevant long-wavelength degrees of freedom, much attention has been drawn in recent years to two classes of continuous quantum phase transitions at which the purely bosonic LGW approach fails. The first, fermionic quantum critical points, comprises phase transitions at which gapless fermionic degrees of freedom couple to order parameter fluctuations via a Yukawa-type coupling. In cases of interest this fermion-boson coupling is relevant at the purely bosonic (e.g., Wilson-Fisher) critical point and drives the system towards a new universality class with coexisting, and in many cases strongly coupled, bosonic and fermionic degrees of freedom. The prime example in this category is systems with Dirac fermion excitations at low energies, such as graphene or the surface of three-dimensional (3D) topological insulators, coupled with real~\cite{herbut2006,grover2014}, complex~\cite{lee2007,roy2013,nandkishore2013,grover2014,ponte2014,zhou2016,li2017b,classen2017}, or vectorial~\cite{sorella1992,herbut2006,honerkamp2008,sorella2012} order parameters. The corresponding critical points are described by the Gross-Neveu-Yukawa (GNY) model~\cite{zinn-justin1991}, which can be studied by perturbative RG in $d=4-\epsilon$ dimensions, or its purely fermionic equivalent, the Gross-Neveu (GN) model~\cite{gross1974}, which can be studied in $d=2+\epsilon$ dimensions. The critical exponents of the GNY model have recently been calculated at three-loop~\cite{mihaila2017} and four-loop~\cite{zerf2017} orders, and those of the GN model have been determined at four-loop order~\cite{gracey2016}. Interesting critical phenomena outside the reach of the purely bosonic $O(n)$ vector model include the emergence of $\mathcal{N}=1$~\cite{grover2014,sonoda2011} and $\mathcal{N}=2$~\cite{lee2007,grover2014,ponte2014,zerf2016,fei2016} spacetime supersymmetry in the real (chiral Ising) and complex (chiral XY) GNY universality classes, respectively.

The second category of critical phenomena not captured by the standard $O(n)$ universality classes are phase transitions involving dynamical gauge fields. In the condensed matter context these occur as a result of the fractionalization of microscopic degrees of freedom, under the influence of strong correlations, into slave particles with fractional quantum numbers. The paradigmatic example is the fractionalization of bosonic local moments into neutral fermionic spinons~\cite{WenBook}. One may further distinguish two subclasses of critical points in this category: those for which the gauge field deconfines only at the critical point itself, dubbed deconfined quantum critical points~\cite{senthil2004,senthil2004b}, and those for which the gauge field deconfines in at least one of the two phases separated by the critical point. While the former subclass corresponds to (LGW-forbidden) transitions between conventional ordered phases, the latter describes transitions involving at least one fractionalized phase. When fermionic spinons acquire a Dirac dispersion~\cite{affleck1988}, one obtains a theory of Dirac fermions interacting with a $U(1)$ gauge field as well as with a bosonic order parameter. The transitions of interest take place in 2+1 dimensions, and thus we will refer to such models as QED$_3$-GNY models, since the fermion-gauge sector is described by massless quantum electrodynamics (QED). Examples of transitions recently studied in this way include transitions from an algebraic spin liquid to either a chiral spin liquid~\cite{he2015,janssen2017}, a $\mathbb{Z}_2$ spin liquid~\cite{boyack2018}, or a N{\'e}el antiferromagnet~\cite{xu2018}, which are described by the chiral Ising, XY, or Heisenberg QED$_3$-GNY models, respectively. Interestingly, it was recently conjectured~\cite{wang2017} that a critical point in the first subclass, the deconfined quantum critical point between a N\'eel antiferromagnet and a valence bond solid on the 2D square lattice~\cite{senthil2004,senthil2004b}, is dual to a critical point in the second subclass, that of the chiral Ising QED$_3$-GNY model with two flavors of two-component Dirac fermions~\cite{janssen2017}. By varying the number of flavors of Dirac fermions in the theory, one may obtain an infinite family of new universality classes, distinct from both the purely bosonic $O(n)$ and GN/GNY universality classes.

Motivated by these recent developments, in this paper we study the critical properties of the chiral Ising QED$_3$-GNY model as a function of the number $N$ of flavors of four-component Dirac fermions. The study of the critical properties of this model in $d=4-\epsilon$ dimensions via the $\epsilon$ expansion was initiated in Ref.~\cite{janssen2017}, where calculations at leading (one-loop) order were performed; here we study this model up to four-loop order. The paper is structured as follows. In Sec.~\ref{sec:model} we define the model. In Sec.~\ref{sec:RG} we discuss basic aspects of the RG procedure and give our results for the beta functions and anomalous dimensions. Results up to three-loop order are given in the main text; four-loop contributions are given separately in Appendices~\ref{app:4Lbeta} and \ref{app:4Lgamma}. In Sec.~\ref{sec:exponents} and \ref{sec:BilinearDim} we present our $\epsilon$-expansion results up to $\mathcal{O}(\epsilon^4)$ for the usual thermodynamic critical exponents as well as the scaling dimensions of certain fermion bilinear operators; Pad\'e approximants are then used to obtain rough estimates in $d=3$. The procedure for the calculation of the stability critical exponent $\omega$ is briefly explained in Appendix~\ref{app:omega}. In addition to our results for the chiral Ising QED$_3$-GNY model, we also compute the scaling dimension of fermion bilinears at the pure QED$_3$ (see Appendix~\ref{app:5LDeltaQED3}) and GNY fixed points. In Sec.~\ref{sec:discussion} we discuss the application of our results to the duality mentioned above. In Sec.~\ref{sec:technical} we discuss technical aspects of the automated procedure employed for the determination of renormalization constants at four-loop order. A brief conclusion is presented in Sec.~\ref{sec:conclusion}.

\section{Model}
\label{sec:model}

We study the chiral Ising QED$_3$-GNY model with $N$ flavors of four-component Dirac fermions $\Psi_i$, $i=1,\ldots,N$, with Lagrangian given by
\begin{align}\label{L}
\mathcal{L}=&\sum_{i=1}^{N}\bar{\Psi}_i\gamma_\mu(\partial_\mu-ieA_\mu)\Psi_i+\frac{1}{4}F_{\mu\nu}^2+\frac{1}{2\xi}(\partial_\mu A_\mu)^2\nn\\
&+\frac{1}{2}(\partial_\mu\phi)^2+\frac{1}{2}m^2\phi^2+\lambda^2\phi^4+g\phi\sum_{i=1}^{N}\bar{\Psi}_i\Psi_i,
\end{align}
where the $\gamma_\mu$ are $4\times 4$ Euclidean gamma matrices, $F_{\mu\nu}=\partial_\mu A_\nu-\partial_\nu A_\mu$ is the field-strength tensor, $\xi$ is a gauge-fixing parameter, and $\phi$ is a real scalar field. In the rest of the paper we will simply refer to this model as the QED$_3$-GNY model. The model has a global discrete $\mathbb{Z}_2$ chiral symmetry,
\begin{align}\label{Z2chiralsymm}
\Psi_i\rightarrow\gamma_5\Psi_i,\hspace{5mm}
\bar{\Psi}_i=\Psi_i^\dag\gamma_0\rightarrow-\bar{\Psi}_i\gamma_5,\hspace{5mm}
\phi\rightarrow-\phi,
\end{align}
where $\gamma_5^2=1$ and $\{\gamma_5,\gamma_\mu\}=0$, under which the fermion mass bilinear $\bar{\Psi}\Psi\equiv\sum_{i=1}^N\bar{\Psi}_i\Psi_i$ changes sign. The scalar mass squared $m^2$ tunes a quantum phase transition from a symmetric phase ($m^2>0$) with massive scalars and massless fermions, described for momenta $p^2\ll m^2$ by pure massless QED, to a phase with spontaneously broken $\mathbb{Z}_2$ symmetry ($m^2<0$) where the scalar acquires a vacuum expectation value and a fermion mass is dynamically generated. 

As mentioned in Sec.~\ref{sec:intro}, when extrapolated to $d=3$ dimensions this model has been argued to be relevant to two problems of current interest in quantum magnetism. For $N=1$, it has been suggested as a possible fermionic dual~\cite{wang2017} to the bosonic $SU(2)$-symmetric noncompact $\mathbb{C}$P$^1$ (NC$\mathbb{C}$P$^1$) model, which describes a deconfined quantum critical point between a N\'eel antiferromagnet and a valence bond solid on the 2D square lattice~\cite{senthil2004,senthil2004b}. For $N=2$, it describes a putative time-reversal breaking quantum phase transition between an algebraic spin liquid and a chiral spin liquid in a spin-1/2 kagom\'e antiferromagnet~\cite{he2015,he2015b}.

\section{Renormalization group analysis}
\label{sec:RG}

To perform an RG analysis of the model (\ref{L}) we use the standard field-theoretic approach with dimensional regularization and modified minimal subtraction ($\overline{\text{MS}}$)~\cite{ZJ}. Comparing the bare Lagrangian
\begin{align}
\mathcal{L}_0=&\sum_{i=1}^{N}\bar{\Psi}_i^0\gamma_\mu(\partial_\mu-ie_0A_\mu^0)\Psi_i^0+\frac{1}{4}\left(F_{\mu\nu}^0\right)^2+\frac{1}{2\xi_0}(\partial_\mu A_\mu^0)^2\nn\\
&+\frac{1}{2}(\partial_\mu\phi_0)^2+\frac{1}{2}m^2_0\phi^2_0+\lambda^2_0\phi^4_0+g_0\phi_0\sum_{i=1}^{N}\bar{\Psi}_i^0\Psi_i^0,
\end{align}
and the renormalized Lagrangian
\begin{align}
\mathcal{L}_R=&\sum_{i=1}^{N}Z_\Psi\bar{\Psi}_i\gamma_\mu(\partial_\mu-ie\mu^{\epsilon/2}A_\mu)\Psi_i+\frac{1}{4}Z_AF_{\mu\nu}^2\nn\\
&+\frac{1}{2\xi}(\partial_\mu A_\mu)^2+\frac{1}{2}Z_\phi(\partial_\mu\phi)^2+\frac{1}{2}Z_{\phi^2}m^2\mu^2\phi^2\nn\\
&+Z_{\lambda^2}\lambda^2\mu^\epsilon\phi^4+Z_gg\mu^{\epsilon/2}\phi\sum_{i=1}^{N}\bar{\Psi}_i\Psi_i,
\end{align}
where $\mu$ is a renormalization scale, we find that the bare and renormalized fields are related by
\begin{align}
\Psi_i^0=\sqrt{Z_\Psi}\Psi_i,\hspace{5mm}
\phi_0=\sqrt{Z_\phi}\phi,\hspace{5mm}
A_\mu^0=\sqrt{Z_A}A_\mu,
\end{align}
implying that the bare and (dimensionless) renormalized couplings are related by
\begin{align}\label{couplings}
e^2&=e_0^2\mu^{-\epsilon}Z_A,\\
g^2&=g_0^2\mu^{-\epsilon}Z_\Psi^2 Z_\phi Z_g^{-2},\\
\lambda^2&=\lambda_0^2\mu^{-\epsilon}Z_\phi^2 Z_{\lambda^2}^{-1}.
\end{align}
The quantum critical point is found by tuning the renormalized scalar mass squared $m^2$ to zero. To calculate the correlation length exponent, however, one must determine the RG eigenvalue of the scalar mass squared at the critical point, for which we need the relation between bare and renormalized masses,
\begin{align}\label{betam2}
m^2=m_0^2\mu^{-2}Z_\phi Z_{\phi^2}^{-1}.
\end{align}
Finally, one must also track the flow of the gauge-fixing parameter $\xi$, which for the choice of gauge-fixing term in Eq.~(\ref{L}) is controlled by the relation
\begin{align}\label{GaugeFixingRen}
\xi=\xi_0 Z_A^{-1}.
\end{align}
The calculation of the renormalization constants $Z_X$, $X\in\{\Psi,\phi,\phi^2,A,g,\lambda^2\}$ at four-loop order is done using an automated setup; the main technical steps of the computation are explained in Sec.~\ref{sec:technical}.

\subsection{Beta functions}
\label{sec:beta}

The beta functions are defined as
\begin{align}
\beta_{e^2}=\frac{de^2}{d\ln\mu},\hspace{5mm}
\beta_{g^2}=\frac{dg^2}{d\ln\mu},\hspace{5mm}
\beta_{\lambda^2}=\frac{d\lambda^2}{d\ln\mu},
\end{align}
and we work with rescaled couplings $\alpha^2/(4\pi)^2\rightarrow\alpha^2$ for $\alpha=e,g,\lambda$. Using Eq.~(\ref{couplings}) and the fact that the bare couplings are independent of $\mu$, we have
\begin{align}
\beta_{e^2}&=(-\epsilon+\gamma_A)e^2,\label{betae2}\\
\beta_{g^2}&=(-\epsilon+2\gamma_\Psi+\gamma_\phi-2\gamma_g)g^2,\\
\beta_{\lambda^2}&=(-\epsilon+2\gamma_\phi-\gamma_{\lambda^2})\lambda^2,
\end{align}
where we define the anomalous dimension
\begin{align}
\gamma_X=\frac{d\ln Z_X}{d\ln\mu},
\end{align}
associated to the renormalization constant $Z_X$. We express the four-loop beta functions as
\begin{align}
\beta_{e^2}&=-\epsilon e^2+\beta^\text{(1L)}_{e^2}
+\beta^\text{(2L)}_{e^2}+\beta^\text{(3L)}_{e^2}+\beta^\text{(4L)}_{e^2},\label{betae24L}\\
\beta_{g^2}&=-\epsilon g^2+\beta^\text{(1L)}_{g^2}
+\beta^\text{(2L)}_{g^2}+\beta^\text{(3L)}_{g^2}+\beta^\text{(4L)}_{g^2},\label{betag24L}\\
\beta_{\lambda^2}&=-\epsilon \lambda^2+\beta^\text{(1L)}_{\lambda^2}
+\beta^\text{(2L)}_{\lambda^2}+\beta^\text{(3L)}_{\lambda^2}+\beta^\text{(4L)}_{\lambda^2}.\label{betal24L}
\end{align}
Here we display our result up to and including three-loop order; the four-loop contributions are lengthy and given in Appendix~\ref{app:4Lbeta} and also in Ref.~\cite{SuppMat}. The beta function $\beta_{e^2}$ for the gauge coupling is given by
\begin{align}
\beta^\text{(1L)}_{e^2}&=\frac{8N}{3}e^4,\\
\beta^\text{(2L)}_{e^2}&=8Ne^6-4Ne^4g^2,\\
\beta^\text{(3L)}_{e^2}&=-6Ne^6g^2+2N(7N+6)e^4g^4-\frac{4N}{9}(22N+9)e^8.
\end{align}
Likewise, the beta function $\beta_{g^2}$ for the Yukawa coupling is specified by
\begin{align}
\beta^\text{(1L)}_{g^2}&=-12e^2g^2+2(2N+3)g^4,\\
\beta^\text{(2L)}_{g^2}&=-\left(24N+\frac{9}{2}\right)g^6+\left(\frac{40N}{3}-6\right)e^4g^2\nn\\
&\phantom{=}+4(5N+12)e^2g^4-96g^4\lambda^2+96g^2\lambda^4,\\
\beta^\text{(3L)}_{g^2}&=\left[-32N^2+N(49-432\zeta_3)+\frac{327}{2}-504\zeta_3\right]e^4g^4\nn\\
&\phantom{=}+\left[\frac{560N^2}{27}+8N(23-24\zeta_3)-258\right]e^6g^2\nn\\
&\phantom{=}+\left[28N^2+N\left(\frac{67}{4}+108\zeta_3\right)-\frac{697}{8}+114\zeta_3\right]g^8\nn\\
&\phantom{=}+144(5N+7)g^6\lambda^2+24(-30N+91)g^4\lambda^4\nn\\
&\phantom{=}-1728g^2\lambda^6+[2N(-79+48\zeta_3)-348+72\zeta_3]e^2g^6\nn\\
&\phantom{=}+96e^2g^4\lambda^2,
\end{align}
where $\zeta_z$ is the Riemann zeta function, with $\zeta_3=1.202...$ Ap\'ery's constant. Finally, the contributions to the beta function $\beta_{\lambda^2}$ for the quartic scalar coupling are
\begin{align}
\beta^\text{(1L)}_{\lambda^2}&=-2Ng^4+8Ng^2\lambda^2+72\lambda^4,\\
\beta^\text{(2L)}_{\lambda^2}&=16Ng^6+28Ng^4\lambda^2-288Ng^2\lambda^4-3264\lambda^6\nn\\
&\phantom{=}-8Ne^2g^4+40Ne^2g^2\lambda^2,
\\
\beta^\text{(3L)}_{\lambda^2}&=-\frac{N}{4}(628N-5+384\zeta_3)g^8\nn\\
&\phantom{=}+\frac{N}{2}(1736N-4395-1872\zeta_3)g^6\lambda^2\nn\\
&\phantom{=}+12N(-72N+361+648\zeta_3)g^4\lambda^4\nn\\
&\phantom{=}+12384Ng^2\lambda^6+1728(145+96\zeta_3)\lambda^8\nn\\
&\phantom{=}+N(116N+131-96\zeta_3)e^4g^4\nn\\
&\phantom{=}-2N(32N+119-144\zeta_3)e^4g^2\lambda^2\nn\\
&\phantom{=}+2N(-11+96\zeta_3)e^2g^6+6N(217-304\zeta_3)e^2g^4\lambda^2\nn\\
&\phantom{=}+216N(-17+16\zeta_3)e^2g^2\lambda^4.
\end{align}

The beta functions (\ref{betae24L})-(\ref{betal24L}) can be checked against known results in various limits. Setting $e=0$ and $g=0$, the model reduces to the bosonic Ising universality class; our result for $\beta_{\lambda^2}$ in that limit agrees with the known four-loop result~\cite{vladimirov1979}. Setting $g=0$ and $\lambda=0$, $\beta_{e^2}$ reproduces the four-loop QED beta function~\cite{gorishny1991}. Setting $e=0$ only, our expressions for $\beta_{g^2}$ and $\beta_{\lambda^2}$ agree with those for the pure GNY model in the chiral Ising class, which were recently computed at four-loop order~\cite{zerf2017}. Finally, for the full QED$_3$-GNY theory with all three couplings nonzero we recover the one-loop beta functions recently obtained in Ref.~\cite{janssen2017}.

\subsection{Anomalous dimensions}
\label{sec:gamma}

The anomalous dimensions of the fields $\phi$, $\phi^2$, $A_\mu$, evaluated at the quantum critical point,
\begin{align}\label{gammas}
\eta_\phi&\equiv\gamma_\phi(e_*^2,g_*^2,\lambda_*^2),\\
\eta_{\phi^2}&\equiv\gamma_{\phi^2}(e_*^2,g_*^2,\lambda_*^2),\\
\eta_A&\equiv\gamma_A(e_*^2,g_*^2,\lambda_*^2),
\end{align}
are universal, gauge-invariant quantities. Considering $\eta_A$ first, from Eq.~(\ref{betae2}) we see that at a fixed point with nonzero gauge coupling $e_*^2\neq 0$ one must have the exact relation $\eta_A=\epsilon$~\cite{IgorBook}. This is a consequence of gauge invariance, since Eq.~(\ref{betae2}) follows from the fact that the gauge coupling and gauge-field wave function renormalizations are related by a Ward identity. As shown at one-loop order in Ref.~\cite{janssen2017}, and confirmed at four-loop order in Sec.~\ref{sec:exponents}, the QED$_3$-GNY critical point indeed has $e_*^2\neq 0$, implying that $\eta_A=1$ exactly in $d=3$ dimensions. In this section we give expressions at four-loop order for $\gamma_\phi$ and $\gamma_{\phi^2}$, which will then be evaluated at the quantum critical point in Sec.~\ref{sec:exponents} to yield the universal exponents $\eta_\phi$ and $\eta_{\phi^2}$. As for the beta functions, we express the anomalous dimensions as a sum of contributions at fixed loop order,
\begin{align}
\gamma_\phi&=\gamma_\phi^\text{(1L)}
+\gamma_\phi^\text{(2L)}+\gamma_\phi^\text{(3L)}+\gamma_\phi^\text{(4L)},\label{gammaphi4L}\\
\gamma_{\phi^2}&=\gamma_{\phi^2}^\text{(1L)}
+\gamma_{\phi^2}^\text{(2L)}+\gamma_{\phi^2}^\text{(3L)}+\gamma_{\phi^2}^\text{(4L)}.\label{gammaphi24L}
\end{align}

To calculate the anomalous dimensions, we make use of the chain rule when taking the derivative with respect to $\ln\mu$, as well as of the fact that the renormalization constants $Z_\phi$ and $Z_{\phi^2}$ have no $\xi$ dependence since the associated fields are gauge invariant~\cite{ZJ},
\begin{align}
\gamma_X=\frac{1}{Z_X}\sum_{\alpha=e,g,\lambda}\frac{\partial Z_X}{\partial\alpha^2}\beta_{\alpha^2},
\end{align}
for $X\in\{\phi,\phi^2\}$. As for the beta functions, here we list the contributions only up to three-loop order and provide the four-loop contributions in Appendix~\ref{app:4Lgamma} and Ref.~\cite{SuppMat}. The anomalous dimension of the scalar field $\phi$ is given by
\begin{align}
\gamma_\phi^\text{(1L)}&=4Ng^2,\\
\gamma_\phi^\text{(2L)}&=20Ne^2g^2-10Ng^4+96\lambda^4,\\
\gamma_\phi^\text{(3L)}&=-3N(7+16\zeta_3)e^2g^4+\frac{N}{4}(200N+21+48\zeta_3)g^6\nn\\
&\phantom{=}+N(-32N-119+144\zeta_3)e^4g^2+240Ng^4\lambda^2\nn\\
&\phantom{=}-720Ng^2\lambda^4-1728\lambda^6,
\end{align}
while for the scalar mass operator $\phi^2$, we find
\begin{align}
\gamma_{\phi^2}^\text{(1L)}&=-24\lambda^2,\\
\gamma_{\phi^2}^\text{(2L)}&=-8Ng^4+96Ng^2\lambda^2+576\lambda^4,\\
\gamma_{\phi^2}^\text{(3L)}&=-32N(4N-9+3\zeta_3)g^6\nn\\
&\phantom{=}+12N(24N-11-120\zeta_3)g^4\lambda^2-2304Ng^2\lambda^4\nn\\
&\phantom{=}-50112\lambda^6+32N(-7+9\zeta_3)e^2g^4\nn\\
&\phantom{=}+72N(17-16\zeta_3)e^2g^2\lambda^2.
\end{align}

Our expressions for $\gamma_\phi$ and $\gamma_{\phi^2}$ can be checked in two limits. Setting $e=0$ and $g=0$, our results agree at four-loop order with those for the Ising universality class~\cite{vladimirov1979}. Setting $e=0$ only, our results agree at that same order with those for the chiral Ising GNY model~\cite{zerf2017}.

\subsection{Fermion bilinears}
\label{sec:FermionBi}

Besides $\phi$ and $\phi^2$, another class of gauge-invariant operators one can consider are fermion bilinears. Restricting ourselves to Lorentz scalars, i.e., mass terms, a generic fermion bilinear can be expressed as a linear combination of an $SU(N)$ flavor-singlet mass $\bar{\Psi}\Psi$, which appears in the Yukawa interaction in Eq.~(\ref{L}), and an $SU(N)$ flavor-adjoint mass
\begin{align}
\bar{\Psi}T_A\Psi\equiv\sum_{i,j=1}^N\bar{\Psi}_iT_A^{ij}\Psi_j,\hspace{5mm}A=1,\ldots,N^2-1,
\end{align}
where the generators $T_A$ of $SU(N)$ are linearly independent traceless Hermitian $N\times N$ matrices. The scaling dimensions of the singlet and adjoint bilinears are in general different. To calculate the scaling dimension $\Delta_{\bar{\Psi}\Gamma\Psi}$ of a fermion bilinear $\bar{\Psi}\Gamma\Psi$ where $\Gamma\in\{1,T_A\}$, we add it to the bare and renormalized Lagrangians,
\begin{align}\label{insertion}
\delta\mathcal{L}_0=\hat{M}_0\bar{\Psi}^0\Gamma\Psi^0,\hspace{5mm}
\delta\mathcal{L}_R=Z_{\hat{M}}\hat{M}\mu\bar{\Psi}\Gamma\Psi,
\end{align}
where we use the shorthand $\hat{M}=M_{\Gamma} \in\{ M_1,M_{T_A}\}\equiv \{M,\tilde{M}\}$.
This implies the relations
\begin{align}
\hat{M}=\hat{M}_{0}\mu^{-1}Z_{\hat{M}}^{-1}Z_\Psi ,
\end{align}
and thus the beta functions
\begin{align}\label{betaM}
\beta_{\hat{M}}=\frac{d\hat{M}}{d\ln\mu}=(-1-\gamma_{\hat{M}}+\gamma_\Psi)\hat{M},
\end{align}
where $\gamma_{\hat{M}}=d\ln Z_{\hat{M}}/d\ln\mu$. Note that $\gamma_\Psi$ and $\gamma_{\hat{M}}$ are not separately gauge invariant, i.e., they depend on the gauge-fixing parameter $\xi$, but all the gauge-dependent terms must cancel out in Eq.~(\ref{betaM}), since the fermion bilinears are gauge-invariant operators. Taking into account its gauge dependence, the fermion field anomalous dimension is given by
\begin{align}
\gamma_\Psi=\frac{1}{Z_\Psi}\sum_{\alpha=e,g,\lambda}\left(\frac{\partial Z_\Psi}{\partial\alpha^2}-\xi\frac{\partial Z_\Psi}{\partial\xi}\frac{\partial\ln Z_A}{\partial\alpha^2}\right)\beta_{\alpha^2},
\end{align}
where we have used Eq.~(\ref{GaugeFixingRen}) to express $d\xi/d\ln\mu=-\gamma_A\xi$. We find that up to four-loop order $\gamma_\Psi$ depends on $\xi$ only	in the one-loop term, as in pure QED~\cite{gracey2007,kisler2017}.

To calculate $Z_{\hat{M}}$ we compute the fermion two-point function at four-loop order with all possible single fermion bilinear and fermion bilinear counterterm insertions. 
Flavor-adjoint bilinear insertions preserve the $\mathbb{Z}_2$ chiral symmetry of the massless theory (for a proof of this statement, see Appendix~\ref{app:GenChiralZ2}). In this case the scaling dimension of the bilinear is simply determined by the slope of the beta function (\ref{betaM}) evaluated at the $\tilde{M}=0$ fixed point,\begin{align}\label{AdjMassScalingDim}
d-\Delta_{\bar{\Psi}T_A\Psi}=1+\eta_{\tilde{M}}-\eta_\Psi,
\end{align}
where $\eta_\Psi$ and $\eta_{\tilde{M}}$ are the anomalous dimensions of the fermion field and adjoint bilinear evaluated at the quantum critical point,
\begin{align}
\eta_{\tilde{M}}&\equiv\gamma_{\tilde{M}}(e_*^2,g_*^2,\lambda_*^2),\label{etaM}\\
\eta_\Psi&\equiv\gamma_\Psi(e_*^2,g_*^2,\lambda_*^2).\label{etaPsi}
\end{align}
Accounting for the $\xi$ dependence of $\gamma_{\tilde{M}}$, one has
\begin{align}\label{gammaMeq}
\gamma_{\tilde{M}}=\frac{1}{Z_{\tilde{M}}}\sum_{\alpha=e,g,\lambda}\left(\frac{\partial Z_{\tilde{M}}}{\partial\alpha^2}-\xi\frac{\partial Z_{\tilde{M}}}{\partial\xi}\frac{\partial\ln Z_A}{\partial\alpha^2}\right)\beta_{\alpha^2}.
\end{align}
At four-loop order, we obtain
\begin{align}\label{gammaM4L}
\gamma_{\tilde{M}}-\gamma_\Psi\equiv\gamma_{\bar{\Psi}T_A\Psi}=\gamma^\text{(1L)}_{\bar{\Psi}T_A\Psi}+\gamma^\text{(2L)}_{\bar{\Psi}T_A\Psi}+\gamma^\text{(3L)}_{\bar{\Psi}T_A\Psi}
+\gamma^\text{(4L)}_{\bar{\Psi}T_A\Psi},
\end{align}
with
\begin{align}
\gamma_{\bar{\Psi}T_A\Psi}^\text{(1L)}&=6e^2-3g^2,\label{gammatilde1L}\\
\gamma_{\bar{\Psi}T_A\Psi}^\text{(2L)}&=-\left(\frac{20 N}{3}-3\right)e^4-24e^2g^2+\left(\frac{28N+9}{4}\right)g^4,\\
\gamma^\text{(3L)}_{\bar{\Psi}T_A\Psi}&=\left(11N^2-\frac{151N}{4}+\frac{697}{16}-57\zeta_3\right)g^6\nn\\
&\phantom{=}+\frac{1}{2}[N(137-144\zeta_3)+348-72\zeta_3]e^2g^4\nn\\
&\phantom{=}+\frac{3}{4}(80N-109+336\zeta_3)e^4g^2\nn\\
&\phantom{=}+\left(-\frac{280N^2}{27}+N(-92+96\zeta_3)+129\right)e^6\nn\\
&\phantom{=}-240g^4\lambda^2+348g^2\lambda^4.\label{gammatildeTA3L}
\end{align}
The four-loop contribution $\gamma^\text{(4L)}_{\bar{\Psi}T_A\Psi}$ is given in Eq.~(\ref{gammatilde4LTA}) of Appendix~\ref{app:4Lgamma}. 
The absence of gauge dependence up to four-loop order is an additional consistency check on the calculation.
Furthermore, we checked that our result for $\gamma_{\bar{\Psi}T_A\Psi}$ agrees with the results available in the literature in the pure QED limit~\cite{Chetyrkin:1997dh,Vermaseren:1997fq,Chetyrkin:2004mf,Czakon:2004bu}, using the fact (discussed in Sec.~\ref{sec:cQED3}) that in the pure QED limit the singlet and adjoint bilinear scaling dimensions are identical.

\begin{figure}[t]
\includegraphics[width=0.8\columnwidth]{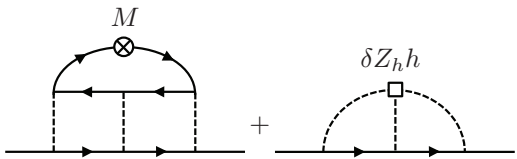}
\caption{A subdivergence coming from the fermion loop with an $SU(N)$ singlet mass insertion $M$ in the first diagram is cancelled by a counterterm insertion $\delta Z_hh$ of the cubic scalar vertex, where $\delta Z_h=Z_h-1$.}
\label{fig:3loopdiv}
\end{figure}

By contrast with adjoint bilinear insertions, singlet bilinear insertions explicitly break the $\mathbb{Z}_2$ chiral symmetry and thus symmetry-breaking interactions will be radiatively induced.
The only relevant (or marginal) such interaction below four dimensions is a $\phi^3$ interaction, which must be kept to preserve renormalizability of the theory. We must therefore additionally include the bare $h_0\phi_0^3$ and renormalized $Z_hh\mu^{1+\epsilon/2}\phi^3$ couplings in the Lagrangian, 
with $Z_h$ a new renormalization constant. This implies the additional relation
\begin{align}
h=h_0\mu^{-1-\epsilon/2}Z_h^{-1}Z_\phi^{3/2},
\end{align}
and the corresponding beta function,
\begin{align}
\beta_h=\frac{dh}{d\ln\mu}=\left(-1-\frac{\epsilon}{2}-\gamma_h+\frac{3}{2}\gamma_\phi\right)h,
\end{align}
with $\gamma_h=d\ln Z_h/d\ln\mu$. Note that one has to introduce this extra coupling in order to obtain a finite/local result for $\gamma_M$ starting at three-loop order.
This is because radiative corrections to the cubic scalar vertex arise for the first time
in the fermion two-point function with single mass insertions in three-loop diagrams (first diagram of Fig.~\ref{fig:3loopdiv}). To calculate $\gamma_M$ we use an equation analogous to Eq.~(\ref{gammaMeq}), but the sum over $\alpha^2$ must additionally include the couplings $M$ and $h$.

In fact, as soon as a flavor-singlet fermion bilinear insertion is present, 
the theory is already nonrenormalizable at one-loop order without a cubic scalar vertex,
because the scalar three-point function becomes divergent through fermion loop diagrams containing a single flavor-singlet bilinear insertion.
Exactly these diagrams reappear as subdiagrams at three-loop order (first diagram in Fig.~\ref{fig:3loopdiv})
and carry a subdivergence which renders the corresponding mass renormalization constant nonlocal (containing logarithms of $\mu$),
if one does not subtract their subdivergence via a corresponding $\phi^3$ vertex counterterm insertion (second diagram in Fig.~\ref{fig:3loopdiv}).

Moreover, even the scalar one-point function becomes divergent at the same loop order, when one allows for a $\phi^2$ operator insertion.
So one has to introduce a tadpole counterterm  $\sim t_0\phi_0$ into the Lagrangian in order to be able to render the one-point function finite.
In our case we renormalize the parameter $t$ in a full subtraction scheme, meaning we have $t=0$ to all orders.
This means effectively we do not need to consider any diagrams containing a $\phi$-tadpole insertion,
because for each such diagram there is a corresponding counterterm diagram which exactly cancels its contribution.

Since we are interested in the scaling dimension of $\bar{\Psi}\Psi$ in the massless, symmetric theory, it is sufficient to calculate $Z_M$ and $Z_h$ up to linear order in $M$ and $h$. 
Note that $Z_M$ will contain the singular ratio $h/M$ as a $\mathbb{Z}_2$ symmetry-breaking mass term can be radiatively generated at two-loop order by the cubic scalar vertex (i.e., the second diagram of Fig.~\ref{fig:3loopdiv} but with the counterterm insertion replaced by the cubic scalar vertex $h$ itself). 
Likewise, $Z_h$ will contain the ratio $M/h$ as the cubic scalar vertex can be radiatively generated at one-loop order by a closed fermion loop with a single $SU(N)$ singlet mass insertion and three external scalar legs (i.e., the fermion loop subdiagram in the first diagram of Fig.~\ref{fig:3loopdiv}). These singular ratios lead to terms linear in $h$ in $\beta_M$ and terms linear in $M$ in $\beta_h$, i.e., mixing between the operators $\bar{\Psi}\Psi$ and $\phi^3$. To linear order one thus obtains the linear system
\begin{align}
\beta_M&=K_{11}M+K_{12}h,\\
\beta_h&=K_{21}M+K_{22}h.
\end{align}
As in Sec.~\ref{sec:beta} we find full cancellation of the gauge dependence in those beta functions~\cite{SuppMat}, which is a strong consistency check of the calculation. The scaling dimensions of the new eigenoperators are given by $\Delta_\pm=d+\Lambda_\pm$ where
\begin{align}
\Lambda_\pm=\frac{K_{11}+K_{22}}{2}\pm\sqrt{\left(\frac{K_{11}-K_{22}}{2}\right)^2+K_{12}K_{21}},
\end{align}
are the eigenvalues of the matrix $K$, evaluated at the quantum critical point. By inspecting the corresponding eigenvectors we find that $\Delta_-$ can be associated with $\Delta_{\bar{\Psi}\Psi}$, and likewise $\Delta_+=\Delta_{\phi^3}$.

\section{Critical exponents}
\label{sec:exponents}

From the knowledge of the beta functions [Eq.~(\ref{betae24L})-(\ref{betal24L})] and anomalous dimensions [Eq.~(\ref{gammaphi4L})-(\ref{gammaphi24L})] one can calculate the usual critical exponents. We begin by searching for fixed points with couplings $(e_*^2,g_*^2,\lambda_*^2)$ at one-loop order~\cite{janssen2017}. At that order one finds eight fixed points: the Gaussian fixed point $(0,0,0)$, the conformal QED~\cite{dipietro2016} fixed point $(\frac{3\epsilon}{8N},0,0)$, the Wilson-Fisher fixed point $(0,0,\frac{\epsilon}{72})$, a conformal QED $\times$ Wilson-Fisher fixed point $(\frac{3\epsilon}{8N},0,\frac{\epsilon}{72})$, two GNY-type fixed points with $e^2_*=0$ and $g^2_*\neq 0$, $\lambda^2_*\neq 0$, and two fixed points with all three couplings nonzero. In agreement with Ref.~\cite{janssen2017}, of all those fixed points only one of the latter two is stable, the so-called QED$_3$-GNY fixed point:
\begin{align}
e_*^2&=\frac{3}{8N}\epsilon+\mathcal{O}(\epsilon^2)\label{QED3GNY1Le2},\\
g_*^2&=\frac{2N+9}{4N(2N+3)}\epsilon+\mathcal{O}(\epsilon^2),\\
\lambda_*^2&=\frac{-2N-15+X}{144N(2N+3)}\epsilon+\mathcal{O}(\epsilon^2),\label{QED3GNY1Ll2}
\end{align}
defining
\begin{align}\label{defX}
X\equiv\sqrt{4N^4+204N^3+1521N^2+2916N}.
\end{align}
Furthermore, all three couplings are positive for all $N$. In the following we study this fixed point at four-loop order, looking for a zero of the beta functions in the form
\begin{align}\label{4LFP}
e_*^2=\sum_{n=1}^4e_n\epsilon^n,\hspace{5mm}
g_*^2=\sum_{n=1}^4g_n\epsilon^n,\hspace{5mm}
\lambda_*^2=\sum_{n=1}^4\lambda_n\epsilon^n,
\end{align}
with the one-loop coefficients $e_1,g_1,\lambda_1$ given in Eqs.~(\ref{QED3GNY1Le2}-\ref{QED3GNY1Ll2}).

Besides the previously determined exponent $\eta_A$, the critical exponents we compute here are the scalar field anomalous dimension $\eta_\phi$, the inverse correlation length exponent $\nu^{-1}$, and the stability critical exponent $\omega$. The exponent $\nu^{-1}$ is defined as the RG eigenvalue associated with the (relevant) scalar mass term,
\begin{align}
\left.\frac{dm^2}{d\ln\mu}\right|_{(e_*^2,g_*^2,\lambda_*^2)}=-\nu^{-1}m^2.
\end{align}
From Eq.~(\ref{betam2}) and Eq.~(\ref{gammas}), one obtains~\cite{ZJ}
\begin{align}
\nu^{-1}=2+\eta_{\phi^2}-\eta_\phi.
\end{align}
The exponent $\omega$ is defined as the RG eigenvalue associated with the least irrelevant operator in the basin of attraction of the fixed point (i.e., the critical hypersurface $m^2=0$), and controls the leading corrections to scaling. In practical terms, it is given by the smallest eigenvalue of the Jacobian (stability) matrix
\begin{align}\label{M}
{\cal J}=\left.\left(\begin{array}{ccc}
\displaystyle\frac{\partial\beta_{e^2}}{\partial e^2} & \displaystyle\frac{\partial\beta_{e^2}}{\partial g^2} & \displaystyle\frac{\partial\beta_{e^2}}{\partial \lambda^2} \\
\displaystyle\frac{\partial\beta_{g^2}}{\partial e^2} & \displaystyle\frac{\partial\beta_{g^2}}{\partial g^2} & \displaystyle\frac{\partial\beta_{g^2}}{\partial \lambda^2} \\
\displaystyle\frac{\partial\beta_{\lambda^2}}{\partial e^2} & \displaystyle\frac{\partial\beta_{\lambda^2}}{\partial g^2} & \displaystyle\frac{\partial\beta_{\lambda^2}}{\partial \lambda^2}
\end{array}\right)\right|_{(e_*^2,g_*^2,\lambda_*^2)}\,.
\end{align}
The approach utilized to diagonalize $\cal{J}$ order by order in $\epsilon$ is tantamount to ordinary quantum-mechanical perturbation theory, and is briefly summarized in Appendix~\ref{app:omega}.

At one-loop order, we find
\begin{align}
\eta_\phi&=\frac{2N+9}{2N+3}\epsilon+\mathcal{O}(\epsilon^2),\\
\nu^{-1}&=2-\frac{10N^2+39N+X}{6N(2N+3)}\epsilon+\mathcal{O}(\epsilon^2),\\
\omega&=\epsilon+\mathcal{O}(\epsilon^2),
\end{align}
with $\eta_\phi$ and $\nu^{-1}$ in agreement with Ref.~\cite{janssen2017}. At higher loop order, analytical expressions for the critical exponents with general $N$ can be obtained but are extremely cumbersome~\cite{SuppMat}. As a nontrivial check on the calculation, we have verified that our four-loop result for $\eta_\phi$, when expanded in inverse powers of $N$ to $\mathcal{O}(1/N)$, agrees with the corresponding $1/N$ expansion result for the QED$_3$-Gross-Neveu (QED$_3$-GN) model in $d$ dimensions, when expanded to $\mathcal{O}(\epsilon^4)$~\cite{gracey1992,NoteEtaPsi}. Here we only give explicit expressions for the critical exponents for the $N=1$ case, relevant for the conjectured duality with the $SU(2)$ NC$\mathbb{C}$P$^1$ model,
\begin{align}
\eta_\phi&\approx 2.2\epsilon-0.2227\epsilon^2+16.88\epsilon^3-205.1\epsilon^4,\\
\nu^{-1}&\approx 2-3.905\epsilon+7.471\epsilon^2-90.6\epsilon^3+1154\epsilon^4,\\
\omega&\approx\epsilon+0.3\epsilon^2+4.294\epsilon^3-119.1\epsilon^4,
\end{align}
and for $N=2$ case, appropriate for the spin-1/2 kagom\'e antiferromagnet:
\begin{align}
\eta_\phi&\approx 1.857\epsilon-0.03989\epsilon^2+4.142\epsilon^3-22.28\epsilon^4,\\
\nu^{-1}&\approx 2-2.794\epsilon+2.444\epsilon^2-16.11\epsilon^3+98.75\epsilon^4,\\
\omega&\approx \epsilon+0.2143\epsilon^2+0.9148\epsilon^3-16.76\epsilon^4.
\end{align}
In both cases coefficients are given numerically to four significant digits.

\subsection{Pad\'e approximants}
\label{sec:Pade}

\begin{figure}[t]
\includegraphics[width=\columnwidth]{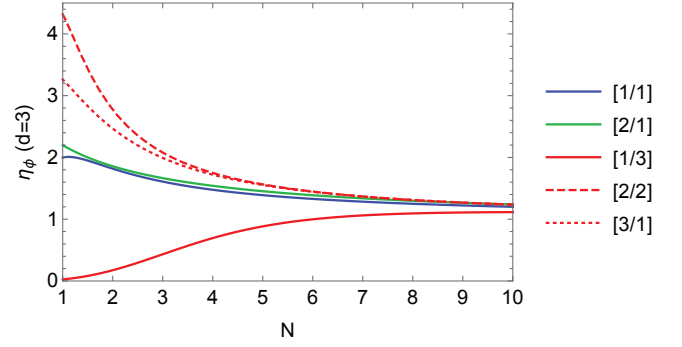}
\caption{Two-loop (blue), three-loop (green), and four-loop (red) Pad\'e approximants to the scalar field anomalous dimension $\eta_\phi$ in $d=3$, as a function of the number $N$ of flavors of four-component Dirac fermions.}
\label{fig:PadeEtaPhi}
\end{figure}

\begin{figure}[t]
\includegraphics[width=\columnwidth]{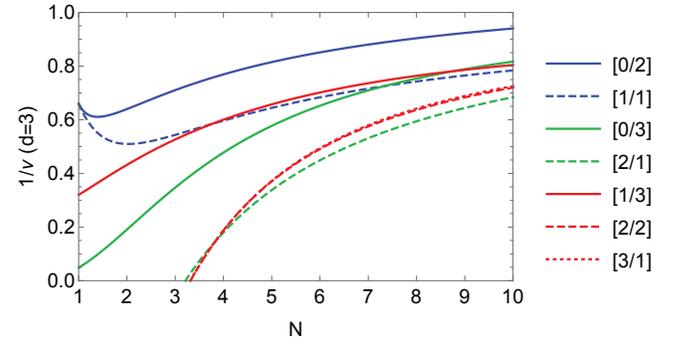}
\caption{Pad\'e approximants to the inverse correlation length exponent $1/\nu$ in $d=3$, as a function of $N$ (color scheme as in Fig.~\ref{fig:PadeEtaPhi}).}
\label{fig:PadeNuInverse}
\end{figure}

\begin{figure}[t]
\includegraphics[width=\columnwidth]{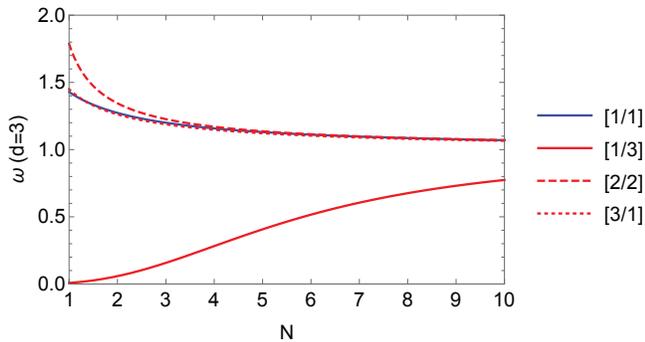}
\caption{Pad\'e approximants to the stability critical exponent $\omega$ in $d=3$, as a function of $N$ (color scheme as in Fig.~\ref{fig:PadeEtaPhi}).}
\label{fig:PadeOmega}
\end{figure}

To obtain approximate values of the critical exponents in physical $d=3$ dimensions, corresponding to $\epsilon=1$, we employ standard one-sided Pad\'e approximants (see, e.g., Ref.~\cite{dipietro2017}), defined as
\begin{align}\label{OP}
[m/n](\epsilon)\equiv\frac{\sum_{i=0}^m a_i\epsilon^i}{1+\sum_{j=1}^n b_j\epsilon^j},
\end{align}
with $m+n=L$, where $L$ is the desired loop order. They reproduce the $\epsilon$-expansion results when expanded to $\mathcal{O}(\epsilon^L)$ and constitute an extrapolation from $d=4$ down to $d=3$. We have also attempted to use two-sided approximants (see, e.g., Ref.~\cite{fei2016}) by combining information from the $4-\epsilon$ expansion of the QED$_3$-GNY model and the $2+\epsilon$ expansion of the fermionic QED$_3$-GN model,
\begin{align}\label{QED3GN}
\tilde{\mathcal{L}}=&\sum_{i=1}^{2N}\bar{\psi}_i\gamma_\mu'(\partial_\mu-ieA_\mu)\psi_i+\frac{1}{4}F_{\mu\nu}^2+\frac{1}{2\xi}(\partial_\mu A_\mu)^2\nn\\
&+u\left(\sum_{i=1}^{2N}\bar{\psi}_i\psi_i\right)^2+v\left(\sum_{i=1}^{2N}\bar{\psi}_i\gamma_\mu'\psi_i\right)^2,
\end{align}
where $\psi_i$ are two-component spinors and the $\gamma_\mu'$ are $2\times 2$ gamma matrices. Just like the standard GNY fixed point in $4-\epsilon$ dimensions is believed to be in the same universality class as the fixed point of the purely fermionic GN model in $2+\epsilon$ dimensions~\cite{zinn-justin1991} when extrapolating $\epsilon\rightarrow 1$, so too is the QED$_3$-GNY fixed point in $4-\epsilon$ dimensions believed to extrapolate to the same universality class as one of the two charged critical points of Eq.~(\ref{QED3GN})~\cite{janssen2017,janssen2016}. Two-sided Pad\'e extrapolation indeed appears to give accurate results for the pure GNY/GN fixed point~\cite{fei2016}. However, since the gauge coupling in Eq.~(\ref{QED3GN}) is strongly relevant near two dimensions, the charged critical point of interest, the QED$_3$-GN fixed point, is not perturbatively accessible at finite $N$ in a strict $2+\epsilon$ expansion, by contrast with the neutral fixed point of the pure GN theory~\cite{zinn-justin1991}. Thus one is forced to proceed in a combined $1/N$ and $2+\epsilon$ expansion~\cite{janssen2017,janssen2016}, which is not expected to be very accurate for small $N$. Unsurprisingly, we have found that two-sided Pad\'e approximants that take into account the leading, i.e., $\mathcal{O}(1/N^0,\epsilon)$ term in the combined $1/N$ and $2+\epsilon$ expansion---the only term known so far~\cite{janssen2017}---produce a large spread of extrapolated values for the critical exponents at small $N$. We thus do not expect the estimates obtained this way to be reliable, and discuss only the one-sided approximants (\ref{OP}) in the rest of the paper.

The results of one-sided Pad\'e extrapolation for the critical exponents are shown as a function of $N$ in Fig.~\ref{fig:PadeEtaPhi}, Fig.~\ref{fig:PadeNuInverse}, and Fig.~\ref{fig:PadeOmega} at two-loop (blue), three-loop (green), and four-loop (red) orders. We use only Pad\'e approximants that do not contain poles in the extrapolation region $0<\epsilon<1$. We observe sizeable variations in the extrapolated exponents for small $N$; such variations have also been seen in the Pad\'e extrapolation of operator scaling dimensions in conformal QED$_3$~\cite{dipietro2017}. Smaller variations seen at large $N$ are expected as the theory becomes weakly coupled in the large-$N$ limit for all $2<d<4$. For $\eta_\phi$ and $\omega$ the $[1/3]$ approximant appears to be an outlier; reasonably good agreement is obtained between the other approximants except at small $N$. A much larger spread of extrapolated values is obtained for $\nu^{-1}$, even at relatively large $N$; again the $[1/3]$ approximant deviates significantly from the two other pole-free approximants at four loops, $[2/2]$ and $[3/1]$.

The results of Pad\'e extrapolation can be compared with unitarity bounds in conformal field theory~\cite{ferrara1974,mack1977}. 
The scaling dimension $\Delta$ of a Lorentz scalar should obey $\Delta\geq\frac{d}{2}-1$; since $\Delta_\phi=(d-2+\eta_\phi)/2$ and $\Delta_{\phi^2}=d-\nu^{-1}$ this implies that $\eta_\phi\geq 0$ and $\nu^{-1}\leq 5/2$ in three dimensions. 
Additionally, by definition $\nu^{-1}$ and $\omega$ should be positive. In Fig.~\ref{fig:PadeNuInverse} the three-loop $[2/1]$ and four-loop $[2/2]$, $[3/1]$ approximants, in close agreement with each other, violate those bounds for $N\leq 3$. 
While these results suggest the possible loss of conformal invariance for sufficiently small $N$ 
--- due either to the loss of conformal invariance in pure QED$_3$ itself, or to the phase transition becoming first order --- 
they should be taken with caution, given the large variations between different approximants at small $N$.

\section{Scaling dimensions of fermion bilinears}
\label{sec:BilinearDim}

The scaling dimensions $\Delta_{\bar{\Psi}\Psi}$ and $\Delta_{\bar{\Psi}T_A\Psi}$ of the $SU(N)$ singlet/adjoint fermion bilinears are obtained from the analysis in Sec.~\ref{sec:FermionBi}. At one-loop order we obtain
\begin{align}\label{FermionMassQED3GNY1L}
\Delta_{\bar{\Psi}\Gamma\Psi}&=3-\left(\frac{2N+6}{2N+3}\right)\epsilon+\mathcal{O}(\epsilon^2),
\end{align}
for both singlet and adjoint bilinears, the latter in agreement with Ref.~\cite{janssen2017}. Starting at two-loop order the two scaling dimensions differ due to mixing of the singlet bilinear with the $\phi^3$ operator. The full expressions at four-loop order for general $N$ are extremely cumbersome~\cite{SuppMat}; here we give expressions for $N=1$ only,
\begin{align}
\Delta_{\bar{\Psi}\Psi}&\approx 
3-1.6\epsilon+0.1114\epsilon^2
-8.442\epsilon^3
+102.5\epsilon^4,\\
\Delta_{\bar\Psi T_A\Psi}&\approx 3-1.6\epsilon+1.987\epsilon^2-17.46\epsilon^3+215.7\epsilon^4,
\end{align}
and $N=2$:
\begin{align}
\Delta_{\bar{\Psi}\Psi}&\approx 3-1.429\epsilon+0.01995\epsilon^2
-2.071\epsilon^3
+11.14\epsilon^4,\\
\Delta_{\bar\Psi T_A\Psi}&\approx 3-1.429\epsilon+0.4548\epsilon^2-2.069\epsilon^3+11.33\epsilon^4.
\end{align}
Strictly speaking, the adjoint bilinear only exists for $N\geq 2$, since $SU(1)$ is trivial. However, since the scaling dimension obtained for $N\geq 2$ is an analytic function of $N$, one can analytically continue the result to $N=1$. We note that in the large-$N$ limit, our four-loop result for $\Delta_{\bar{\Psi}T_A\Psi}$ agrees with the corresponding quantity for the QED$_3$-GN model computed in the $1/N$ expansion at $\mathcal{O}(1/N^2)$~\cite{GraceyPrivateComm}.

\begin{figure}[t]
\includegraphics[width=\columnwidth]{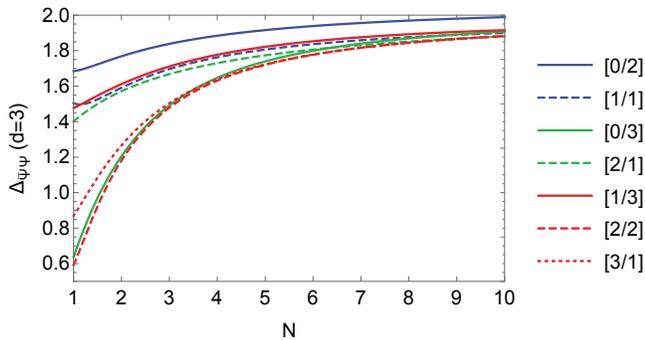}
\caption{Pad\'e approximants to the scaling dimension of the singlet fermion bilinear $\bar{\Psi}\Psi$ in $d=3$, as a function of $N$ (color scheme as in Fig.~\ref{fig:PadeEtaPhi}).}
\label{fig:PadeSinglet}
\end{figure}

\begin{figure}[t]
\includegraphics[width=\columnwidth]{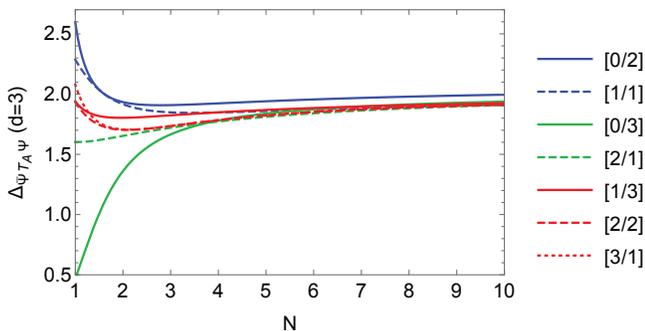}
\caption{Pad\'e approximants to the scaling dimension of the adjoint fermion bilinear $\bar{\Psi}T_A\Psi$ in $d=3$, as a function of $N$ (color scheme as in Fig.~\ref{fig:PadeEtaPhi}).}
\label{fig:PadeAdjoint}
\end{figure}

As in Sec.~\ref{sec:exponents} we perform both one-sided and two-sided Pad\'e extrapolation, but discard the two-sided approximants due to their large spread in numerical values, which itself stems from the additional large-$N$ approximation required near the lower critical dimension. The results of Pad\'e extrapolation for pole-free approximants are given in Fig.~\ref{fig:PadeSinglet} and Fig.~\ref{fig:PadeAdjoint}. As for the critical exponents there is a relatively large spread in the extrapolated values for small $N$, which was also seen in the Pad\'e extrapolation of fermion bilinear and quadrilinear operators in conformal QED$_3$~\cite{dipietro2017}. Unitarity bounds require $\Delta_{\bar{\Psi}\Gamma\Psi}\geq 1/2$. This is satisfied for all $N$ by all approximants except the three-loop $[0/3]$ approximant, which predicts the breakdown of conformal invariance at $N=1$ with $\Delta_{\bar{\Psi}T_A\Psi}\approx 0.467$. By contrast with the thermodynamic exponents $\eta_\phi$, $\nu^{-1}$, and $\omega$ in Fig.~\ref{fig:PadeEtaPhi}-\ref{fig:PadeOmega}, a better convergence of the approximants with increasing loop order seems to be achieved for the fermion bilinear scaling dimensions. In particular, in Fig.~\ref{fig:PadeAdjoint} the four-loop result for the adjoint bilinear is sandwiched between the two-loop and three-loop results at small $N$, and the three four-loop approximants (red lines) are in close agreement with each other. Taking the mean of the three four-loop approximants, we arrive at the estimates
\begin{align}
\Delta_{\bar{\Psi}T_A\Psi}&\approx 1.98\pm 0.08\text{ for }N=1,\label{N=1DeltaAdjEstimate}\\
\Delta_{\bar{\Psi}T_A\Psi}&\approx 1.74\pm 0.06\text{ for }N=2,
\end{align}
Where the indicated uncertainties correspond to one standard deviation on either side of the mean. For both the singlet and adjoint bilinears, the $[1/3]$ approximant deviates noticeably from the other two four-loop approximants ($[2/2]$ and $[3/1]$).
This deviation is similar to, but less significant than that observed for the critical exponents in Sec.~\ref{sec:exponents}.

\subsection{Conformal QED$_3$}
\label{sec:cQED3}

When setting $g=0$, the Lagrangian (\ref{L}) reduces to decoupled copies of massless QED and scalar $\phi^4$ theory. In the loop expansion of the fermion two-point function, one important difference between the singlet and adjoint bilinears comes from closed fermion loops with a single bilinear insertion, which vanish for the adjoint bilinear due to the tracelessness of the flavor matrix $T_A$ but are generically nonzero for the singlet bilinear. In pure QED such closed fermion loops with bilinear insertions always involve a trace over an odd number of gamma matrices, and vanish regardless of the choice of flavor matrix $\Gamma$. Since the issue of mixing with the $\phi^3$ operator is absent in pure QED, the difference in $Z_{\hat{M}}$ for the singlet and adjoint bilinears only comes from closed fermion loops, and thus those two bilinears have the same scaling dimension in QED$_3$~\cite{dipietro2017}. Evaluating the adjoint anomalous dimension (\ref{gammaM4L}) at the QED$_3$ fixed point $g_*^2=\lambda_*^2=0$, $e_*^2=e^2_{*,\text{QED}_3}$, where $e^2_{*,\text{QED}_3}$ can be determined to $\mathcal{O}(\epsilon^4)$ from the beta function (\ref{betae2}) in the QED limit $g\rightarrow 0$, 
we can obtain $\Delta_{\bar{\Psi}\Gamma\Psi}$ at that fixed point. In fact, using the known five-loop QED/quantum chromodynamics (QCD) results~\cite{Baikov:2008cp,Baikov:2010je,Baikov:2012zm,Baikov:2014qja,Chetyrkin:2016uhw,Baikov:2016tgj,Luthe:2016ima,Luthe:2016xec,Herzog:2017ohr,Luthe:2017ttc,Baikov:2017ujl,Luthe:2017ttg,Chetyrkin:2017bjc} for $\beta_{e^2}$ and $\gamma_{\bar{\Psi}\Psi}$ we can calculate $\Delta_{\bar{\Psi}\Gamma\Psi}$ at five-loop order (see Appendix~\ref{app:5LDeltaQED3} and Ref.~\cite{SuppMat}), which agrees at three-loop order with Ref.~\cite{dipietro2017}. As another nontrivial check on the calculation, we have also verified that the large-$N$ expansion of Eq.~(\ref{DeltaQED3}) to $\mathcal{O}(1/N^2)$ precisely matches the result of the large-$N$ expansion in fixed $2<d<4$ carried out to $\mathcal{O}(1/N^2)$ in Ref.~\cite{gracey1993}, when expanded to $\mathcal{O}(\epsilon^5)$ in $d=4-\epsilon$ dimensions.

\begin{figure}[t]
\includegraphics[width=\columnwidth]{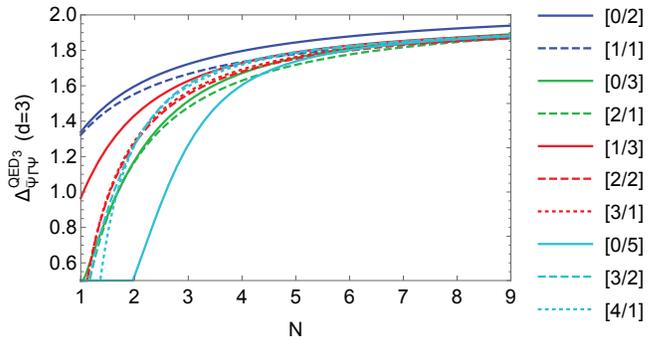}
\caption{Pad\'e approximants to the scaling dimension of the (singlet or adjoint) fermion bilinear $\bar{\Psi}\Gamma\Psi$ at the conformal QED$_3$ fixed point in $d=3$, as a function of $N$ (color scheme as in Fig.~\ref{fig:PadeEtaPhi}, with five-loop approximants in cyan).}
\label{fig:PadeDeltaQED3}
\end{figure}

One-sided Pad\'e approximants up to five-loop order are shown in Fig.~\ref{fig:PadeDeltaQED3}. We plot the results only up to $N=9$, as the four-loop $[2/2]$ approximant has a pole in the extrapolation region at $N=10$. Excluding the $[1/3]$ and $[0/5]$ approximants, relatively good convergence is obtained with increasing loop order. The unitarity bound $\Delta_{\bar{\Psi}\Gamma\Psi}\geq 1/2$ is violated at $N=1$ by the three-loop $[0/3]$, $[2/1]$, four-loop $[2/2]$, $[3/1]$, and five-loop $[3/2]$, $[4/1]$ approximants (see Table~\ref{table:XSB}, from which we have excluded $[0/5]$ which strongly deviates from the other approximants). 
One can thus use the approximants to extract a critical value $N_c$ of the fermion flavor number below which conformal invariance is lost, presumably due to the dynamical generation of an $SU(N)$ singlet fermion mass $\langle\bar{\Psi}\Psi\rangle\neq 0$, which translates in $d=3$ 
to the spontaneous breaking of chiral symmetry $SU(2N)\rightarrow SU(N)\times SU(N)\times U(1)$~\cite{vafa1984}. 
The estimates obtained this way (Table~\ref{table:XSB}) are relatively close to the estimate $N_c\approx 1.02$ obtained from an entirely different condition, that of unitary bound violation for monopole operators~\cite{pufu2014}. 
The implied breakdown of chiral symmetry for $N=1$ but not for $N\geq 2$ is consistent with lattice gauge theory results~\cite{hands2002,hands2004,strouthos2009}, except the most recent ones~\cite{karthik2016,karthik2016b,karthik2017} which predict the absence of chiral symmetry breaking even at $N=1$.

\begin{table}[t]
\centering
\begin{tabular}{ |c||c|c|c|c|c|c| } 
 \hline
  & $[0/3]$ & $[2/1]$ & $[2/2]$ & $[3/1]$ & $[3/2]$ & $[4/1]$ \\ 
  \hline
$\Delta_{\bar{\Psi}\Gamma\Psi}^\text{QED$_3$}(N=1)$ & $0.452$ & $0.238$ & $0.265$ & $0.266$ & $0.238$ & $-0.960$ \\ 
$N_c$ & $1.05$ & $1.16$ & $1.12$ & $1.12$ & $1.16$ & $1.36$  \\ 
 \hline
\end{tabular}
\caption{Pad\'e estimates for unitarity bound violations in QED$_3$.}
\label{table:XSB}
\end{table}

\subsection{Chiral Ising GNY model}

Finally, when setting $e=0$ in Eq.~(\ref{L}) the model reduces to the pure GNY model in the chiral Ising universality class, and thus we can also calculate the scaling dimensions of the singlet and adjoint bilinears at the GNY fixed point at four-loop order. At one-loop order we obtain the same scaling dimension for the singlet and adjoint bilinears,
\begin{align}\label{FermionMassDeltaGNY}
\Delta_{\bar{\Psi}\Gamma\Psi}^\text{GNY}=3-\left(\frac{4N+3}{4N+6}\right)\epsilon+\mathcal{O}(\epsilon^2).
\end{align}
Starting at two-loop order, $\Delta_{\bar\Psi\Psi}$ differs from $\Delta_{\bar\Psi T_A\Psi}$ due to mixing between the $\bar{\Psi}\Psi$ and $\phi^3$ operators. At one-loop order, our result for the singlet dimension disagrees with Ref.~\cite{fei2016}, while the adjoint dimension agrees with Ref.~\cite{giombi2017}. However, our full four-loop results~\cite{SuppMat} agree with the corresponding large-$N$ results at $\mathcal{O}(1/N^2)$ for both the singlet~\cite{gracey1992b} and adjoint~\cite{GraceyPrivateComm} mass dimensions in the GN model. Furthermore, the dimension $\Delta_{\phi^3}$ of the $\phi^3$ operator~\cite{SuppMat}, determined from the other eigenvalue of the mixing matrix $K$ in Sec.~\ref{sec:FermionBi}, agrees at one-loop order with Ref.~\cite{fei2016}, and at four-loop order with the corresponding large-$N$ result in the GN model, which has been determined at $\mathcal{O}(1/N^2)$ only recently~\cite{manashov2018}.

\begin{figure}[t]
\includegraphics[width=\columnwidth]{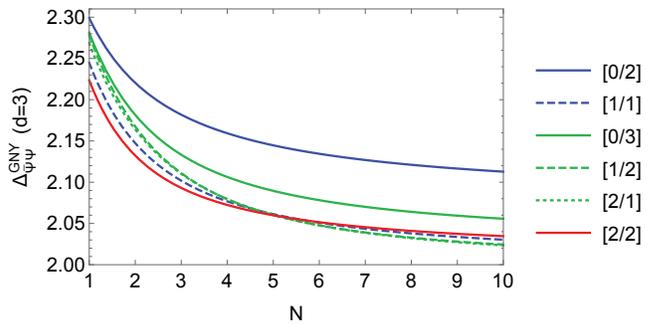}
\caption{Pad\'e approximants to the scaling dimension of the singlet fermion bilinear $\bar{\Psi}\Psi$ at the chiral Ising GNY fixed point in $d=3$, as a function of $N$ (color scheme as in Fig.~\ref{fig:PadeEtaPhi}).}
\label{fig:PadeSingletGNY}
\end{figure}

\begin{figure}[t]
\includegraphics[width=\columnwidth]{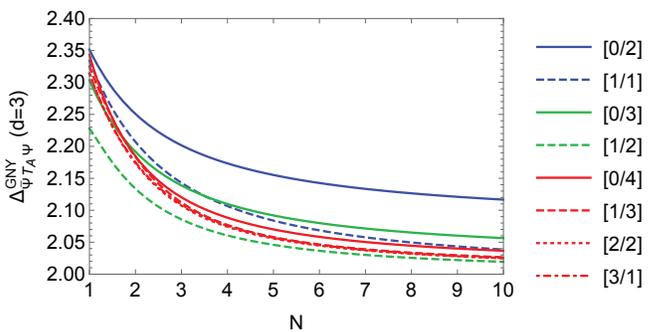}
\caption{Pad\'e approximants to the scaling dimension of the adjoint fermion bilinear $\bar{\Psi}T_A\Psi$ at the chiral Ising GNY fixed point in $d=3$, as a function of $N$ (color scheme as in Fig.~\ref{fig:PadeEtaPhi}).}
\label{fig:PadeAdjointGNY}
\end{figure}

We show the results of one-sided Pad\'e extrapolation in Fig.~\ref{fig:PadeSingletGNY} and Fig.~\ref{fig:PadeAdjointGNY}; the spread of values is significantly smaller than for the QED$_3$-GNY model, suggesting that gauge fluctuations tend to worsen the convergence of the $\epsilon$ expansion. In particular, for the adjoint bilinear all four approximants at four loops agree closely with each other.

\section{Discussion}
\label{sec:discussion}

We now discuss some applications of our results. We have already mentioned the $N=2$ case, which describes a putative quantum phase transition between a gapless Dirac spin liquid and a gapped chiral spin liquid in a spin-1/2 kagom\'e antiferromagnet~\cite{he2015,he2015b}.
The $N=1$ case corresponds in $d=3$ to the QED$_3$-GNY model with two flavors of two-component Dirac fermions, 
which has been proposed to be dual to the critical point of the $SU(2)$-symmetric NC$\mathbb{C}$P$^1$ model~\cite{wang2017}. 
According to this conjectured duality, the parity-even, flavor-symmetry-breaking bilinear $\bar{\psi}\sigma^z\psi=\bar{\psi}_1\psi_1-\bar{\psi}_2\psi_2$ in the QED$_3$-GNY theory should be dual to the mass term $z^\dag z$ for the bosonic $\mathbb{C}$P$^1$ field $z=(z_1,z_2)$ in the $SU(2)$ NC$\mathbb{C}$P$^1$ model. 
Furthermore, the duality requires an emergent $SO(5)$ symmetry in the infrared under which $\bar{\psi}\sigma^z\psi$ and the QED$_3$-GNY scalar mass operator $\phi^2$ are predicted to transform as different components of the same traceless symmetric tensor $X_{ab}^{(2)}$, $a,b,=1,\ldots,5$. 
As a result, the duality implies that in three dimensions $\bar{\psi}\sigma^z\psi$ and $\phi^2$ should have the same scaling dimension. 
But which bilinear in the four-dimensional theory should one use for this comparison? 
Since $\sigma^z$ is traceless in $SU(2)$ flavor space, in fixed $d=3$ 
the loop expansion with single $\bar{\psi}\sigma^z\psi$ insertions leads to a vanishing contribution of bilinear insertions into closed fermion loops. 
Therefore, the $d=4-\epsilon$ bilinear whose loop expansion behaves like that of the $d=3$ flavor-symmetry-breaking bilinear, 
in the sense that closed fermion loops with bilinear insertions do not contribute, is the adjoint bilinear $\bar{\Psi}T_A\Psi$, analytically continued to $N=1$. With this prescription one is thus led to compare $\Delta_{\bar{\Psi}T_A\Psi}$ with the scaling dimension of the $\phi^2$ operator, which equals $3-\nu^{-1}$. 

\begin{figure}[t]
\includegraphics[width=\columnwidth]{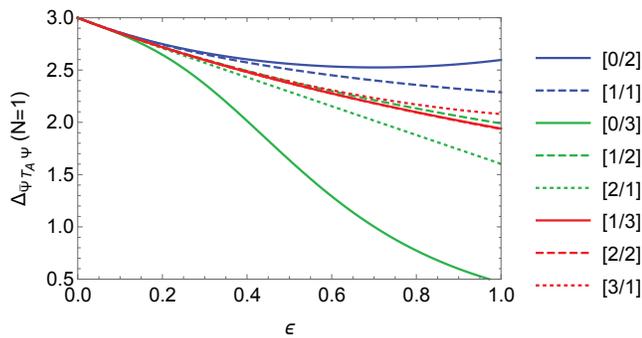}
\caption{Pad\'e extrapolation of the adjoint bilinear scaling dimension for $N=1$ (color scheme as in Fig.~\ref{fig:PadeEtaPhi}).}
\label{fig:adjointepsilon}
\end{figure}

\begin{figure}[t]
\includegraphics[width=\columnwidth]{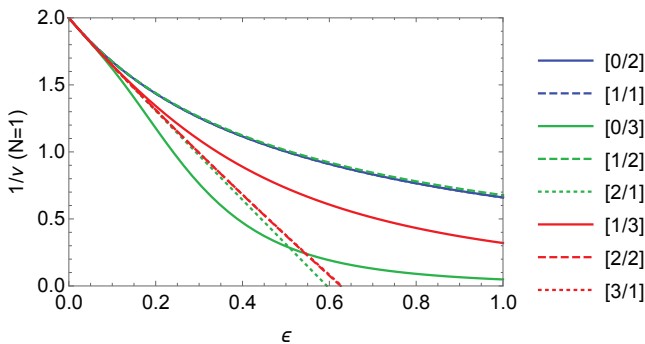}
\caption{Pad\'e extrapolation of the inverse correlation length exponent for $N=1$ (color scheme as in Fig.~\ref{fig:PadeEtaPhi}); the [2/2] and [3/1] curves are essentially superimposed.}
\label{fig:nuinvepsilon}
\end{figure}

\begin{figure}[t]
\includegraphics[width=\columnwidth]{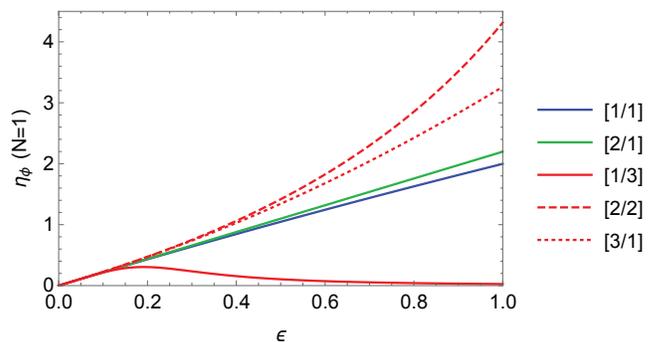}
\caption{Pad\'e extrapolation of the scalar field anomalous dimension for $N=1$ (color scheme as in Fig.~\ref{fig:PadeEtaPhi}).}
\label{fig:etaphiepsilon}
\end{figure}

While the four-loop Pad\'e approximants give a reasonably consistent extrapolated value for $\Delta_{\bar{\Psi}T_A\Psi}$ [see Fig.~\ref{fig:adjointepsilon} and Eq.~(\ref{N=1DeltaAdjEstimate})], a much higher degree of uncertainty remains for $\nu^{-1}$ (Fig.~\ref{fig:nuinvepsilon}), preventing an unambiguous verification of the duality. 
The same can be said for the value of $\nu^{-1}$ itself which, according to the duality, should be the same as that at the N\'eel-valence-bond-solid transition. 
The latter has been studied numerically in lattice spin systems by Monte Carlo methods~\cite{sandvik2007,melko2008,nahum2015}, with estimates for $\nu^{-1}$ ranging from 1.3 to 2.0. 
As another prediction of the duality, the scalar field $\phi$ should be dual to the $\mathbb{C}$P$^1$ bilinear $z^\dag\sigma^zz=|z_1|^2-|z_2|^2$, an element of the N\'eel order parameter $\b{N}=z^\dag\bsigma z$, which itself is predicted to form a vector under the emergent $SO(5)$ symmetry when combined with the $\mathbb{C}$P$^1$ monopole operator, a complex scalar that has the physical interpretation of a valence-bond-solid order parameter. 
As a result, the scalar field anomalous dimension $\eta_\phi$ should be equal to that of the N\'eel and valence-bond-solid order parameters. These order parameter anomalous dimensions have also been determined numerically, with values ranging from 0.25 to 0.35~\cite{sandvik2007,melko2008,nahum2015}. 
Apart from the $[1/3]$ approximant, our simple one-sided Pad\'e estimates yield extrapolated values of $\eta_\phi$ an order of magnitude larger than this (Fig.~\ref{fig:etaphiepsilon}). One can also attempt to improve the naive Pad\'e estimates by the Pad\'e-Borel method~\cite{KleinertBook}, in which Pad\'e extrapolation is applied to the Borel sum $B_\Delta(\epsilon)$ of a critical exponent $\Delta(\epsilon)=\sum_k\Delta_k\epsilon^k$ known in the $\epsilon$ expansion,
\begin{align}
B_\Delta(\epsilon)\equiv\sum_k\frac{\Delta_k}{k!}\epsilon^k,
\end{align}
rather than to the exponent itself. An estimate for the exponent is then obtained by computing the Borel transform,
\begin{align}
\int_0^\infty dt\,e^{-t}B_\Delta(\epsilon t)=\Delta(\epsilon).
\end{align}
Pad\'e-Borel estimates for the $N=1$ exponents $1/\nu$, $\eta_\phi$, and $\Delta_{\bar{\Psi}T_A\Psi}$ are given in Table~\ref{tab: Pade Borel approx}, alongside with the ordinary Pad\'e estimates for comparison; a significant spread of extrapolated values remains even with this method.

\begin{table}[t]
\centering
\begin{tabular}{ | l || c | c | c | }
      \hline
      $N = 1$            & $1/\nu$                 & $\eta_\phi$                 & $\Delta_{\bar{\Psi}T_{A}\Psi}$ \\
          \hline
      $[0/2]$             & 0.660                    & $\times$                      & 2.60 \\
      $[0/2]_{\textrm{PB}}$ 
                              & 0.748                    & $\times$                      & 2.20 \\
      \hline
      $[1/1]$             & 0.660                    & 2.00                             & 2.29 \\
      $[1/1]_{\textrm{PB}}$   
                              & 0.387                    & 2.01                             & 2.19 \\
    
      \hline
      $[0/3]$             & 0.0486                    & $\times$                     & 0.467 \\
      $[0/3]_{\textrm{PB}}$  
                              & 0.597                    & $\times$                       & 1.69 \\
      \hline
      $[1/2]$             &  0.677                   & $\times$                       & 1.99 \\
      $[1/2]_{\textrm{PB}}$         
                             & $\times$                 & $\times$                      & 2.14\\
       \hline
      $[2/1]$            &  $\times$                & 2.20                            & 1.60 \\
      $[2/1]_{\textrm{PB}}$                 
                             & $\times$                & 2.20                             & 1.67 \\
   
       \hline
      $[0/4]$            &  $\times$               & $\times$                     & $\times$ \\
      $[0/4]_{\textrm{PB}}$   
                             & 0.584                     & $\times$                     & $\times$ \\
       \hline
      $[1/3]$            &  0.320                    & 0.0259                        & 1.94 \\
      $[1/3]_{\textrm{PB}}$     
                             & 0.580                     & 0.494                         & 1.97 \\
      \hline 
      $[2/2]$            & $\times$                & 4.32                           & 1.94 \\
      $[2/2]_{\textrm{PB}}$     
                             & $\times$                & $\times$                    & 1.74 \\
      \hline
      $[3/1]$            & $\times$                & 3.26                          & 2.08 \\
      $[3/1]_{\textrm{PB}}$    
                             & $\times$                & 3.59                          & 1.75 \\
      \hline
\end{tabular}
\caption{Pad\'e and Pad\'e-Borel resummed estimates of the inverse correlation length $1/\nu$, the boson anomalous dimension $\eta_\phi$, and the adjoint bilinear scaling dimension $\Delta_{\bar{\Psi}T_{A}\Psi}$ to three significant digits, for $d=3$ and $N=1$. 
      The values for which the approximant either has a pole in the domain $\epsilon \in [0,1]$, is undefined, or is negative, are denoted by $\times$.}
      \label{tab: Pade Borel approx}
\end{table}


\section{Calculation of the renormalization constants: technical aspects}
\label{sec:technical}

In order to extract the renormalization constants $Z_X$ for the QED$_3$-GNY model up to and including four loops we use a highly automated setup that has gone through several nontrivial checks. It has already been used to obtain the renormalization constants for the pure chiral Ising, XY, and Heisenberg GNY models in Ref.~\cite{zerf2017}, and was able to reproduce the four-loop QCD beta function~\cite{vanRitbergen:1997va,Chetyrkin:2004mf,Czakon:2004bu}. Furthermore, because the QED$_3$-GNY model is an Abelian gauge theory we were able to keep the full dependence of the amplitudes on the gauge parameter $\xi$, in order to explicitly verify the cancellation of the $\xi$ dependence in gauge-invariant quantities.

Our setup uses  {\sc QGRAF}~\cite{Nogueira:1991ex} to automatically generate all diagrams, and further uses {\sc q2e} and {\sc exp}~\cite{Harlander:1997zb,Seidensticker:1999bb} to transform the output of {\sc QGRAF} into {\sc FORM}-readable source files. The amplitude reduction of each one-particle irreducible Green's function --
including traces over Dirac gamma matrices -- is then performed within {\sc FORM}~\cite{Vermaseren:2000nd,Kuipers:2012rf}. A listing of the numbers of Feynman diagrams computed for each specific 
Green's function and at a given loop order can be found in Table~\ref{TAB:DiagramListing}.
All numbers are given for a vanishing $\phi^3$ coupling.

\renewcommand{\picscalefactor}{0.15}
\begin{table}[t]
\centering
\begin{tabular}{c|c|c|c|c}
Loops & 1 & 2 & 3 & 4 \\
\hline
 \vcenteredhbox{\includegraphics[scale=\picscalefactor]{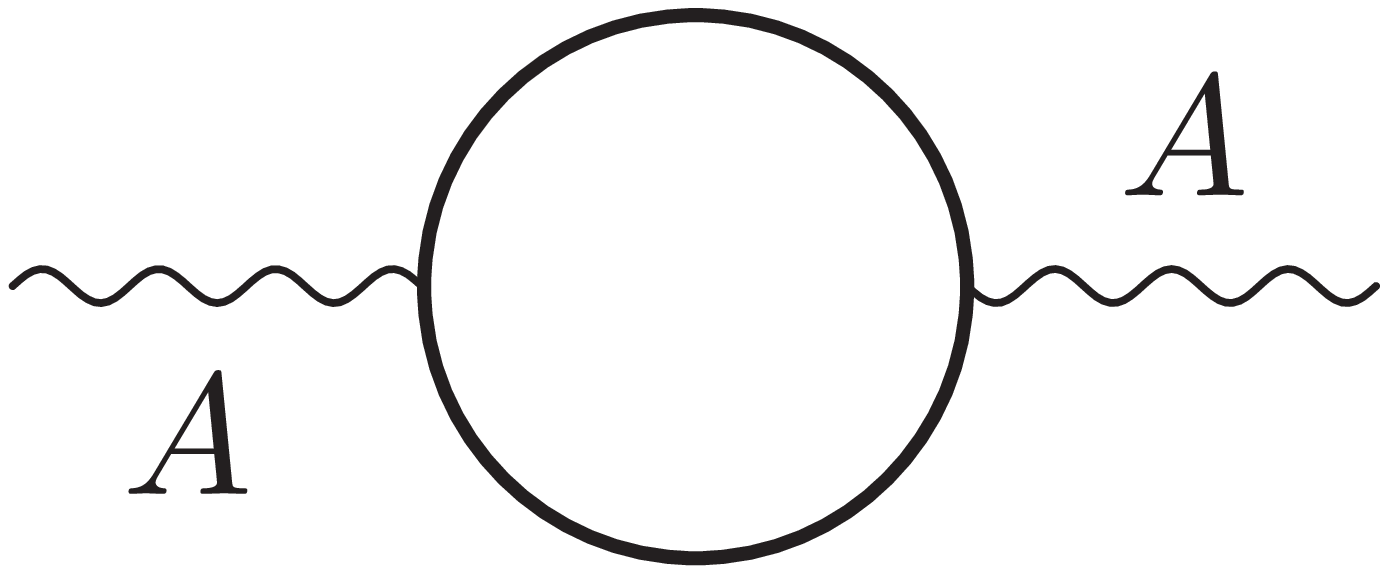}} & \vcenteredhbox{$1$} & \vcenteredhbox{$6$} & \vcenteredhbox{$83$} & \vcenteredhbox{$1610$}\\
\hline
 \vcenteredhbox{\includegraphics[scale=\picscalefactor]{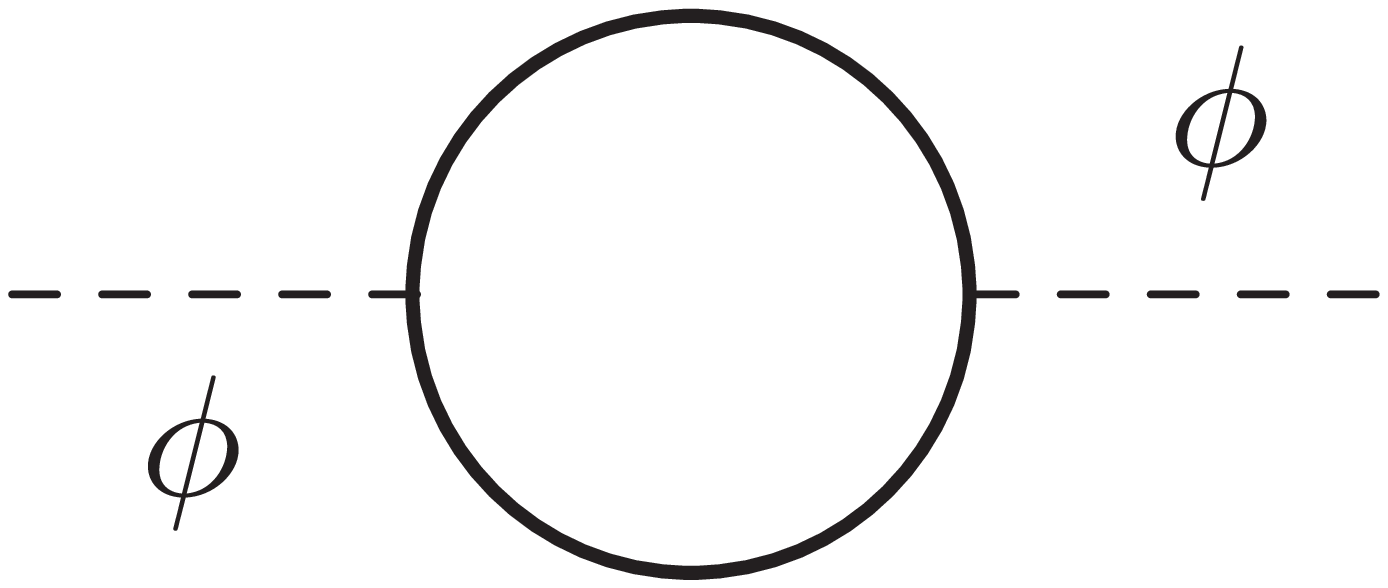}}  & \vcenteredhbox{$2$} & \vcenteredhbox{$9$} & \vcenteredhbox{$99$}& \vcenteredhbox{$1808$}\\
\hline
 \vcenteredhbox{\includegraphics[scale=\picscalefactor]{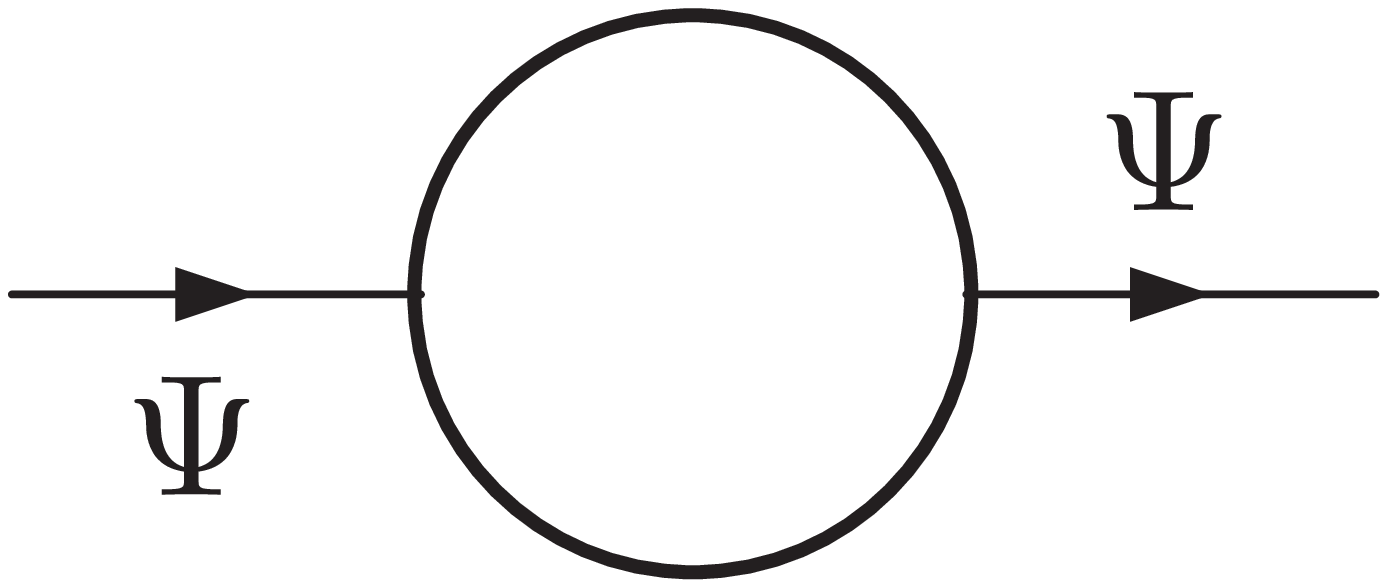}} & \vcenteredhbox{$2$} & \vcenteredhbox{$13$} & \vcenteredhbox{$177$}& \vcenteredhbox{$3387$}\\
\hline
 \vcenteredhbox{\includegraphics[scale=\picscalefactor]{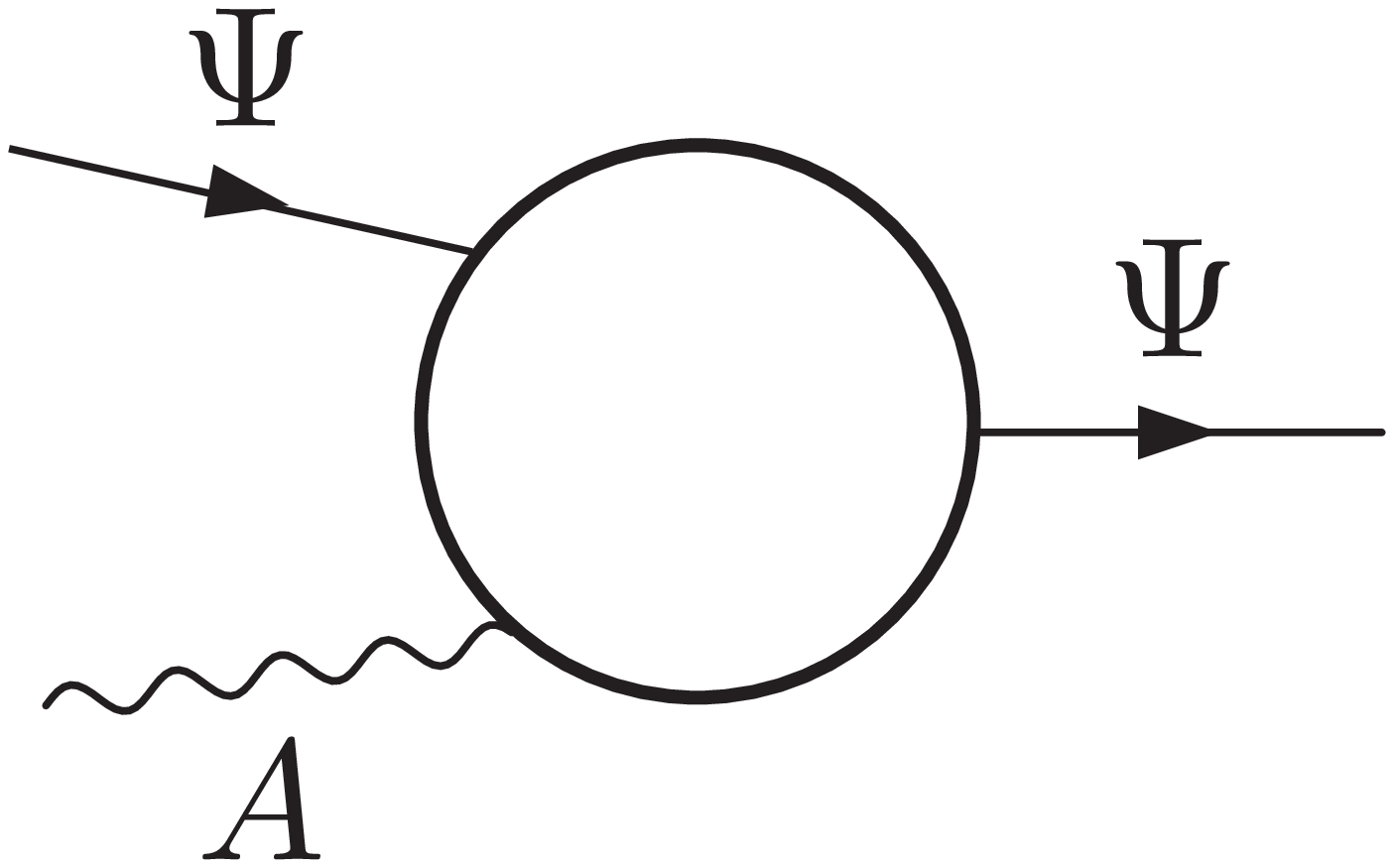}} & \vcenteredhbox{$2$} & \vcenteredhbox{$37$} & \vcenteredhbox{$844$}& \vcenteredhbox{$22818$}\\
\hline
 \vcenteredhbox{\includegraphics[scale=\picscalefactor]{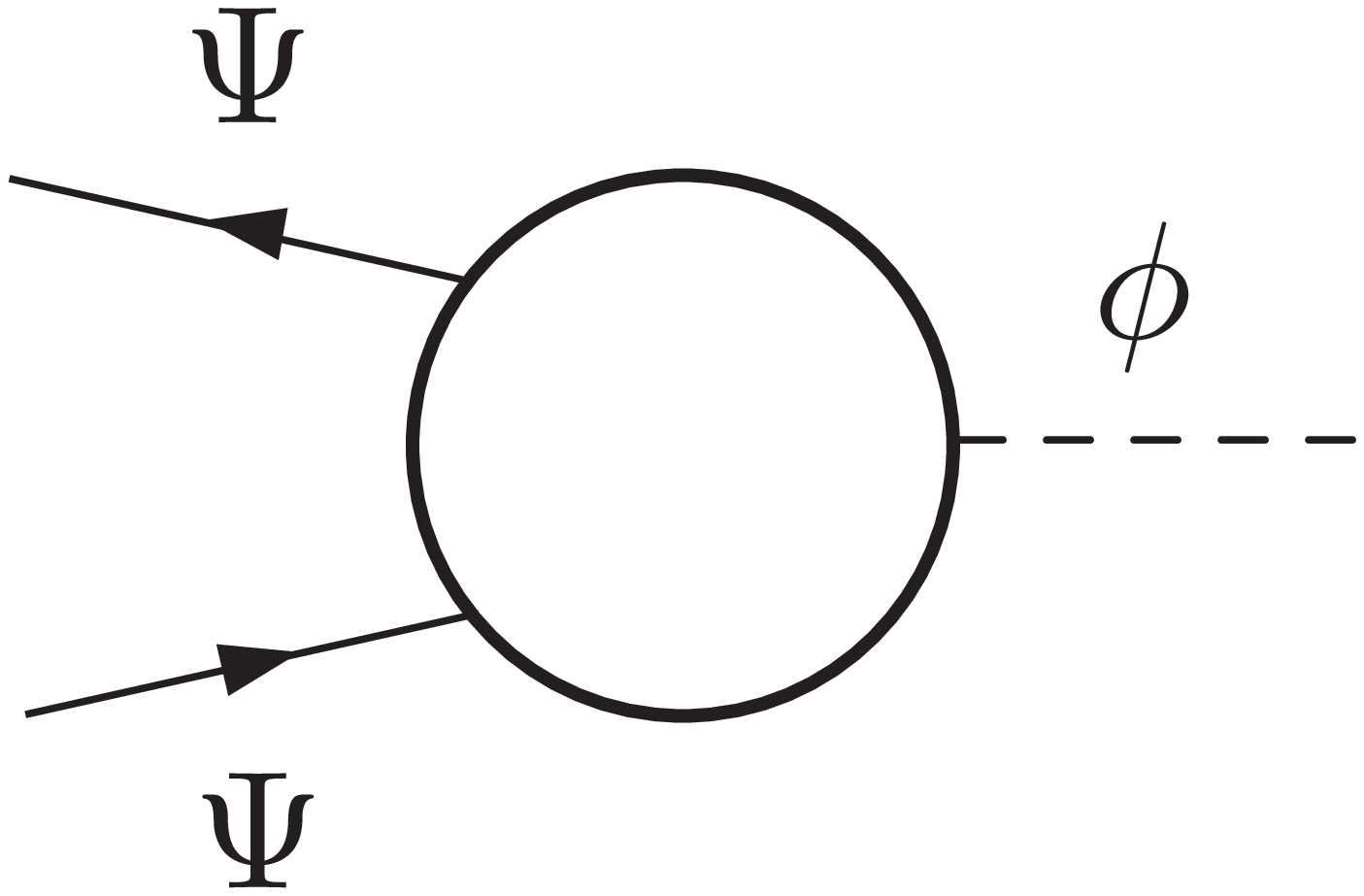}}  & \vcenteredhbox{$2$} & \vcenteredhbox{$38$} & \vcenteredhbox{$876$}& \vcenteredhbox{$23767$}\\
\hline
 \vcenteredhbox{\includegraphics[scale=\picscalefactor]{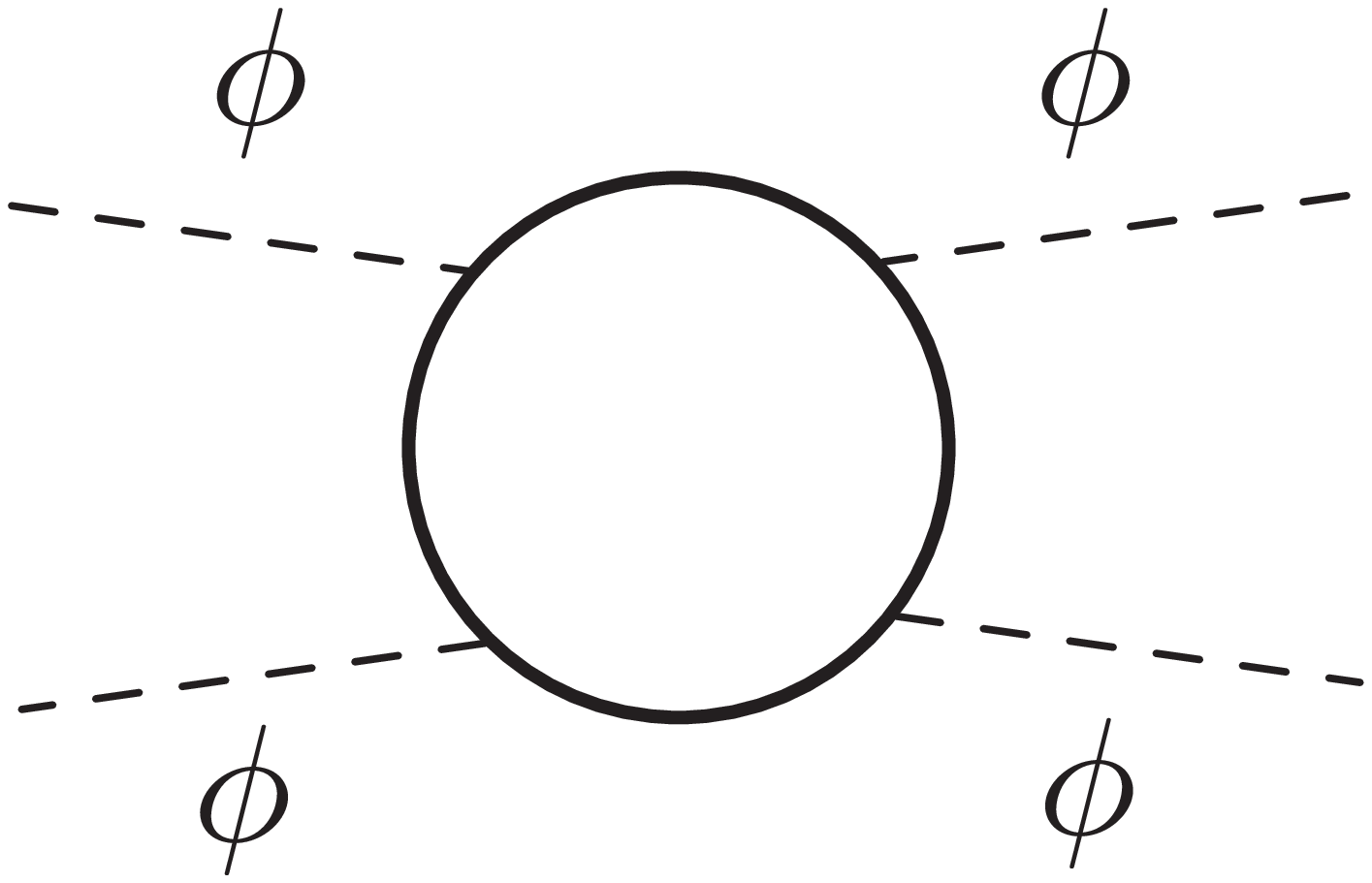}} & \vcenteredhbox{$9$} & \vcenteredhbox{$153$} & \vcenteredhbox{$4248$}&  \vcenteredhbox{$138849$}\\
\end{tabular}
 \caption{List of all relevant $n$-point functions and the associated number of Feynman diagrams evaluated for vanishing $\phi^3$ coupling, as a function of the number of loops ($A$: gauge field, $\phi$: scalar field, $\Psi$: Dirac fermion).\label{TAB:DiagramListing}}
\end{table}

In order to be able to reduce all appearing integrals to tadpole integrals only,
we treat all relevant external momenta as being small and keep a much larger
common regulator mass in all propagators in order to avoid infrared singularities.
This then allows one to expand naively in any external momentum that appears.
The systematic treatment of the unphysical regulator mass is called infrared rearrangement~\cite{Misiak:1994zw,Chetyrkin:1997fm}
and is based on an exact decomposition of a massless propagator into massive ones.
It requires an immediate cancellation of all subdivergences, and thus one 
has to perform renormalization via explicit counterterm insertions.
The infrared rearrangement method has been recently applied to the computation of the five-loop QCD beta function and anomalous dimensions ~\cite{Luthe:2016ima,Luthe:2016xec}
and is in full agreement with the results of completely different approaches~\cite{Baikov:2016tgj,Herzog:2017ohr}.

All appearing tadpole integrals are reduced to a well-known finite set of master integrals (see appendix of Ref.~\cite{Czakon:2004bu}) employing an integral reduction table. The latter was created with the program {\sc CRUSHER}~\cite{crusher}, which first generates integration-by-parts identities for all appearing integrals 
in order to obtain a coupled system of linear equations for them.
This system of equations is then solved by an implementation of Laporta's algorithm~\cite{Laporta:2001dd}.
The solutions of this system yield a decomposition of all appearing integrals in terms of master integrals.
The reduction table itself has been conveniently implemented using {\sc FORM}'s {\tt TableBase} functionality.

Because we kept the full $\xi$ dependence, it turned out that the most involved $n$-point function at four-loop order was the photon polarization function. Here the coefficient $\sim N g^8 $ alone generated about 53 million terms at maximum expression size which made {\sc FORM} use approximately $261~$GB of RAM.

In order to extract the renormalization constants from all bare subdivergence-subtracted amplitudes,
we rely on a generic renormalization program written in {\sc FORM}.
It performs the renormalization order by order and ensures the correct insertions of all relevant counterterm combinations.

\section{Conclusion}
\label{sec:conclusion}

In summary, we have studied the critical properties of the QED$_3$-GNY model in the $\epsilon$ expansion below four dimensions at four-loop order, expanding upon the existing results at one-loop order. 
Besides the usual thermodynamic critical exponents $\eta_\phi$, $\nu^{-1}$, and $\omega$, we have also calculated the scaling dimensions of $SU(N)$ flavor singlet and adjoint fermion bilinears. 
The latter calculations were also performed in the pure QED$_3$ and GNY limits, expanding upon existing results at lower loop orders. Agreement was found with all available large-$N$ results for the corresponding GN-type models. 
In an effort to access the critical properties of the corresponding $d=3$ theories we performed Pad\'e (and Pad\'e-Borel, for $N=1$) extrapolation. 
While substantial uncertainties remained at small $N$, reasonably good convergence with increasing loop order was achieved at sufficiently large $N$.
However, due to the large spread of the Pad\'e extrapolated values for the inverse correlation length exponent $\nu^{-1}$ at small $N$, a sharp statement concerning the validity of the conjectured duality at $N=1$ could not be made.

It is conceivable that computing higher-order $1/N$ corrections to critical exponents in the $d=2+\epsilon$ QED$_3$-GN model might improve the quality of two-sided Pad\'e approximants even at small $N$, and perhaps yield a narrower range of extrapolated values than the one-sided approximants. More sophisticated methods such as the use of conformal mappings in Borel resummation~\cite{ZJ,KleinertBook}, which has recently been applied to the GN and GNY models~\cite{ihrig2018}, could also be employed here, and would benefit from a careful study of the large-order behavior of the $\epsilon$-expansion coefficients in Yukawa-type 
theories, which is not currently known. 
Finally, calculations of critical exponents in the $1/N$ expansion in fixed $d=3$ dimensions would provide an alternative test of the proposed duality.

\acknowledgements

We gratefully acknowledge I. Affleck, L. Di Pietro, S. Giombi, J. A. Gracey, L. Janssen, M. M. Scherer, and K. Wamer for helpful discussions. We would also like to thank P. Uwer and the administrators of the PEP working group cluster at HU Berlin for providing high-quality computing resources.
JM was supported by NSERC grant \#RGPIN-2014-4608, the CRC Program, CIFAR, and the University of Alberta.

{\it Note added.} During the preparation of this manuscript we became aware of Ref.~\cite{ihrig2018b}, where an analysis similar to ours was performed at three-loop order. Our results agree wherever they overlap.

\appendix
\numberwithin{equation}{section}
\numberwithin{figure}{section}

\section{Chiral $\mathbb{Z}_2$ symmetry and fermion bilinear operators}
\label{app:GenChiralZ2}

In order to see that an $SU(N)$ flavor adjoint fermion bilinear operator is even under a chiral $\mathbb{Z}_2$ transformation,
first note that an arbitrary matrix $\Gamma$ in the Lie algebra $\mathfrak{g}$ of $SU(N)$ can always be brought to diagonal form by an $SU(N)$ transformation on $\Psi$, $\bar{\Psi}$, and can thus be expanded as $\Gamma=\b{\lambda}\cdot\b{H}$ where $H_i$, $i=1,\ldots,N-1$ are generators of the Cartan subalgebra. Modulo an overall prefactor to be absorbed in $M$, the expansion coefficients $(\lambda_1,\ldots,\lambda_{N-1})\equiv\b{\lambda}$ can be normalized such that $\b{\lambda}^2=1$. Since the action of the orthogonal group $O(N-1)$ is transitive on the unit sphere in $\mathbb{R}^{N-1}$, one can always perform a change of basis of the Cartan subalgebra $H_i\rightarrow R_{ij}H_j$, $R\in O(N-1)$ such that $\Gamma=\b{\alpha}\cdot\b{H}$ where $\b{\alpha}$ is a simple root of $\mathfrak{g}$. We further note that the discrete chiral symmetry in Eq.~(\ref{Z2chiralsymm}) can be more generally defined as involving an $SU(N)$ transformation on the fermions,
\begin{align}\label{modifiedZ2chiralsymm}
\Psi\rightarrow e^{-iW}\gamma_5\Psi,\hspace{5mm}
\bar{\Psi}\rightarrow-\bar{\Psi}\gamma_5 e^{iW},\hspace{5mm}
\phi\rightarrow-\phi,
\end{align}
with $W$ an element of $\mathfrak{g}$. Under this generalized transformation the flavor singlet bilinear remains odd, but the flavor adjoint bilinear transforms as
\begin{align}
\bar{\Psi}\b{\alpha}\cdot\b{H}\Psi\rightarrow-\bar{\Psi}e^{iW}\b{\alpha}\cdot\b{H} e^{-iW}\Psi.
\end{align}
Defining the non-Cartan generators $T_\b{\alpha}^1=(E_\b{\alpha}+E_\b{\alpha}^\dag)/\sqrt{2}$, $T_\b{\alpha}^2=-i(E_\b{\alpha}-E_\b{\alpha}^\dag)/\sqrt{2}$, where $E_\b{\alpha}$ ($E_\b{\alpha}^\dag$) is the raising (lowering) operator for the $SU(2)$ subalgebra associated with the simple root $\b{\alpha}$, one can show that
\begin{align}
e^{i\theta T_\b{\alpha}^1}\b{\alpha}\cdot\b{H}e^{-i\theta T_\b{\alpha}^1}=\cos\theta\,\b{\alpha}\cdot\b{H}+\sin\theta\,T_\b{\alpha}^2.
\end{align}
Thus if one chooses $W=\pi T_\b{\alpha}^1$ in Eq.~(\ref{modifiedZ2chiralsymm}), the adjoint bilinear preserves a generalized $\mathbb{Z}_2$ chiral symmetry,
because the additional minus sign is absorbed due to $\cos\pi=-1$.

\section{Four-loop contributions to the beta functions}
\label{app:4Lbeta}

The four-loop contributions to the beta functions in Sec.~\ref{sec:beta} are given here explicitly (see also Ref.~\cite{SuppMat}). For the gauge coupling, we have
\begin{widetext}
\begin{align}
\beta^\text{(4L)}_{e^2}&=\frac{2N}{3}[N(267-432\zeta_3)+43-336\zeta_3]e^6g^4
+\frac{2N}{9}(322N-27+648\zeta_3)e^8g^2-240Ne^4g^4\lambda^2+576Ne^4g^2\lambda^4\nn\\
&\phantom{=}-\frac{4N}{243}[616N^2+36N(-95+312\zeta_3)+5589]e^{10}-\frac{4N}{9}[27N^2+N(299-24\zeta_3)+105+9\zeta_3]e^4g^6.
\end{align}
For the Yukawa coupling, we have
\begin{align}
\beta^\text{(4L)}_{g^2}&=\biggl[\frac{88N^3}{3}-\frac{2N^2}{3}(899+1500\zeta_3-324\zeta_4)+N\left(\frac{9907}{4}-2648\zeta_3+552\zeta_4-1680\zeta_5\right)+\frac{30529}{32}+10\zeta_3+342\zeta_4\nn\\
&\phantom{=}-1720\zeta_5\biggr]g^{10}+\biggl[\frac{32N^3}{81}(83-144\zeta_3)+\frac{64N^2}{27}(-19+270\zeta_3-162\zeta_4)+N\left(\frac{352}{3}-288\zeta_3+1920\zeta_5\right)+\frac{1261}{2}\nn\\
&\phantom{=}+1344\zeta_3\biggr]e^8g^2
+\biggl[\frac{16N^3}{243}(-1625+1296\zeta_3)+\frac{4N^2}{81}(-27739+35856\zeta_3-7776\zeta_4)\nn\\
&\phantom{=}+N\left(\frac{35}{3}-\frac{9856\zeta_3}{3}+912\zeta_4+10080\zeta_5\right)+\frac{9899}{2}-7464\zeta_3+1512\zeta_4+10800\zeta_5\biggr]e^6g^4\nn\\
&\phantom{=}+2[96N^2-8N(683+648\zeta_3)-3(943+1008\zeta_3)]g^8\lambda^2
-8[24N^2+4N(635-324\zeta_3)+135(33+40\zeta_3)]g^6\lambda^4\nn\\
&\phantom{=}+576(8N-455+144\zeta_3)g^4\lambda^6+224640g^2\lambda^8+\biggl[N^2\left(-\frac{1022}{9}+928\zeta_3+64\zeta_4-640\zeta_5\right)\nn\\
&\phantom{=}+N\left(-\frac{38065}{9}+\frac{22264\zeta_3}{3}-528\zeta_4+1920\zeta_5\right)-\frac{16949}{4}+6024\zeta_3-432\zeta_4+2640\zeta_5\biggr]e^4g^6\nn\\
&\phantom{=}+\left(\frac{14944N}{3}-456+5760\zeta_3\right)e^4g^4\lambda^2+\biggl[N^2(216-640\zeta_3+96\zeta_4)+N\left(\frac{27133}{12}+2056\zeta_3-756\zeta_4-1400\zeta_5\right)\nn\\
&\phantom{=}+\frac{19659}{8}+3386\zeta_3-918\zeta_4-1460\zeta_5\biggr]e^2g^8
+16[N(250-432\zeta_3)+291+360\zeta_3]e^2g^6\lambda^2\nn\\
&\phantom{=}+24[6N(-67+48\zeta_3)-815]e^2g^4\lambda^4.
\end{align}
Finally, for the quartic scalar coupling we have
\begin{align}
\beta^\text{(4L)}_{\lambda^2}&=\frac{N}{24}\biggl[64N^2(-193+252\zeta_3)-32N(-1289+540\zeta_3-288\zeta_4-1920\zeta_5)+3(-4473+8864\zeta_3-3696\zeta_4
+10400\zeta_5)\biggr]\nn\\
&\phantom{=}\times g^{10}-\frac{N}{3}\biggl[4N^2(-1685+2736\zeta_3)+N(-69220-49872\zeta_3+32400\zeta_4+40320\zeta_5)-3(9745+18708\zeta_3
-984\zeta_4\nn\\
&\phantom{=}+12560\zeta_5)\biggr]g^8\lambda^2+N[2304N^2(-1+2\zeta_3)-8N(15649+8784\zeta_3-3888\zeta_4)+211565+7296\zeta_3
+40176\zeta_4\nn\\
&\phantom{=}+167040\zeta_5]g^6\lambda^4+64N[6N(263+72\zeta_3)-14521-7452\zeta_3-5184\zeta_4-17280\zeta_5]g^4\lambda^6\nn\\
&\phantom{=}-576N(1355+3456\zeta_3-1728\zeta_4)g^2\lambda^8-6912(3499+3744\zeta_3-864\zeta_4+5760\zeta_5)\lambda^{10}\nn\\
&\phantom{=}-\frac{4N}{243}[16N^2(-125+324\zeta_3)-81N(185+16\zeta_3-240\zeta_4)-81(-1471+72\zeta_3+432\zeta_4+240\zeta_5)]e^6g^4\nn\\
&\phantom{=}+N\biggl[\frac{32N^2}{243}(-1625+1296\zeta_3)+8N(-125+16\zeta_3+96\zeta_4)+\frac{9302}{3}+1856\zeta_3-864\zeta_4-2880\zeta_5\biggr]e^6g^2\lambda^2\nn\\
&\phantom{=}-\frac{N}{6}[8N(-275+680\zeta_3+336\zeta_4+1920\zeta_5)+3(1747-8928\zeta_3+5040\zeta_4+3680\zeta_5)]e^4g^6\nn\\
&\phantom{=}+\frac{2N}{9}[2N(-26873+18912\zeta_3-144\zeta_4+5760\zeta_5)-9(7459+15272\zeta_3-7920\zeta_4-12000\zeta_5)]e^4g^4\lambda^2\nn\\
&\phantom{=}-12N[4N(-155+64\zeta_3-48\zeta_4)+3(-479-1056\zeta_3+432\zeta_4+960\zeta_5)]e^4g^2\lambda^4\nn\\
&\phantom{=}-\frac{N}{3}[3N(335+2464\zeta_3-1200\zeta_4)+4(845+3246\zeta_3-1566\zeta_4+2220\zeta_5)]e^2g^8\nn\\
&\phantom{=}+\frac{N}{6}[16N(2725+5376\zeta_3-2736\zeta_4)+30419+138720\zeta_3-24624\zeta_4-30240\zeta_5]e^2g^6\lambda^2\nn\\
&\phantom{=}-8N[9N(199+288\zeta_3-144\zeta_4)+19661-27216\zeta_3+1080\zeta_4+6480\zeta_5]e^2g^4\lambda^4\nn\\
&\phantom{=}+288N(1109-1104\zeta_3)e^2g^2\lambda^6.
\end{align}

\section{Four-loop contributions to the anomalous dimensions}
\label{app:4Lgamma}

In this Appendix we give the four-loop contributions to the anomalous dimensions $\gamma_\phi$, $\gamma_{\phi^2}$ [Eq.~(\ref{gammaphi4L})-(\ref{gammaphi24L})], and $\gamma_{\bar{\Psi}T_A\Psi}$ [Eq.~(\ref{gammaM4L})]; see also Ref.~\cite{SuppMat}. For the scalar field $\phi$, we have
\begin{align}
\gamma_\phi^\text{(4L)}&=\frac{N}{243}[16N^2(-1625+1296\zeta_3)+972N(-125+16\zeta_3+96\zeta_4)+81(4651+2784\zeta_3-1296\zeta_4-4320\zeta_5)]e^6g^2\nn\\
&\phantom{=}-\frac{2N}{3}[N^2(-101+144\zeta_3)+N(-211+636\zeta_3+108\zeta_4)-435+369\zeta_3+90\zeta_4+120\zeta_5]g^8\nn\\
&\phantom{=}+\frac{N}{9}[N(-4570+3264\zeta_3+4896\zeta_4-5760\zeta_5)-9(673+1064\zeta_3-1008\zeta_4+480\zeta_5)]e^4g^4\nn\\
&\phantom{=}-16N(76N+249-48\zeta_3)g^6\lambda^2-64N(3N+182-162\zeta_3)g^4\lambda^4+4608Ng^2\lambda^6+224640\lambda^8\nn\\
&\phantom{=}+\frac{N}{12}[32N(161+96\zeta_3-72\zeta_4)+11363+2784\zeta_3-3888\zeta_4-1440\zeta_5]e^2g^6-944Ne^2g^4\lambda^2\nn\\
&\phantom{=}+144N(-67+48\zeta_3)e^2g^2\lambda^4,
\end{align}
and for the scalar mass operator $\phi^2$ we obtain
\begin{align}
\gamma_{\phi^2}^\text{(4L)}&=\frac{N}{2}[64N^2(-11+18\zeta_3)+8N(-651+40\zeta_3+108\zeta_4+560\zeta_5)-1423-2688\zeta_3-2016\zeta_4+5040\zeta_5]g^8\nn\\
&\phantom{=}-3N[256N^2(-1+2\zeta_3)-8N(809+336\zeta_3-240\zeta_4)+12989-5120\zeta_3+3312\zeta_4-5760\zeta_5]g^6\lambda^2\nn\\
&\phantom{=}-96N[16N(11+3\zeta_3)-949-1440\zeta_3-216\zeta_4]g^4\lambda^4+576N(313+96\zeta_3)g^2\lambda^6
+27648(187+18\zeta_3+36\zeta_4)\lambda^8\nn\\
&\phantom{=}-\frac{2N}{3}[4N(-683+480\zeta_3-72\zeta_4+240\zeta_5)-3393-7104\zeta_3+3456\zeta_4+6240\zeta_5]e^4g^4\nn\\
&\phantom{=}+4N[4N(-155+64\zeta_3-48\zeta_4)+3(-479-1056\zeta_3+432\zeta_4+960\zeta_5)]e^4g^2\lambda^2\nn\\
&\phantom{=}-4N[3N(89+192\zeta_3-96\zeta_4)-177+1808\zeta_3-720\zeta_4-200\zeta_5]e^2g^6\nn\\
&\phantom{=}+8N[N(597+864\zeta_3-432\zeta_4)+3019-2928\zeta_3-504\zeta_4+1200\zeta_5]e^2g^4\lambda^2
+1152N(-49+48\zeta_3)e^2g^2\lambda^4.
\end{align}
For the $SU(N)$ adjoint bilinear $\bar{\Psi}T_A\Psi$ we obtain
\begin{align}
\gamma^\text{(4L)}_{\bar{\Psi}T_A\Psi}&=\left[\frac{4N^2}{81}(2183-1728\zeta_3)+\frac{4N}{3}(457-868\zeta_3+144\zeta_4)+\frac{1}{4}(-9899+14928\zeta_3-3024\zeta_4-21600\zeta_5)\right]e^6g^2\nn\\
&\phantom{=}+\biggl[\frac{2N^2}{9}(-983+456\zeta_3-216\zeta_4)+\frac{4N}{9}(1013-5586\zeta_3+756\zeta_4+900\zeta_5)\nn\\
&\phantom{=}+\frac{1}{8}(16949-24096\zeta_3+1728\zeta_4-10560\zeta_5)\biggr]e^4g^4+\biggl[\frac{16N^3}{81}(-83+144\zeta_3)+\frac{32N^2}{27}(19-270\zeta_3+162\zeta_4)\nn\\
&\phantom{=}+\frac{16N}{3}(-11+27\zeta_3-180\zeta_5)-\frac{1}{4}(1261+2688\zeta_3)\biggr]e^8
+\biggl[N^3(19-48\zeta_3)+98N^2\nn\\
&\phantom{=}+N\left(-\frac{2475}{8}+689\zeta_3-90\zeta_4+160\zeta_5\right)-\frac{30529}{64}-5\zeta_3-171\zeta_4+860\zeta_5\biggr]g^8
+\biggl[\frac{16N^2}{3}(20+84\zeta_3-27\zeta_4)\nn\\
&\phantom{=}+N\left(-\frac{9805}{12}+496\zeta_3-216\zeta_4\right)
-\frac{19659}{16}-1693\zeta_3+459\zeta_4+730\zeta_5\biggr]e^2g^6+(2456N+2046+1440\zeta_3)g^6\lambda^2\nn\\
&\phantom{=}+(-3344N+11484-6048\zeta_3)g^4\lambda^4-13536g^2\lambda^6-192(2+9\zeta_3)e^2g^4\lambda^2
-588e^2g^2\lambda^4.\label{gammatilde4LTA}
\end{align}

\section{Scaling dimension of fermion bilinear in conformal QED$_3$ at five-loop order}
\label{app:5LDeltaQED3}

The scaling dimension of the $SU(N)$ singlet/adjoint fermion mass bilinear $\bar{\Psi}\Gamma\Psi$ at the QED$_3$ fixed point is~\cite{SuppMat}
\begin{align}\label{DeltaQED3}
\Delta_{\bar{\Psi}\Gamma\Psi}^\text{QED$_3$}&=3-\epsilon-\frac{9}{4N}\epsilon+\frac{15(4N+9)}{64N^2}\epsilon^2
+\frac{140N^2+81N(5-16\zeta_3)-3078}{256N^3}\epsilon^3+\frac{1}{16384N^4}\nn\\
&\phantom{=}\times\Bigl[64N^3(83-144\zeta_3)
-288N^2(101-360\zeta_3+216\zeta_4)
-432N(11-228\zeta_3-720\zeta_5)
+567(1183+384\zeta_3)\Bigr]\epsilon^4\nn\\
&\phantom{=}+\frac{1}{65536 N^5}\Bigl[192 N^4 (65 + 80 \zeta_3 - 144 \zeta_4) + 
 32 N^3 (-3607 + 1944 \zeta_3 + 9720 \zeta_4 - 5184 \zeta_5) \nn\\
 &\phantom{=} -  1728 N^2 (56 - 663 \zeta_3 + 504 \zeta_3^2 - 171 \zeta_4 + 1620 \zeta_5 - 900 \zeta_6)
  -  972 (14087 + 4592 \zeta_3 + 2080 \zeta_5)  \nn\\
&\phantom{=}  - 27 N (55915 + 140256 \zeta_3 - 24192 \zeta_4 - 30720 \zeta_5 + 241920 \zeta_7)\Bigr]\epsilon^5
+\mathcal{O}(\epsilon^6).
\end{align}

\end{widetext}

\section{Calculation of the stability critical exponent $\omega$}
\label{app:omega}

Here we explain the procedure used to calculate the stability critical exponent $\omega$, defined as the smallest eigenvalue of the stability matrix $\mathcal{J}$ defined in Eq.~(\ref{M}). For a generic number $N$ of fermion flavors the matrix elements of $\mathcal{J}$ at four-loop order are extremely lengthy, and direct diagonalization, which involves analytically finding the roots of a cubic secular equation involving those matrix elements, is needlessly complicated. Since $\omega$ must be computed only to order $\epsilon^4$, one can simply proceed as in ordinary (Rayleigh-Schr\"odinger) perturbation theory. We expand the eigenvalues $\omega_i$, $i=1,2,3$ and corresponding eigenvectors $\b{u}_i$ of $\mathcal{J}$, as well as $\mathcal{J}$ itself, in powers of $\epsilon$:
\begin{align}
\mathcal{J}&=\sum_{L=1}^4\mathcal{J}^{(L)}\epsilon^L,\\
\omega_i&=\sum_{L=1}^4\omega_i^{(L)}\epsilon^L,\\
\b{u}_i&=\sum_{L=1}^4\b{u}_i^{(L)}\epsilon^{L-1}.
\end{align}
Solving the (right) eigenvalue problem $\mathcal{J}\cdot\b{u}_i=\omega_i\b{u}_i$ order by order in $\epsilon$, one obtains the equation
\begin{align}\label{OmegaEqPT}
\sum_{n+n'=L+1}\mathcal{J}^{(n)}\cdot\b{u}_i^{(n')}=\sum_{n+n'=L+1}\omega_i^{(n)}\b{u}_i^{(n')},
\end{align}
at each loop order $L=1,\ldots,4$, where $n,n'=1,\ldots,4$.

At one-loop order, $\omega_i^{(1)}$ are simply given by the eigenvalues of $\mathcal{J}^{(1)}$; one can check that this matrix is (lower) triangular, thus its right and left eigenvalues are equal. However, it is not symmetric, thus its right eigenvectors $\b{u}_i^{(1)}$ and left eigenvectors $\tilde{\b{u}}_i^{(1)}$, defined by
\begin{align}
\tilde{\b{u}}_i^{(1)}\cdot \mathcal{J}^{(1)}=\tilde{\b{u}}_i^{(1)}\omega_i^{(1)},
\end{align}
are not equal. At two-loop order, left-multiplying the $L=2$ equation in Eq.~(\ref{OmegaEqPT}) by the left eigenvector $\tilde{\b{u}}_i^{(1)}$, one obtains
\begin{align}
\omega_i^{(2)}=\frac{\hat{\mathcal{J}}^{(2)}_{ii}}{S_{ii}},
\end{align}
where we define the matrix elements $\hat{\mathcal{J}}_{ij}^{(L)}$ at loop order $L$ and the overlap matrix $S_{ij}$ by
\begin{align}
\hat{\mathcal{J}}_{ij}^{(L)}=\tilde{\b{u}}_i^{(1)}\cdot \mathcal{J}^{(L)}\cdot\b{u}_j^{(1)},\hspace{5mm}S_{ij}=\tilde{\b{u}}_i^{(1)}\cdot\b{u}_j^{(1)}.
\end{align}
One can check that $S$ is in fact diagonal, $S_{ij}=S_{ii}\delta_{ij}$. At loop orders three and four, one proceeds as in second- and third-order perturbation theory, respectively, expanding the eigenvector contributions for $L=2$ and $L=3$ on the basis of one-loop eigenvectors,
\begin{align}\label{1storderPT}
\b{u}_i^{(2,3)}=\sum_{k\neq i}c_{ik}^{(2,3)}\b{u}_k^{(1)}.
\end{align}
One can check that the eigenvalues $\omega_i^{(1)}$ of $\mathcal{J}^{(1)}$ are distinct for all (finite) $N$, thus the eigenvectors $\b{u}_i^{(1)}$ are linearly independent. As in ordinary perturbation theory, the diagonal coefficients $c_{ii}^{(2,3)}$ are arbitrary and can be set to zero. Substituting the expansion (\ref{1storderPT}) into the $L=2$ and $L=3$ equations in Eq.~(\ref{OmegaEqPT}), and left-multiplying by $\tilde{\b{u}}_j^{(1)}$ with $j\neq i$, one can solve for the expansion coefficients:
\begin{align}
c_{ij}^{(2)}&=\frac{\hat{\mathcal{J}}_{ji}^{(2)}}{S_{jj}\left(\omega_i^{(1)}-\omega_j^{(1)}\right)},\\
c_{ij}^{(3)}&=\frac{\hat{\mathcal{J}}_{ji}^{(3)}+\tilde{\b{u}}_j^{(1)}\cdot \mathcal{J}^{(2)}\cdot\b{u}_i^{(2)}
-\omega_i^{(2)}\tilde{\b{u}}_j^{(1)}\cdot \b{u}_i^{(2)}}{S_{jj}\left(\omega_i^{(1)}-\omega_j^{(1)}\right)}.
\end{align}
Finally, substituting Eq.~(\ref{1storderPT}) into the $L=3$ and $L=4$ equations in Eq.~(\ref{OmegaEqPT}), we obtain
\begin{align}
\omega_i^{(3)}&=\frac{1}{S_{ii}}\left(\hat{\mathcal{J}}_{ii}^{(3)}+\sum_{j\neq i}c_{ij}^{(2)}\hat{\mathcal{J}}_{ij}^{(2)}\right),\\
\omega_i^{(4)}&=\frac{1}{S_{ii}}\left[\hat{\mathcal{J}}_{ii}^{(4)}+\sum_{j\neq i}\left(c_{ij}^{(3)}\hat{\mathcal{J}}_{ij}^{(2)}+c_{ij}^{(2)}\hat{\mathcal{J}}_{ij}^{(3)}\right)\right].
\end{align}

\bibliography{QED3GNY4L}

\begin{thebibliography}{97}%
\makeatletter
\providecommand \@ifxundefined [1]{%
 \@ifx{#1\undefined}
}%
\providecommand \@ifnum [1]{%
 \ifnum #1\expandafter \@firstoftwo
 \else \expandafter \@secondoftwo
 \fi
}%
\providecommand \@ifx [1]{%
 \ifx #1\expandafter \@firstoftwo
 \else \expandafter \@secondoftwo
 \fi
}%
\providecommand \natexlab [1]{#1}%
\providecommand \enquote  [1]{``#1''}%
\providecommand \bibnamefont  [1]{#1}%
\providecommand \bibfnamefont [1]{#1}%
\providecommand \citenamefont [1]{#1}%
\providecommand \href@noop [0]{\@secondoftwo}%
\providecommand \href [0]{\begingroup \@sanitize@url \@href}%
\providecommand \@href[1]{\@@startlink{#1}\@@href}%
\providecommand \@@href[1]{\endgroup#1\@@endlink}%
\providecommand \@sanitize@url [0]{\catcode `\\12\catcode `\$12\catcode
  `\&12\catcode `\#12\catcode `\^12\catcode `\_12\catcode `\%12\relax}%
\providecommand \@@startlink[1]{}%
\providecommand \@@endlink[0]{}%
\providecommand \url  [0]{\begingroup\@sanitize@url \@url }%
\providecommand \@url [1]{\endgroup\@href {#1}{\urlprefix }}%
\providecommand \urlprefix  [0]{URL }%
\providecommand \Eprint [0]{\href }%
\providecommand \doibase [0]{http://dx.doi.org/}%
\providecommand \selectlanguage [0]{\@gobble}%
\providecommand \bibinfo  [0]{\@secondoftwo}%
\providecommand \bibfield  [0]{\@secondoftwo}%
\providecommand \translation [1]{[#1]}%
\providecommand \BibitemOpen [0]{}%
\providecommand \bibitemStop [0]{}%
\providecommand \bibitemNoStop [0]{.\EOS\space}%
\providecommand \EOS [0]{\spacefactor3000\relax}%
\providecommand \BibitemShut  [1]{\csname bibitem#1\endcsname}%
\let\auto@bib@innerbib\@empty
\bibitem [{\citenamefont {Fisher}(1974)}]{fisher1974}%
  \BibitemOpen
  \bibfield  {author} {\bibinfo {author} {\bibfnamefont {M.~E.}\ \bibnamefont
  {Fisher}},\ }\href {\doibase 10.1103/RevModPhys.46.597} {\bibfield  {journal}
  {\bibinfo  {journal} {Rev. Mod. Phys.}\ }\textbf {\bibinfo {volume} {46}},\
  \bibinfo {pages} {597} (\bibinfo {year} {1974})}\BibitemShut {NoStop}%
\bibitem [{\citenamefont {Wilson}\ and\ \citenamefont
  {Fisher}(1972)}]{wilson1972}%
  \BibitemOpen
  \bibfield  {author} {\bibinfo {author} {\bibfnamefont {K.~G.}\ \bibnamefont
  {Wilson}}\ and\ \bibinfo {author} {\bibfnamefont {M.~E.}\ \bibnamefont
  {Fisher}},\ }\href {\doibase 10.1103/PhysRevLett.28.240} {\bibfield
  {journal} {\bibinfo  {journal} {Phys. Rev. Lett.}\ }\textbf {\bibinfo
  {volume} {28}},\ \bibinfo {pages} {240} (\bibinfo {year} {1972})}\BibitemShut
  {NoStop}%
\bibitem [{\citenamefont {Batkovich}\ \emph {et~al.}(2016)\citenamefont
  {Batkovich}, \citenamefont {Chetyrkin},\ and\ \citenamefont
  {Kompaniets}}]{batkovich2016}%
  \BibitemOpen
  \bibfield  {author} {\bibinfo {author} {\bibfnamefont {D.~V.}\ \bibnamefont
  {Batkovich}}, \bibinfo {author} {\bibfnamefont {K.~G.}\ \bibnamefont
  {Chetyrkin}}, \ and\ \bibinfo {author} {\bibfnamefont {M.~V.}\ \bibnamefont
  {Kompaniets}},\ }\href {\doibase 10.1016/j.nuclphysb.2016.03.009} {\bibfield
  {journal} {\bibinfo  {journal} {Nucl. Phys. B}\ }\textbf {\bibinfo {volume}
  {906}},\ \bibinfo {pages} {147} (\bibinfo {year} {2016})}\BibitemShut
  {NoStop}%
\bibitem [{\citenamefont {Kompaniets}\ and\ \citenamefont
  {Panzer}(2017)}]{kompaniets2017}%
  \BibitemOpen
  \bibfield  {author} {\bibinfo {author} {\bibfnamefont {M.~V.}\ \bibnamefont
  {Kompaniets}}\ and\ \bibinfo {author} {\bibfnamefont {E.}~\bibnamefont
  {Panzer}},\ }\href {\doibase 10.1103/PhysRevD.96.036016} {\bibfield
  {journal} {\bibinfo  {journal} {Phys. Rev. D}\ }\textbf {\bibinfo {volume}
  {96}},\ \bibinfo {pages} {036016} (\bibinfo {year} {2017})}\BibitemShut
  {NoStop}%
\bibitem [{\citenamefont {Schnetz}(2018)}]{schnetz2018}%
  \BibitemOpen
  \bibfield  {author} {\bibinfo {author} {\bibfnamefont {O.}~\bibnamefont
  {Schnetz}},\ }\href {\doibase 10.1103/PhysRevD.97.085018} {\bibfield
  {journal} {\bibinfo  {journal} {Phys. Rev. D}\ }\textbf {\bibinfo {volume}
  {97}},\ \bibinfo {pages} {085018} (\bibinfo {year} {2018})}\BibitemShut
  {NoStop}%
\bibitem [{\citenamefont {Guida}\ and\ \citenamefont
  {Zinn-Justin}(1998)}]{guida1998}%
  \BibitemOpen
  \bibfield  {author} {\bibinfo {author} {\bibfnamefont {R.}~\bibnamefont
  {Guida}}\ and\ \bibinfo {author} {\bibfnamefont {J.}~\bibnamefont
  {Zinn-Justin}},\ }\href {\doibase 10.1088/0305-4470/31/40/006} {\bibfield
  {journal} {\bibinfo  {journal} {J. Phys. A}\ }\textbf {\bibinfo {volume}
  {31}},\ \bibinfo {pages} {8103} (\bibinfo {year} {1998})}\BibitemShut
  {NoStop}%
\bibitem [{\citenamefont {Herbut}(2006)}]{herbut2006}%
  \BibitemOpen
  \bibfield  {author} {\bibinfo {author} {\bibfnamefont {I.~F.}\ \bibnamefont
  {Herbut}},\ }\href {\doibase 10.1103/PhysRevLett.97.146401} {\bibfield
  {journal} {\bibinfo  {journal} {Phys. Rev. Lett.}\ }\textbf {\bibinfo
  {volume} {97}},\ \bibinfo {pages} {146401} (\bibinfo {year}
  {2006})}\BibitemShut {NoStop}%
\bibitem [{\citenamefont {Grover}\ \emph {et~al.}(2014)\citenamefont {Grover},
  \citenamefont {Sheng},\ and\ \citenamefont {Vishwanath}}]{grover2014}%
  \BibitemOpen
  \bibfield  {author} {\bibinfo {author} {\bibfnamefont {T.}~\bibnamefont
  {Grover}}, \bibinfo {author} {\bibfnamefont {D.~N.}\ \bibnamefont {Sheng}}, \
  and\ \bibinfo {author} {\bibfnamefont {A.}~\bibnamefont {Vishwanath}},\
  }\href {\doibase 10.1126/science.1248253} {\bibfield  {journal} {\bibinfo
  {journal} {Science}\ }\textbf {\bibinfo {volume} {344}},\ \bibinfo {pages}
  {280} (\bibinfo {year} {2014})}\BibitemShut {NoStop}%
\bibitem [{\citenamefont {Lee}(2007)}]{lee2007}%
  \BibitemOpen
  \bibfield  {author} {\bibinfo {author} {\bibfnamefont {S.-S.}\ \bibnamefont
  {Lee}},\ }\href {\doibase 10.1103/PhysRevB.76.075103} {\bibfield  {journal}
  {\bibinfo  {journal} {Phys. Rev. B}\ }\textbf {\bibinfo {volume} {76}},\
  \bibinfo {pages} {075103} (\bibinfo {year} {2007})}\BibitemShut {NoStop}%
\bibitem [{\citenamefont {Roy}\ \emph {et~al.}(2013)\citenamefont {Roy},
  \citenamefont {{Juri\v{c}i\'c}},\ and\ \citenamefont {Herbut}}]{roy2013}%
  \BibitemOpen
  \bibfield  {author} {\bibinfo {author} {\bibfnamefont {B.}~\bibnamefont
  {Roy}}, \bibinfo {author} {\bibfnamefont {V.}~\bibnamefont
  {{Juri\v{c}i\'c}}}, \ and\ \bibinfo {author} {\bibfnamefont {I.~F.}\
  \bibnamefont {Herbut}},\ }\href {\doibase 10.1103/PhysRevB.87.041401}
  {\bibfield  {journal} {\bibinfo  {journal} {Phys. Rev. B}\ }\textbf {\bibinfo
  {volume} {87}},\ \bibinfo {pages} {041401} (\bibinfo {year}
  {2013})}\BibitemShut {NoStop}%
\bibitem [{\citenamefont {Nandkishore}\ \emph {et~al.}(2013)\citenamefont
  {Nandkishore}, \citenamefont {Maciejko}, \citenamefont {Huse},\ and\
  \citenamefont {Sondhi}}]{nandkishore2013}%
  \BibitemOpen
  \bibfield  {author} {\bibinfo {author} {\bibfnamefont {R.}~\bibnamefont
  {Nandkishore}}, \bibinfo {author} {\bibfnamefont {J.}~\bibnamefont
  {Maciejko}}, \bibinfo {author} {\bibfnamefont {D.~A.}\ \bibnamefont {Huse}},
  \ and\ \bibinfo {author} {\bibfnamefont {S.~L.}\ \bibnamefont {Sondhi}},\
  }\href {\doibase 10.1103/PhysRevB.87.174511} {\bibfield  {journal} {\bibinfo
  {journal} {Phys. Rev. B}\ }\textbf {\bibinfo {volume} {87}},\ \bibinfo
  {pages} {174511} (\bibinfo {year} {2013})}\BibitemShut {NoStop}%
\bibitem [{\citenamefont {Ponte}\ and\ \citenamefont {Lee}(2014)}]{ponte2014}%
  \BibitemOpen
  \bibfield  {author} {\bibinfo {author} {\bibfnamefont {P.}~\bibnamefont
  {Ponte}}\ and\ \bibinfo {author} {\bibfnamefont {S.-S.}\ \bibnamefont
  {Lee}},\ }\href {\doibase 10.1088/1367-2630/16/1/013044} {\bibfield
  {journal} {\bibinfo  {journal} {New J. Phys.}\ }\textbf {\bibinfo {volume}
  {16}},\ \bibinfo {pages} {013044} (\bibinfo {year} {2014})}\BibitemShut
  {NoStop}%
\bibitem [{\citenamefont {Zhou}\ \emph {et~al.}(2016)\citenamefont {Zhou},
  \citenamefont {Wang}, \citenamefont {Meng}, \citenamefont {Wang},\ and\
  \citenamefont {Wu}}]{zhou2016}%
  \BibitemOpen
  \bibfield  {author} {\bibinfo {author} {\bibfnamefont {Z.}~\bibnamefont
  {Zhou}}, \bibinfo {author} {\bibfnamefont {D.}~\bibnamefont {Wang}}, \bibinfo
  {author} {\bibfnamefont {Z.~Y.}\ \bibnamefont {Meng}}, \bibinfo {author}
  {\bibfnamefont {Y.}~\bibnamefont {Wang}}, \ and\ \bibinfo {author}
  {\bibfnamefont {C.}~\bibnamefont {Wu}},\ }\href {\doibase
  10.1103/PhysRevB.93.245157} {\bibfield  {journal} {\bibinfo  {journal} {Phys.
  Rev. B}\ }\textbf {\bibinfo {volume} {93}},\ \bibinfo {pages} {245157}
  (\bibinfo {year} {2016})}\BibitemShut {NoStop}%
\bibitem [{\citenamefont {Li}\ \emph {et~al.}(2017)\citenamefont {Li},
  \citenamefont {Jiang}, \citenamefont {Jian},\ and\ \citenamefont
  {Yao}}]{li2017b}%
  \BibitemOpen
  \bibfield  {author} {\bibinfo {author} {\bibfnamefont {Z.-X.}\ \bibnamefont
  {Li}}, \bibinfo {author} {\bibfnamefont {Y.-F.}\ \bibnamefont {Jiang}},
  \bibinfo {author} {\bibfnamefont {S.-K.}\ \bibnamefont {Jian}}, \ and\
  \bibinfo {author} {\bibfnamefont {H.}~\bibnamefont {Yao}},\ }\href {\doibase
  10.1038/s41467-017-00167-6} {\bibfield  {journal} {\bibinfo  {journal} {Nat.
  Commun.}\ }\textbf {\bibinfo {volume} {8}},\ \bibinfo {pages} {314} (\bibinfo
  {year} {2017})}\BibitemShut {NoStop}%
\bibitem [{\citenamefont {Classen}\ \emph {et~al.}(2017)\citenamefont
  {Classen}, \citenamefont {Herbut},\ and\ \citenamefont
  {Scherer}}]{classen2017}%
  \BibitemOpen
  \bibfield  {author} {\bibinfo {author} {\bibfnamefont {L.}~\bibnamefont
  {Classen}}, \bibinfo {author} {\bibfnamefont {I.~F.}\ \bibnamefont {Herbut}},
  \ and\ \bibinfo {author} {\bibfnamefont {M.~M.}\ \bibnamefont {Scherer}},\
  }\href {\doibase 10.1103/PhysRevB.96.115132} {\bibfield  {journal} {\bibinfo
  {journal} {Phys. Rev. B}\ }\textbf {\bibinfo {volume} {96}},\ \bibinfo
  {pages} {115132} (\bibinfo {year} {2017})}\BibitemShut {NoStop}%
\bibitem [{\citenamefont {Sorella}\ and\ \citenamefont
  {Tosatti}(1992)}]{sorella1992}%
  \BibitemOpen
  \bibfield  {author} {\bibinfo {author} {\bibfnamefont {S.}~\bibnamefont
  {Sorella}}\ and\ \bibinfo {author} {\bibfnamefont {E.}~\bibnamefont
  {Tosatti}},\ }\href {\doibase 10.1209/0295-5075/19/8/007} {\bibfield
  {journal} {\bibinfo  {journal} {Europhys. Lett.}\ }\textbf {\bibinfo {volume}
  {19}},\ \bibinfo {pages} {699} (\bibinfo {year} {1992})}\BibitemShut
  {NoStop}%
\bibitem [{\citenamefont {Honerkamp}(2008)}]{honerkamp2008}%
  \BibitemOpen
  \bibfield  {author} {\bibinfo {author} {\bibfnamefont {C.}~\bibnamefont
  {Honerkamp}},\ }\href {\doibase 10.1103/PhysRevLett.100.146404} {\bibfield
  {journal} {\bibinfo  {journal} {Phys. Rev. Lett.}\ }\textbf {\bibinfo
  {volume} {100}},\ \bibinfo {pages} {146404} (\bibinfo {year}
  {2008})}\BibitemShut {NoStop}%
\bibitem [{\citenamefont {Sorella}\ \emph {et~al.}(2012)\citenamefont
  {Sorella}, \citenamefont {Otsuka},\ and\ \citenamefont
  {Yunoki}}]{sorella2012}%
  \BibitemOpen
  \bibfield  {author} {\bibinfo {author} {\bibfnamefont {S.}~\bibnamefont
  {Sorella}}, \bibinfo {author} {\bibfnamefont {Y.}~\bibnamefont {Otsuka}}, \
  and\ \bibinfo {author} {\bibfnamefont {S.}~\bibnamefont {Yunoki}},\ }\href
  {\doibase 10.1038/srep00992} {\bibfield  {journal} {\bibinfo  {journal} {Sci.
  Rep.}\ }\textbf {\bibinfo {volume} {2}},\ \bibinfo {pages} {992} (\bibinfo
  {year} {2012})}\BibitemShut {NoStop}%
\bibitem [{\citenamefont {Zinn-Justin}(1991)}]{zinn-justin1991}%
  \BibitemOpen
  \bibfield  {author} {\bibinfo {author} {\bibfnamefont {J.}~\bibnamefont
  {Zinn-Justin}},\ }\href {\doibase 10.1016/0550-3213(91)90043-W} {\bibfield
  {journal} {\bibinfo  {journal} {Nucl. Phys. B}\ }\textbf {\bibinfo {volume}
  {367}},\ \bibinfo {pages} {105} (\bibinfo {year} {1991})}\BibitemShut
  {NoStop}%
\bibitem [{\citenamefont {Gross}\ and\ \citenamefont
  {Neveu}(1974)}]{gross1974}%
  \BibitemOpen
  \bibfield  {author} {\bibinfo {author} {\bibfnamefont {D.~J.}\ \bibnamefont
  {Gross}}\ and\ \bibinfo {author} {\bibfnamefont {A.}~\bibnamefont {Neveu}},\
  }\href {\doibase 10.1103/PhysRevD.10.3235} {\bibfield  {journal} {\bibinfo
  {journal} {Phys. Rev. D}\ }\textbf {\bibinfo {volume} {10}},\ \bibinfo
  {pages} {3235} (\bibinfo {year} {1974})}\BibitemShut {NoStop}%
\bibitem [{\citenamefont {Mihaila}\ \emph {et~al.}(2017)\citenamefont
  {Mihaila}, \citenamefont {Zerf}, \citenamefont {Ihrig}, \citenamefont
  {Herbut},\ and\ \citenamefont {Scherer}}]{mihaila2017}%
  \BibitemOpen
  \bibfield  {author} {\bibinfo {author} {\bibfnamefont {L.~N.}\ \bibnamefont
  {Mihaila}}, \bibinfo {author} {\bibfnamefont {N.}~\bibnamefont {Zerf}},
  \bibinfo {author} {\bibfnamefont {B.}~\bibnamefont {Ihrig}}, \bibinfo
  {author} {\bibfnamefont {I.~F.}\ \bibnamefont {Herbut}}, \ and\ \bibinfo
  {author} {\bibfnamefont {M.~M.}\ \bibnamefont {Scherer}},\ }\href {\doibase
  10.1103/PhysRevB.96.165133} {\bibfield  {journal} {\bibinfo  {journal} {Phys.
  Rev. B}\ }\textbf {\bibinfo {volume} {96}},\ \bibinfo {pages} {165133}
  (\bibinfo {year} {2017})}\BibitemShut {NoStop}%
\bibitem [{\citenamefont {Zerf}\ \emph {et~al.}(2017)\citenamefont {Zerf},
  \citenamefont {Mihaila}, \citenamefont {Marquard}, \citenamefont {Herbut},\
  and\ \citenamefont {Scherer}}]{zerf2017}%
  \BibitemOpen
  \bibfield  {author} {\bibinfo {author} {\bibfnamefont {N.}~\bibnamefont
  {Zerf}}, \bibinfo {author} {\bibfnamefont {L.~N.}\ \bibnamefont {Mihaila}},
  \bibinfo {author} {\bibfnamefont {P.}~\bibnamefont {Marquard}}, \bibinfo
  {author} {\bibfnamefont {I.~F.}\ \bibnamefont {Herbut}}, \ and\ \bibinfo
  {author} {\bibfnamefont {M.~M.}\ \bibnamefont {Scherer}},\ }\href {\doibase
  10.1103/PhysRevD.96.096010} {\bibfield  {journal} {\bibinfo  {journal} {Phys.
  Rev. D}\ }\textbf {\bibinfo {volume} {96}},\ \bibinfo {pages} {096010}
  (\bibinfo {year} {2017})}\BibitemShut {NoStop}%
\bibitem [{\citenamefont {Gracey}\ \emph {et~al.}(2016)\citenamefont {Gracey},
  \citenamefont {Luthe},\ and\ \citenamefont {{Schr\"oder}}}]{gracey2016}%
  \BibitemOpen
  \bibfield  {author} {\bibinfo {author} {\bibfnamefont {J.~A.}\ \bibnamefont
  {Gracey}}, \bibinfo {author} {\bibfnamefont {T.}~\bibnamefont {Luthe}}, \
  and\ \bibinfo {author} {\bibfnamefont {Y.}~\bibnamefont {{Schr\"oder}}},\
  }\href {\doibase 10.1103/PhysRevD.94.125028} {\bibfield  {journal} {\bibinfo
  {journal} {Phys. Rev. D}\ }\textbf {\bibinfo {volume} {94}},\ \bibinfo
  {pages} {125028} (\bibinfo {year} {2016})}\BibitemShut {NoStop}%
\bibitem [{\citenamefont {Sonoda}(2011)}]{sonoda2011}%
  \BibitemOpen
  \bibfield  {author} {\bibinfo {author} {\bibfnamefont {H.}~\bibnamefont
  {Sonoda}},\ }\href {\doibase 10.1143/PTP.126.57} {\bibfield  {journal}
  {\bibinfo  {journal} {Prog. Theor. Phys.}\ }\textbf {\bibinfo {volume}
  {126}},\ \bibinfo {pages} {57} (\bibinfo {year} {2011})}\BibitemShut
  {NoStop}%
\bibitem [{\citenamefont {Zerf}\ \emph {et~al.}(2016)\citenamefont {Zerf},
  \citenamefont {Lin},\ and\ \citenamefont {Maciejko}}]{zerf2016}%
  \BibitemOpen
  \bibfield  {author} {\bibinfo {author} {\bibfnamefont {N.}~\bibnamefont
  {Zerf}}, \bibinfo {author} {\bibfnamefont {C.-H.}\ \bibnamefont {Lin}}, \
  and\ \bibinfo {author} {\bibfnamefont {J.}~\bibnamefont {Maciejko}},\ }\href
  {\doibase 10.1103/PhysRevB.94.205106} {\bibfield  {journal} {\bibinfo
  {journal} {Phys. Rev. B}\ }\textbf {\bibinfo {volume} {94}},\ \bibinfo
  {pages} {205106} (\bibinfo {year} {2016})}\BibitemShut {NoStop}%
\bibitem [{\citenamefont {Fei}\ \emph {et~al.}(2016)\citenamefont {Fei},
  \citenamefont {Giombi}, \citenamefont {Klebanov},\ and\ \citenamefont
  {Tarnopolsky}}]{fei2016}%
  \BibitemOpen
  \bibfield  {author} {\bibinfo {author} {\bibfnamefont {L.}~\bibnamefont
  {Fei}}, \bibinfo {author} {\bibfnamefont {S.}~\bibnamefont {Giombi}},
  \bibinfo {author} {\bibfnamefont {I.~R.}\ \bibnamefont {Klebanov}}, \ and\
  \bibinfo {author} {\bibfnamefont {G.}~\bibnamefont {Tarnopolsky}},\ }\href
  {\doibase 10.1093/ptep/ptw120} {\bibfield  {journal} {\bibinfo  {journal}
  {Prog. Theor. Exp. Phys.}\ }\textbf {\bibinfo {volume} {2016}},\ \bibinfo
  {pages} {12C105} (\bibinfo {year} {2016})},\ \Eprint
  {http://arxiv.org/abs/arXiv:1607.05316} {arXiv:1607.05316} \BibitemShut
  {NoStop}%
\bibitem [{\citenamefont {Wen}(2004)}]{WenBook}%
  \BibitemOpen
  \bibfield  {author} {\bibinfo {author} {\bibfnamefont {X.~G.}\ \bibnamefont
  {Wen}},\ }\href@noop {} {\emph {\bibinfo {title} {Quantum Field Theory of
  Many-Body Systems: From the Origin of Sound to an Origin of Light and
  Electrons}}}\ (\bibinfo  {publisher} {Oxford University Press},\ \bibinfo
  {address} {Oxford},\ \bibinfo {year} {2004})\BibitemShut {NoStop}%
\bibitem [{\citenamefont {Senthil}\ \emph
  {et~al.}(2004{\natexlab{a}})\citenamefont {Senthil}, \citenamefont
  {Vishwanath}, \citenamefont {Balents}, \citenamefont {Sachdev},\ and\
  \citenamefont {Fisher}}]{senthil2004}%
  \BibitemOpen
  \bibfield  {author} {\bibinfo {author} {\bibfnamefont {T.}~\bibnamefont
  {Senthil}}, \bibinfo {author} {\bibfnamefont {A.}~\bibnamefont {Vishwanath}},
  \bibinfo {author} {\bibfnamefont {L.}~\bibnamefont {Balents}}, \bibinfo
  {author} {\bibfnamefont {S.}~\bibnamefont {Sachdev}}, \ and\ \bibinfo
  {author} {\bibfnamefont {M.~P.~A.}\ \bibnamefont {Fisher}},\ }\href {\doibase
  10.1126/science.1091806} {\bibfield  {journal} {\bibinfo  {journal}
  {Science}\ }\textbf {\bibinfo {volume} {303}},\ \bibinfo {pages} {1490}
  (\bibinfo {year} {2004}{\natexlab{a}})}\BibitemShut {NoStop}%
\bibitem [{\citenamefont {Senthil}\ \emph
  {et~al.}(2004{\natexlab{b}})\citenamefont {Senthil}, \citenamefont {Balents},
  \citenamefont {Sachdev}, \citenamefont {Vishwanath},\ and\ \citenamefont
  {Fisher}}]{senthil2004b}%
  \BibitemOpen
  \bibfield  {author} {\bibinfo {author} {\bibfnamefont {T.}~\bibnamefont
  {Senthil}}, \bibinfo {author} {\bibfnamefont {L.}~\bibnamefont {Balents}},
  \bibinfo {author} {\bibfnamefont {S.}~\bibnamefont {Sachdev}}, \bibinfo
  {author} {\bibfnamefont {A.}~\bibnamefont {Vishwanath}}, \ and\ \bibinfo
  {author} {\bibfnamefont {M.~P.~A.}\ \bibnamefont {Fisher}},\ }\href {\doibase
  10.1103/PhysRevB.70.144407} {\bibfield  {journal} {\bibinfo  {journal} {Phys.
  Rev. B}\ }\textbf {\bibinfo {volume} {70}},\ \bibinfo {pages} {144407}
  (\bibinfo {year} {2004}{\natexlab{b}})}\BibitemShut {NoStop}%
\bibitem [{\citenamefont {Affleck}\ and\ \citenamefont
  {Marston}(1988)}]{affleck1988}%
  \BibitemOpen
  \bibfield  {author} {\bibinfo {author} {\bibfnamefont {I.}~\bibnamefont
  {Affleck}}\ and\ \bibinfo {author} {\bibfnamefont {J.~B.}\ \bibnamefont
  {Marston}},\ }\href {\doibase 10.1103/PhysRevB.37.3774} {\bibfield  {journal}
  {\bibinfo  {journal} {Phys. Rev. B}\ }\textbf {\bibinfo {volume} {37}},\
  \bibinfo {pages} {3774} (\bibinfo {year} {1988})}\BibitemShut {NoStop}%
\bibitem [{\citenamefont {He}\ \emph {et~al.}(2015)\citenamefont {He},
  \citenamefont {Fuji},\ and\ \citenamefont {Bhattacharjee}}]{he2015}%
  \BibitemOpen
  \bibfield  {author} {\bibinfo {author} {\bibfnamefont {Y.-C.}\ \bibnamefont
  {He}}, \bibinfo {author} {\bibfnamefont {Y.}~\bibnamefont {Fuji}}, \ and\
  \bibinfo {author} {\bibfnamefont {S.}~\bibnamefont {Bhattacharjee}},\ }\href
  {http://arxiv.org/abs/1512.05381} {\bibfield  {journal} {\bibinfo  {journal}
  {arXiv:1512.05381}\ } (\bibinfo {year} {2015})}\BibitemShut {NoStop}%
\bibitem [{\citenamefont {Janssen}\ and\ \citenamefont
  {He}(2017)}]{janssen2017}%
  \BibitemOpen
  \bibfield  {author} {\bibinfo {author} {\bibfnamefont {L.}~\bibnamefont
  {Janssen}}\ and\ \bibinfo {author} {\bibfnamefont {Y.-C.}\ \bibnamefont
  {He}},\ }\href {\doibase 10.1103/PhysRevB.96.205113} {\bibfield  {journal}
  {\bibinfo  {journal} {Phys. Rev. B}\ }\textbf {\bibinfo {volume} {96}},\
  \bibinfo {pages} {205113} (\bibinfo {year} {2017})}\BibitemShut {NoStop}%
\bibitem [{\citenamefont {Boyack}\ \emph {et~al.}(2018)\citenamefont {Boyack},
  \citenamefont {Lin}, \citenamefont {Zerf}, \citenamefont {Rayyan},\ and\
  \citenamefont {Maciejko}}]{boyack2018}%
  \BibitemOpen
  \bibfield  {author} {\bibinfo {author} {\bibfnamefont {R.}~\bibnamefont
  {Boyack}}, \bibinfo {author} {\bibfnamefont {C.-H.}\ \bibnamefont {Lin}},
  \bibinfo {author} {\bibfnamefont {N.}~\bibnamefont {Zerf}}, \bibinfo {author}
  {\bibfnamefont {A.}~\bibnamefont {Rayyan}}, \ and\ \bibinfo {author}
  {\bibfnamefont {J.}~\bibnamefont {Maciejko}},\ }\href {\doibase
  10.1103/PhysRevB.98.035137} {\bibfield  {journal} {\bibinfo  {journal} {Phys.
  Rev. B}\ }\textbf {\bibinfo {volume} {98}},\ \bibinfo {pages} {035137}
  (\bibinfo {year} {2018})}\BibitemShut {NoStop}%
\bibitem [{\citenamefont {Xu}\ \emph {et~al.}(2018)\citenamefont {Xu},
  \citenamefont {Qi}, \citenamefont {Zhang}, \citenamefont {Assaad},
  \citenamefont {Xu},\ and\ \citenamefont {Meng}}]{xu2018}%
  \BibitemOpen
  \bibfield  {author} {\bibinfo {author} {\bibfnamefont {X.~Y.}\ \bibnamefont
  {Xu}}, \bibinfo {author} {\bibfnamefont {Y.}~\bibnamefont {Qi}}, \bibinfo
  {author} {\bibfnamefont {L.}~\bibnamefont {Zhang}}, \bibinfo {author}
  {\bibfnamefont {F.~F.}\ \bibnamefont {Assaad}}, \bibinfo {author}
  {\bibfnamefont {C.}~\bibnamefont {Xu}}, \ and\ \bibinfo {author}
  {\bibfnamefont {Z.~Y.}\ \bibnamefont {Meng}},\ }\href
  {http://arxiv.org/abs/1807.07574} {\bibfield  {journal} {\bibinfo  {journal}
  {arXiv:1807.07574}\ } (\bibinfo {year} {2018})}\BibitemShut {NoStop}%
\bibitem [{\citenamefont {Wang}\ \emph {et~al.}(2017)\citenamefont {Wang},
  \citenamefont {Nahum}, \citenamefont {Metlitski}, \citenamefont {Xu},\ and\
  \citenamefont {Senthil}}]{wang2017}%
  \BibitemOpen
  \bibfield  {author} {\bibinfo {author} {\bibfnamefont {C.}~\bibnamefont
  {Wang}}, \bibinfo {author} {\bibfnamefont {A.}~\bibnamefont {Nahum}},
  \bibinfo {author} {\bibfnamefont {M.~A.}\ \bibnamefont {Metlitski}}, \bibinfo
  {author} {\bibfnamefont {C.}~\bibnamefont {Xu}}, \ and\ \bibinfo {author}
  {\bibfnamefont {T.}~\bibnamefont {Senthil}},\ }\href {\doibase
  10.1103/PhysRevX.7.031051} {\bibfield  {journal} {\bibinfo  {journal} {Phys.
  Rev. X}\ }\textbf {\bibinfo {volume} {7}},\ \bibinfo {pages} {031051}
  (\bibinfo {year} {2017})}\BibitemShut {NoStop}%
\bibitem [{\citenamefont {He}\ and\ \citenamefont {Chen}(2015)}]{he2015b}%
  \BibitemOpen
  \bibfield  {author} {\bibinfo {author} {\bibfnamefont {Y.-C.}\ \bibnamefont
  {He}}\ and\ \bibinfo {author} {\bibfnamefont {Y.}~\bibnamefont {Chen}},\
  }\href {\doibase 10.1103/PhysRevLett.114.037201} {\bibfield  {journal}
  {\bibinfo  {journal} {Phys. Rev. Lett.}\ }\textbf {\bibinfo {volume} {114}},\
  \bibinfo {pages} {037201} (\bibinfo {year} {2015})}\BibitemShut {NoStop}%
\bibitem [{\citenamefont {Zinn-Justin}(2002)}]{ZJ}%
  \BibitemOpen
  \bibfield  {author} {\bibinfo {author} {\bibfnamefont {J.}~\bibnamefont
  {Zinn-Justin}},\ }\href@noop {} {\emph {\bibinfo {title} {Quantum Field
  Theory and Critical Phenomena}}},\ \bibinfo {edition} {4th}\ ed.\ (\bibinfo
  {publisher} {Clarendon Press},\ \bibinfo {address} {Oxford},\ \bibinfo {year}
  {2002})\BibitemShut {NoStop}%
\bibitem [{Sup()}]{SuppMat}%
  \BibitemOpen
  \href@noop {} {}\bibinfo {note} {See Supplemental Material at [URL] for the
  full analytical four-loop expressions for general $N$ for the $\gamma$ and
  $\beta$ renormalization group functions, as well as for the critical
  exponents and scaling dimensions.}\BibitemShut {Stop}%
\bibitem [{vla()}]{vladimirov1979}%
  \BibitemOpen
  \href@noop {} {}\bibinfo {note} {A. A. Vladimirov, D. I. Kazakov, and O. V.
  Tarasov, Sov. Phys. JETP {\bf 50}, 521 (1979) [Zh. Eksp. Teor. Fiz. {\bf 77},
  1035 (1979)].}\BibitemShut {Stop}%
\bibitem [{\citenamefont {Gorishny}\ \emph {et~al.}(1991)\citenamefont
  {Gorishny}, \citenamefont {Kataev}, \citenamefont {Larin},\ and\
  \citenamefont {Surguladze}}]{gorishny1991}%
  \BibitemOpen
  \bibfield  {author} {\bibinfo {author} {\bibfnamefont {S.~G.}\ \bibnamefont
  {Gorishny}}, \bibinfo {author} {\bibfnamefont {A.~L.}\ \bibnamefont
  {Kataev}}, \bibinfo {author} {\bibfnamefont {S.~A.}\ \bibnamefont {Larin}}, \
  and\ \bibinfo {author} {\bibfnamefont {L.~R.}\ \bibnamefont {Surguladze}},\
  }\href {\doibase 10.1016/0370-2693(91)90222-C} {\bibfield  {journal}
  {\bibinfo  {journal} {Phys. Lett. B}\ }\textbf {\bibinfo {volume} {256}},\
  \bibinfo {pages} {81} (\bibinfo {year} {1991})}\BibitemShut {NoStop}%
\bibitem [{\citenamefont {Herbut}(2007)}]{IgorBook}%
  \BibitemOpen
  \bibfield  {author} {\bibinfo {author} {\bibfnamefont {I.}~\bibnamefont
  {Herbut}},\ }\href@noop {} {\emph {\bibinfo {title} {A Modern Approach to
  Critical Phenomena}}}\ (\bibinfo  {publisher} {Cambridge University Press},\
  \bibinfo {address} {Cambridge},\ \bibinfo {year} {2007})\BibitemShut
  {NoStop}%
\bibitem [{\citenamefont {Gracey}(2007)}]{gracey2007}%
  \BibitemOpen
  \bibfield  {author} {\bibinfo {author} {\bibfnamefont {J.~A.}\ \bibnamefont
  {Gracey}},\ }\href {\doibase 10.1088/1751-8113/40/46/011} {\bibfield
  {journal} {\bibinfo  {journal} {J. Phys. A}\ }\textbf {\bibinfo {volume}
  {40}},\ \bibinfo {pages} {13989} (\bibinfo {year} {2007})}\BibitemShut
  {NoStop}%
\bibitem [{\citenamefont {Ki{\ss}ler}\ and\ \citenamefont
  {Kreimer}(2017)}]{kisler2017}%
  \BibitemOpen
  \bibfield  {author} {\bibinfo {author} {\bibfnamefont {H.}~\bibnamefont
  {Ki{\ss}ler}}\ and\ \bibinfo {author} {\bibfnamefont {D.}~\bibnamefont
  {Kreimer}},\ }\href {\doibase 10.1016/j.physletb.2016.11.052} {\bibfield
  {journal} {\bibinfo  {journal} {Phys. Lett. B}\ }\textbf {\bibinfo {volume}
  {764}},\ \bibinfo {pages} {318} (\bibinfo {year} {2017})}\BibitemShut
  {NoStop}%
\bibitem [{\citenamefont {Chetyrkin}(1997)}]{Chetyrkin:1997dh}%
  \BibitemOpen
  \bibfield  {author} {\bibinfo {author} {\bibfnamefont {K.~G.}\ \bibnamefont
  {Chetyrkin}},\ }\href {\doibase 10.1016/S0370-2693(97)00535-2} {\bibfield
  {journal} {\bibinfo  {journal} {Phys. Lett. B}\ }\textbf {\bibinfo {volume}
  {404}},\ \bibinfo {pages} {161} (\bibinfo {year} {1997})}\BibitemShut
  {NoStop}%
\bibitem [{\citenamefont {Vermaseren}\ \emph {et~al.}(1997)\citenamefont
  {Vermaseren}, \citenamefont {Larin},\ and\ \citenamefont {van
  Ritbergen}}]{Vermaseren:1997fq}%
  \BibitemOpen
  \bibfield  {author} {\bibinfo {author} {\bibfnamefont {J.~A.~M.}\
  \bibnamefont {Vermaseren}}, \bibinfo {author} {\bibfnamefont {S.~A.}\
  \bibnamefont {Larin}}, \ and\ \bibinfo {author} {\bibfnamefont
  {T.}~\bibnamefont {van Ritbergen}},\ }\href {\doibase
  10.1016/S0370-2693(97)00660-6} {\bibfield  {journal} {\bibinfo  {journal}
  {Phys. Lett. B}\ }\textbf {\bibinfo {volume} {405}},\ \bibinfo {pages} {327}
  (\bibinfo {year} {1997})}\BibitemShut {NoStop}%
\bibitem [{\citenamefont {Chetyrkin}(2005)}]{Chetyrkin:2004mf}%
  \BibitemOpen
  \bibfield  {author} {\bibinfo {author} {\bibfnamefont {K.~G.}\ \bibnamefont
  {Chetyrkin}},\ }\href {\doibase 10.1016/j.nuclphysb.2005.01.011} {\bibfield
  {journal} {\bibinfo  {journal} {Nucl. Phys. B}\ }\textbf {\bibinfo {volume}
  {710}},\ \bibinfo {pages} {499} (\bibinfo {year} {2005})}\BibitemShut
  {NoStop}%
\bibitem [{\citenamefont {Czakon}(2005)}]{Czakon:2004bu}%
  \BibitemOpen
  \bibfield  {author} {\bibinfo {author} {\bibfnamefont {M.}~\bibnamefont
  {Czakon}},\ }\href {\doibase 10.1016/j.nuclphysb.2005.01.012} {\bibfield
  {journal} {\bibinfo  {journal} {Nucl. Phys. B}\ }\textbf {\bibinfo {volume}
  {710}},\ \bibinfo {pages} {485} (\bibinfo {year} {2005})}\BibitemShut
  {NoStop}%
\bibitem [{\citenamefont {Di~Pietro}\ \emph {et~al.}(2016)\citenamefont
  {Di~Pietro}, \citenamefont {Komargodski}, \citenamefont {Shamir},\ and\
  \citenamefont {Stamou}}]{dipietro2016}%
  \BibitemOpen
  \bibfield  {author} {\bibinfo {author} {\bibfnamefont {L.}~\bibnamefont
  {Di~Pietro}}, \bibinfo {author} {\bibfnamefont {Z.}~\bibnamefont
  {Komargodski}}, \bibinfo {author} {\bibfnamefont {I.}~\bibnamefont {Shamir}},
  \ and\ \bibinfo {author} {\bibfnamefont {E.}~\bibnamefont {Stamou}},\ }\href
  {\doibase 10.1103/PhysRevLett.116.131601} {\bibfield  {journal} {\bibinfo
  {journal} {Phys. Rev. Lett.}\ }\textbf {\bibinfo {volume} {116}},\ \bibinfo
  {pages} {131601} (\bibinfo {year} {2016})}\BibitemShut {NoStop}%
\bibitem [{\citenamefont {Gracey}(1992{\natexlab{a}})}]{gracey1992}%
  \BibitemOpen
  \bibfield  {author} {\bibinfo {author} {\bibfnamefont {J.~A.}\ \bibnamefont
  {Gracey}},\ }\href {\doibase 10.1088/0305-4470/25/3/005} {\bibfield
  {journal} {\bibinfo  {journal} {J. Phys. A}\ }\textbf {\bibinfo {volume}
  {25}},\ \bibinfo {pages} {L109} (\bibinfo {year}
  {1992}{\natexlab{a}})}\BibitemShut {NoStop}%
\bibitem [{Not()}]{NoteEtaPsi}%
  \BibitemOpen
  \href@noop {} {}\bibinfo {note} {As discussed briefly in Sec.~\ref{sec:Pade},
  the QED$_3$-GN model corresponds to supplementing QED$_3$ with four-fermion
  interactions, and is expected to be in the same universality class as the
  QED$_3$-GNY model. We have also verified that our four-loop result for the
  (non-gauge-invariant) fermion anomalous dimension $\eta_\psi$, calculated in
  a general $\xi$-gauge, agrees with the large-$N$ Landau gauge ($\xi=0$)
  result computed to $\mathcal{O}(1/N)$ in Ref.~\cite{gracey1992}, and first
  computed to $\mathcal{O}(1/N^2)$ in Ref.~\cite{gracey1993b} but recently
  corrected in Ref.~\cite{GraceyPrivateComm}.}\BibitemShut {Stop}%
\bibitem [{\citenamefont {{Di Pietro}}\ and\ \citenamefont
  {Stamou}(2017)}]{dipietro2017}%
  \BibitemOpen
  \bibfield  {author} {\bibinfo {author} {\bibfnamefont {L.}~\bibnamefont {{Di
  Pietro}}}\ and\ \bibinfo {author} {\bibfnamefont {E.}~\bibnamefont
  {Stamou}},\ }\href {\doibase 10.1007/JHEP12(2017)054} {\bibfield  {journal}
  {\bibinfo  {journal} {JHEP}\ }\textbf {\bibinfo {volume} {12}},\ \bibinfo
  {pages} {054} (\bibinfo {year} {2017})}\BibitemShut {NoStop}%
\bibitem [{\citenamefont {Janssen}(2016)}]{janssen2016}%
  \BibitemOpen
  \bibfield  {author} {\bibinfo {author} {\bibfnamefont {L.}~\bibnamefont
  {Janssen}},\ }\href {\doibase 10.1103/PhysRevD.94.094013} {\bibfield
  {journal} {\bibinfo  {journal} {Phys. Rev. D}\ }\textbf {\bibinfo {volume}
  {94}},\ \bibinfo {pages} {094013} (\bibinfo {year} {2016})}\BibitemShut
  {NoStop}%
\bibitem [{\citenamefont {Ferrara}\ \emph {et~al.}(1974)\citenamefont
  {Ferrara}, \citenamefont {Gatto},\ and\ \citenamefont
  {Grillo}}]{ferrara1974}%
  \BibitemOpen
  \bibfield  {author} {\bibinfo {author} {\bibfnamefont {S.}~\bibnamefont
  {Ferrara}}, \bibinfo {author} {\bibfnamefont {R.}~\bibnamefont {Gatto}}, \
  and\ \bibinfo {author} {\bibfnamefont {A.}~\bibnamefont {Grillo}},\ }\href
  {\doibase 10.1103/PhysRevD.9.3564} {\bibfield  {journal} {\bibinfo  {journal}
  {Phys. Rev. D}\ }\textbf {\bibinfo {volume} {9}},\ \bibinfo {pages} {3564}
  (\bibinfo {year} {1974})}\BibitemShut {NoStop}%
\bibitem [{\citenamefont {Mack}(1977)}]{mack1977}%
  \BibitemOpen
  \bibfield  {author} {\bibinfo {author} {\bibfnamefont {G.}~\bibnamefont
  {Mack}},\ }\href {\doibase 10.1007/BF01613145} {\bibfield  {journal}
  {\bibinfo  {journal} {Commun. Math. Phys.}\ }\textbf {\bibinfo {volume}
  {55}},\ \bibinfo {pages} {1} (\bibinfo {year} {1977})}\BibitemShut {NoStop}%
\bibitem [{\citenamefont {Gracey}(2018)}]{GraceyPrivateComm}%
  \BibitemOpen
  \bibfield  {author} {\bibinfo {author} {\bibfnamefont {J.~A.}\ \bibnamefont
  {Gracey}},\ }\href {http://arxiv.org/abs/1808.07697} {\bibfield  {journal}
  {\bibinfo  {journal} {arXiv:1808.07697}\ } (\bibinfo {year}
  {2018})}\BibitemShut {NoStop}%
\bibitem [{\citenamefont {Baikov}\ \emph {et~al.}(2007)\citenamefont {Baikov},
  \citenamefont {Chetyrkin},\ and\ \citenamefont {Kuhn}}]{Baikov:2008cp}%
  \BibitemOpen
  \bibfield  {author} {\bibinfo {author} {\bibfnamefont {P.~A.}\ \bibnamefont
  {Baikov}}, \bibinfo {author} {\bibfnamefont {K.~G.}\ \bibnamefont
  {Chetyrkin}}, \ and\ \bibinfo {author} {\bibfnamefont {J.~H.}\ \bibnamefont
  {Kuhn}},\ }\bibfield  {booktitle} {\emph {\bibinfo {booktitle} {{Applications
  of quantum field theory to phenomenology. Proceedings, 8th International
  Symposium on Radiative Corrections, RADCOR 2007, Florence, Italy, October
  1-5, 2007}}},\ }\href {\doibase 10.22323/1.048.0023} {\bibfield  {journal}
  {\bibinfo  {journal} {PoS}\ }\textbf {\bibinfo {volume} {RADCOR2007}},\
  \bibinfo {pages} {023} (\bibinfo {year} {2007})},\ \Eprint
  {http://arxiv.org/abs/0810.4048} {0810.4048} \BibitemShut {NoStop}%
\bibitem [{\citenamefont {Baikov}\ \emph {et~al.}(2010)\citenamefont {Baikov},
  \citenamefont {Chetyrkin},\ and\ \citenamefont {{K\"uhn}}}]{Baikov:2010je}%
  \BibitemOpen
  \bibfield  {author} {\bibinfo {author} {\bibfnamefont {P.~A.}\ \bibnamefont
  {Baikov}}, \bibinfo {author} {\bibfnamefont {K.~G.}\ \bibnamefont
  {Chetyrkin}}, \ and\ \bibinfo {author} {\bibfnamefont {J.~H.}\ \bibnamefont
  {{K\"uhn}}},\ }\href {\doibase 10.1103/PhysRevLett.104.132004} {\bibfield
  {journal} {\bibinfo  {journal} {Phys. Rev. Lett.}\ }\textbf {\bibinfo
  {volume} {104}},\ \bibinfo {pages} {132004} (\bibinfo {year}
  {2010})}\BibitemShut {NoStop}%
\bibitem [{\citenamefont {Baikov}\ \emph {et~al.}(2012)\citenamefont {Baikov},
  \citenamefont {Chetyrkin}, \citenamefont {{K\"uhn}},\ and\ \citenamefont
  {Rittinger}}]{Baikov:2012zm}%
  \BibitemOpen
  \bibfield  {author} {\bibinfo {author} {\bibfnamefont {P.~A.}\ \bibnamefont
  {Baikov}}, \bibinfo {author} {\bibfnamefont {K.~G.}\ \bibnamefont
  {Chetyrkin}}, \bibinfo {author} {\bibfnamefont {J.~H.}\ \bibnamefont
  {{K\"uhn}}}, \ and\ \bibinfo {author} {\bibfnamefont {J.}~\bibnamefont
  {Rittinger}},\ }\href {\doibase 10.1007/JHEP07(2012)017} {\bibfield
  {journal} {\bibinfo  {journal} {JHEP}\ }\textbf {\bibinfo {volume} {07}},\
  \bibinfo {pages} {017} (\bibinfo {year} {2012})}\BibitemShut {NoStop}%
\bibitem [{\citenamefont {Baikov}\ \emph {et~al.}(2014)\citenamefont {Baikov},
  \citenamefont {Chetyrkin},\ and\ \citenamefont {{K\"uhn}}}]{Baikov:2014qja}%
  \BibitemOpen
  \bibfield  {author} {\bibinfo {author} {\bibfnamefont {P.~A.}\ \bibnamefont
  {Baikov}}, \bibinfo {author} {\bibfnamefont {K.~G.}\ \bibnamefont
  {Chetyrkin}}, \ and\ \bibinfo {author} {\bibfnamefont {J.~H.}\ \bibnamefont
  {{K\"uhn}}},\ }\href {\doibase 10.1007/JHEP10(2014)076} {\bibfield  {journal}
  {\bibinfo  {journal} {JHEP}\ }\textbf {\bibinfo {volume} {10}},\ \bibinfo
  {pages} {076} (\bibinfo {year} {2014})}\BibitemShut {NoStop}%
\bibitem [{\citenamefont {Chetyrkin}\ \emph {et~al.}(2016)\citenamefont
  {Chetyrkin}, \citenamefont {Baikov},\ and\ \citenamefont
  {Kühn}}]{Chetyrkin:2016uhw}%
  \BibitemOpen
  \bibfield  {author} {\bibinfo {author} {\bibfnamefont {K.}~\bibnamefont
  {Chetyrkin}}, \bibinfo {author} {\bibfnamefont {P.}~\bibnamefont {Baikov}}, \
  and\ \bibinfo {author} {\bibfnamefont {J.}~\bibnamefont {Kühn}},\ }\bibfield
   {booktitle} {\emph {\bibinfo {booktitle} {{Proceedings, 13th DESY Workshop
  on Elementary Particle Physics: Loops and Legs in Quantum Field Theory
  (LL2016): Leipzig, Germany, April 24-29, 2016}}},\ }\href {\doibase
  10.22323/1.260.0010} {\bibfield  {journal} {\bibinfo  {journal} {PoS}\
  }\textbf {\bibinfo {volume} {LL2016}},\ \bibinfo {pages} {010} (\bibinfo
  {year} {2016})}\BibitemShut {NoStop}%
\bibitem [{\citenamefont {Baikov}\ \emph
  {et~al.}(2017{\natexlab{a}})\citenamefont {Baikov}, \citenamefont
  {Chetyrkin},\ and\ \citenamefont {{K\"uhn}}}]{Baikov:2016tgj}%
  \BibitemOpen
  \bibfield  {author} {\bibinfo {author} {\bibfnamefont {P.~A.}\ \bibnamefont
  {Baikov}}, \bibinfo {author} {\bibfnamefont {K.~G.}\ \bibnamefont
  {Chetyrkin}}, \ and\ \bibinfo {author} {\bibfnamefont {J.~H.}\ \bibnamefont
  {{K\"uhn}}},\ }\href {\doibase 10.1103/PhysRevLett.118.082002} {\bibfield
  {journal} {\bibinfo  {journal} {Phys. Rev. Lett.}\ }\textbf {\bibinfo
  {volume} {118}},\ \bibinfo {pages} {082002} (\bibinfo {year}
  {2017}{\natexlab{a}})}\BibitemShut {NoStop}%
\bibitem [{\citenamefont {Luthe}\ \emph {et~al.}(2016)\citenamefont {Luthe},
  \citenamefont {Maier}, \citenamefont {Marquard},\ and\ \citenamefont
  {{Schr\"oder}}}]{Luthe:2016ima}%
  \BibitemOpen
  \bibfield  {author} {\bibinfo {author} {\bibfnamefont {T.}~\bibnamefont
  {Luthe}}, \bibinfo {author} {\bibfnamefont {A.}~\bibnamefont {Maier}},
  \bibinfo {author} {\bibfnamefont {P.}~\bibnamefont {Marquard}}, \ and\
  \bibinfo {author} {\bibfnamefont {Y.}~\bibnamefont {{Schr\"oder}}},\ }\href
  {\doibase 10.1007/JHEP07(2016)127} {\bibfield  {journal} {\bibinfo  {journal}
  {JHEP}\ }\textbf {\bibinfo {volume} {07}},\ \bibinfo {pages} {127} (\bibinfo
  {year} {2016})}\BibitemShut {NoStop}%
\bibitem [{\citenamefont {Luthe}\ \emph
  {et~al.}(2017{\natexlab{a}})\citenamefont {Luthe}, \citenamefont {Maier},
  \citenamefont {Marquard},\ and\ \citenamefont
  {{Schr\"oder}}}]{Luthe:2016xec}%
  \BibitemOpen
  \bibfield  {author} {\bibinfo {author} {\bibfnamefont {T.}~\bibnamefont
  {Luthe}}, \bibinfo {author} {\bibfnamefont {A.}~\bibnamefont {Maier}},
  \bibinfo {author} {\bibfnamefont {P.}~\bibnamefont {Marquard}}, \ and\
  \bibinfo {author} {\bibfnamefont {Y.}~\bibnamefont {{Schr\"oder}}},\ }\href
  {\doibase 10.1007/JHEP01(2017)081} {\bibfield  {journal} {\bibinfo  {journal}
  {JHEP}\ }\textbf {\bibinfo {volume} {01}},\ \bibinfo {pages} {081} (\bibinfo
  {year} {2017}{\natexlab{a}})}\BibitemShut {NoStop}%
\bibitem [{\citenamefont {Herzog}\ \emph {et~al.}(2017)\citenamefont {Herzog},
  \citenamefont {Ruijl}, \citenamefont {Ueda}, \citenamefont {Vermaseren},\
  and\ \citenamefont {Vogt}}]{Herzog:2017ohr}%
  \BibitemOpen
  \bibfield  {author} {\bibinfo {author} {\bibfnamefont {F.}~\bibnamefont
  {Herzog}}, \bibinfo {author} {\bibfnamefont {B.}~\bibnamefont {Ruijl}},
  \bibinfo {author} {\bibfnamefont {T.}~\bibnamefont {Ueda}}, \bibinfo {author}
  {\bibfnamefont {J.~A.~M.}\ \bibnamefont {Vermaseren}}, \ and\ \bibinfo
  {author} {\bibfnamefont {A.}~\bibnamefont {Vogt}},\ }\href {\doibase
  10.1007/JHEP02(2017)090} {\bibfield  {journal} {\bibinfo  {journal} {JHEP}\
  }\textbf {\bibinfo {volume} {02}},\ \bibinfo {pages} {090} (\bibinfo {year}
  {2017})}\BibitemShut {NoStop}%
\bibitem [{\citenamefont {Luthe}\ \emph
  {et~al.}(2017{\natexlab{b}})\citenamefont {Luthe}, \citenamefont {Maier},
  \citenamefont {Marquard},\ and\ \citenamefont
  {{Schr\"oder}}}]{Luthe:2017ttc}%
  \BibitemOpen
  \bibfield  {author} {\bibinfo {author} {\bibfnamefont {T.}~\bibnamefont
  {Luthe}}, \bibinfo {author} {\bibfnamefont {A.}~\bibnamefont {Maier}},
  \bibinfo {author} {\bibfnamefont {P.}~\bibnamefont {Marquard}}, \ and\
  \bibinfo {author} {\bibfnamefont {Y.}~\bibnamefont {{Schr\"oder}}},\ }\href
  {\doibase 10.1007/JHEP03(2017)020} {\bibfield  {journal} {\bibinfo  {journal}
  {JHEP}\ }\textbf {\bibinfo {volume} {03}},\ \bibinfo {pages} {020} (\bibinfo
  {year} {2017}{\natexlab{b}})}\BibitemShut {NoStop}%
\bibitem [{\citenamefont {Baikov}\ \emph
  {et~al.}(2017{\natexlab{b}})\citenamefont {Baikov}, \citenamefont
  {Chetyrkin},\ and\ \citenamefont {{K\"uhn}}}]{Baikov:2017ujl}%
  \BibitemOpen
  \bibfield  {author} {\bibinfo {author} {\bibfnamefont {P.~A.}\ \bibnamefont
  {Baikov}}, \bibinfo {author} {\bibfnamefont {K.~G.}\ \bibnamefont
  {Chetyrkin}}, \ and\ \bibinfo {author} {\bibfnamefont {J.~H.}\ \bibnamefont
  {{K\"uhn}}},\ }\href {\doibase 10.1007/JHEP04(2017)119} {\bibfield  {journal}
  {\bibinfo  {journal} {JHEP}\ }\textbf {\bibinfo {volume} {04}},\ \bibinfo
  {pages} {119} (\bibinfo {year} {2017}{\natexlab{b}})}\BibitemShut {NoStop}%
\bibitem [{\citenamefont {Luthe}\ \emph
  {et~al.}(2017{\natexlab{c}})\citenamefont {Luthe}, \citenamefont {Maier},
  \citenamefont {Marquard},\ and\ \citenamefont
  {{Schr\"oder}}}]{Luthe:2017ttg}%
  \BibitemOpen
  \bibfield  {author} {\bibinfo {author} {\bibfnamefont {T.}~\bibnamefont
  {Luthe}}, \bibinfo {author} {\bibfnamefont {A.}~\bibnamefont {Maier}},
  \bibinfo {author} {\bibfnamefont {P.}~\bibnamefont {Marquard}}, \ and\
  \bibinfo {author} {\bibfnamefont {Y.}~\bibnamefont {{Schr\"oder}}},\ }\href
  {\doibase 10.1007/JHEP10(2017)166} {\bibfield  {journal} {\bibinfo  {journal}
  {JHEP}\ }\textbf {\bibinfo {volume} {10}},\ \bibinfo {pages} {166} (\bibinfo
  {year} {2017}{\natexlab{c}})}\BibitemShut {NoStop}%
\bibitem [{\citenamefont {Chetyrkin}\ \emph {et~al.}(2017)\citenamefont
  {Chetyrkin}, \citenamefont {Falcioni}, \citenamefont {Herzog},\ and\
  \citenamefont {Vermaseren}}]{Chetyrkin:2017bjc}%
  \BibitemOpen
  \bibfield  {author} {\bibinfo {author} {\bibfnamefont {K.~G.}\ \bibnamefont
  {Chetyrkin}}, \bibinfo {author} {\bibfnamefont {G.}~\bibnamefont {Falcioni}},
  \bibinfo {author} {\bibfnamefont {F.}~\bibnamefont {Herzog}}, \ and\ \bibinfo
  {author} {\bibfnamefont {J.~A.~M.}\ \bibnamefont {Vermaseren}},\ }\href
  {\doibase 10.1007/JHEP10(2017)179} {\bibfield  {journal} {\bibinfo  {journal}
  {JHEP}\ }\textbf {\bibinfo {volume} {10}},\ \bibinfo {pages} {179} (\bibinfo
  {year} {2017})}\BibitemShut {NoStop}%
\bibitem [{\citenamefont {Gracey}(1993{\natexlab{a}})}]{gracey1993}%
  \BibitemOpen
  \bibfield  {author} {\bibinfo {author} {\bibfnamefont {J.~A.}\ \bibnamefont
  {Gracey}},\ }\href {\doibase 10.1016/0370-2693(93)91017-H} {\bibfield
  {journal} {\bibinfo  {journal} {Phys. Lett. B}\ }\textbf {\bibinfo {volume}
  {317}},\ \bibinfo {pages} {415} (\bibinfo {year}
  {1993}{\natexlab{a}})}\BibitemShut {NoStop}%
\bibitem [{\citenamefont {Vafa}\ and\ \citenamefont {Witten}(1984)}]{vafa1984}%
  \BibitemOpen
  \bibfield  {author} {\bibinfo {author} {\bibfnamefont {C.}~\bibnamefont
  {Vafa}}\ and\ \bibinfo {author} {\bibfnamefont {E.}~\bibnamefont {Witten}},\
  }\href {http://projecteuclid.org/euclid.cmp/1103941573} {\bibfield  {journal}
  {\bibinfo  {journal} {Commun. Math. Phys.}\ }\textbf {\bibinfo {volume}
  {95}},\ \bibinfo {pages} {257} (\bibinfo {year} {1984})}\BibitemShut
  {NoStop}%
\bibitem [{\citenamefont {Pufu}(2014)}]{pufu2014}%
  \BibitemOpen
  \bibfield  {author} {\bibinfo {author} {\bibfnamefont {S.~S.}\ \bibnamefont
  {Pufu}},\ }\href {\doibase 10.1103/PhysRevD.89.065016} {\bibfield  {journal}
  {\bibinfo  {journal} {Phys. Rev. D}\ }\textbf {\bibinfo {volume} {89}},\
  \bibinfo {pages} {065016} (\bibinfo {year} {2014})}\BibitemShut {NoStop}%
\bibitem [{\citenamefont {Hands}\ \emph {et~al.}(2002)\citenamefont {Hands},
  \citenamefont {Kogut},\ and\ \citenamefont {Strouthos}}]{hands2002}%
  \BibitemOpen
  \bibfield  {author} {\bibinfo {author} {\bibfnamefont {S.~J.}\ \bibnamefont
  {Hands}}, \bibinfo {author} {\bibfnamefont {J.~B.}\ \bibnamefont {Kogut}}, \
  and\ \bibinfo {author} {\bibfnamefont {C.~G.}\ \bibnamefont {Strouthos}},\
  }\href {\doibase 10.1016/S0550-3213(02)00869-6} {\bibfield  {journal}
  {\bibinfo  {journal} {Nucl. Phys. B}\ }\textbf {\bibinfo {volume} {645}},\
  \bibinfo {pages} {321} (\bibinfo {year} {2002})}\BibitemShut {NoStop}%
\bibitem [{\citenamefont {Hands}\ \emph {et~al.}(2004)\citenamefont {Hands},
  \citenamefont {Kogut}, \citenamefont {Scorzato},\ and\ \citenamefont
  {Strouthos}}]{hands2004}%
  \BibitemOpen
  \bibfield  {author} {\bibinfo {author} {\bibfnamefont {S.~J.}\ \bibnamefont
  {Hands}}, \bibinfo {author} {\bibfnamefont {J.~B.}\ \bibnamefont {Kogut}},
  \bibinfo {author} {\bibfnamefont {L.}~\bibnamefont {Scorzato}}, \ and\
  \bibinfo {author} {\bibfnamefont {C.~G.}\ \bibnamefont {Strouthos}},\ }\href
  {\doibase 10.1103/PhysRevB.70.104501} {\bibfield  {journal} {\bibinfo
  {journal} {Phys. Rev. B}\ }\textbf {\bibinfo {volume} {70}},\ \bibinfo
  {pages} {104501} (\bibinfo {year} {2004})}\BibitemShut {NoStop}%
\bibitem [{\citenamefont {Strouthos}\ and\ \citenamefont
  {Kogut}(2009)}]{strouthos2009}%
  \BibitemOpen
  \bibfield  {author} {\bibinfo {author} {\bibfnamefont {C.}~\bibnamefont
  {Strouthos}}\ and\ \bibinfo {author} {\bibfnamefont {J.~B.}\ \bibnamefont
  {Kogut}},\ }\href {\doibase 10.1088/1742-6596/150/5/052247} {\bibfield
  {journal} {\bibinfo  {journal} {J. Phys. Conf. Ser.}\ }\textbf {\bibinfo
  {volume} {150}},\ \bibinfo {pages} {052247} (\bibinfo {year}
  {2009})}\BibitemShut {NoStop}%
\bibitem [{\citenamefont {Karthik}\ and\ \citenamefont
  {Narayanan}(2016{\natexlab{a}})}]{karthik2016}%
  \BibitemOpen
  \bibfield  {author} {\bibinfo {author} {\bibfnamefont {N.}~\bibnamefont
  {Karthik}}\ and\ \bibinfo {author} {\bibfnamefont {R.}~\bibnamefont
  {Narayanan}},\ }\href {\doibase 10.1103/PhysRevD.93.045020} {\bibfield
  {journal} {\bibinfo  {journal} {Phys. Rev. D}\ }\textbf {\bibinfo {volume}
  {93}},\ \bibinfo {pages} {045020} (\bibinfo {year}
  {2016}{\natexlab{a}})}\BibitemShut {NoStop}%
\bibitem [{\citenamefont {Karthik}\ and\ \citenamefont
  {Narayanan}(2016{\natexlab{b}})}]{karthik2016b}%
  \BibitemOpen
  \bibfield  {author} {\bibinfo {author} {\bibfnamefont {N.}~\bibnamefont
  {Karthik}}\ and\ \bibinfo {author} {\bibfnamefont {R.}~\bibnamefont
  {Narayanan}},\ }\href {\doibase 10.1103/PhysRevD.94.065026} {\bibfield
  {journal} {\bibinfo  {journal} {Phys. Rev. D}\ }\textbf {\bibinfo {volume}
  {94}},\ \bibinfo {pages} {065026} (\bibinfo {year}
  {2016}{\natexlab{b}})}\BibitemShut {NoStop}%
\bibitem [{\citenamefont {Karthik}\ and\ \citenamefont
  {Narayanan}(2017)}]{karthik2017}%
  \BibitemOpen
  \bibfield  {author} {\bibinfo {author} {\bibfnamefont {N.}~\bibnamefont
  {Karthik}}\ and\ \bibinfo {author} {\bibfnamefont {R.}~\bibnamefont
  {Narayanan}},\ }\href {\doibase 10.1103/PhysRevD.96.054509} {\bibfield
  {journal} {\bibinfo  {journal} {Phys. Rev. D}\ }\textbf {\bibinfo {volume}
  {96}},\ \bibinfo {pages} {054509} (\bibinfo {year} {2017})}\BibitemShut
  {NoStop}%
\bibitem [{\citenamefont {Giombi}\ \emph {et~al.}(2017)\citenamefont {Giombi},
  \citenamefont {Kirilin},\ and\ \citenamefont {Skvortsov}}]{giombi2017}%
  \BibitemOpen
  \bibfield  {author} {\bibinfo {author} {\bibfnamefont {S.}~\bibnamefont
  {Giombi}}, \bibinfo {author} {\bibfnamefont {V.}~\bibnamefont {Kirilin}}, \
  and\ \bibinfo {author} {\bibfnamefont {E.}~\bibnamefont {Skvortsov}},\ }\href
  {\doibase 10.1007/JHEP05(2017)041} {\bibfield  {journal} {\bibinfo  {journal}
  {JHEP}\ }\textbf {\bibinfo {volume} {05}},\ \bibinfo {pages} {041} (\bibinfo
  {year} {2017})}\BibitemShut {NoStop}%
\bibitem [{\citenamefont {Gracey}(1992{\natexlab{b}})}]{gracey1992b}%
  \BibitemOpen
  \bibfield  {author} {\bibinfo {author} {\bibfnamefont {J.~A.}\ \bibnamefont
  {Gracey}},\ }\href {\doibase 10.1016/0370-2693(92)91265-B} {\bibfield
  {journal} {\bibinfo  {journal} {Phys. Lett. B}\ }\textbf {\bibinfo {volume}
  {297}},\ \bibinfo {pages} {293} (\bibinfo {year}
  {1992}{\natexlab{b}})}\BibitemShut {NoStop}%
\bibitem [{\citenamefont {Manashov}\ and\ \citenamefont
  {Strohmaier}(2018)}]{manashov2018}%
  \BibitemOpen
  \bibfield  {author} {\bibinfo {author} {\bibfnamefont {A.~N.}\ \bibnamefont
  {Manashov}}\ and\ \bibinfo {author} {\bibfnamefont {M.}~\bibnamefont
  {Strohmaier}},\ }\href {\doibase 10.1140/epjc/s10052-018-5902-1} {\bibfield
  {journal} {\bibinfo  {journal} {Eur. Phys. J. C}\ }\textbf {\bibinfo {volume}
  {78}},\ \bibinfo {pages} {454} (\bibinfo {year} {2018})}\BibitemShut
  {NoStop}%
\bibitem [{\citenamefont {Sandvik}(2007)}]{sandvik2007}%
  \BibitemOpen
  \bibfield  {author} {\bibinfo {author} {\bibfnamefont {A.~W.}\ \bibnamefont
  {Sandvik}},\ }\href {\doibase 10.1103/PhysRevLett.98.227202} {\bibfield
  {journal} {\bibinfo  {journal} {Phys. Rev. Lett.}\ }\textbf {\bibinfo
  {volume} {98}},\ \bibinfo {pages} {227202} (\bibinfo {year}
  {2007})}\BibitemShut {NoStop}%
\bibitem [{\citenamefont {Melko}\ and\ \citenamefont {Kaul}(2008)}]{melko2008}%
  \BibitemOpen
  \bibfield  {author} {\bibinfo {author} {\bibfnamefont {R.~G.}\ \bibnamefont
  {Melko}}\ and\ \bibinfo {author} {\bibfnamefont {R.~K.}\ \bibnamefont
  {Kaul}},\ }\href {\doibase 10.1103/PhysRevLett.100.017203} {\bibfield
  {journal} {\bibinfo  {journal} {Phys. Rev. Lett.}\ }\textbf {\bibinfo
  {volume} {100}},\ \bibinfo {pages} {017203} (\bibinfo {year}
  {2008})}\BibitemShut {NoStop}%
\bibitem [{\citenamefont {Nahum}\ \emph {et~al.}(2015)\citenamefont {Nahum},
  \citenamefont {Chalker}, \citenamefont {Serna}, \citenamefont
  {{Ortu{\~n}o}},\ and\ \citenamefont {Somoza}}]{nahum2015}%
  \BibitemOpen
  \bibfield  {author} {\bibinfo {author} {\bibfnamefont {A.}~\bibnamefont
  {Nahum}}, \bibinfo {author} {\bibfnamefont {J.~T.}\ \bibnamefont {Chalker}},
  \bibinfo {author} {\bibfnamefont {P.}~\bibnamefont {Serna}}, \bibinfo
  {author} {\bibfnamefont {M.}~\bibnamefont {{Ortu{\~n}o}}}, \ and\ \bibinfo
  {author} {\bibfnamefont {A.~M.}\ \bibnamefont {Somoza}},\ }\href {\doibase
  10.1103/PhysRevX.5.041048} {\bibfield  {journal} {\bibinfo  {journal} {Phys.
  Rev. X}\ }\textbf {\bibinfo {volume} {5}},\ \bibinfo {pages} {041048}
  (\bibinfo {year} {2015})}\BibitemShut {NoStop}%
\bibitem [{\citenamefont {Kleinert}\ and\ \citenamefont
  {Schulte-Frohlinde}(2001)}]{KleinertBook}%
  \BibitemOpen
  \bibfield  {author} {\bibinfo {author} {\bibfnamefont {H.}~\bibnamefont
  {Kleinert}}\ and\ \bibinfo {author} {\bibfnamefont {V.}~\bibnamefont
  {Schulte-Frohlinde}},\ }\href@noop {} {\emph {\bibinfo {title} {Critical
  Properties of $\phi^4$-Theories}}}\ (\bibinfo  {publisher} {World
  Scientific},\ \bibinfo {year} {2001})\BibitemShut {NoStop}%
\bibitem [{\citenamefont {van Ritbergen}\ \emph {et~al.}(1997)\citenamefont
  {van Ritbergen}, \citenamefont {Vermaseren},\ and\ \citenamefont
  {Larin}}]{vanRitbergen:1997va}%
  \BibitemOpen
  \bibfield  {author} {\bibinfo {author} {\bibfnamefont {T.}~\bibnamefont {van
  Ritbergen}}, \bibinfo {author} {\bibfnamefont {J.~A.~M.}\ \bibnamefont
  {Vermaseren}}, \ and\ \bibinfo {author} {\bibfnamefont {S.~A.}\ \bibnamefont
  {Larin}},\ }\href {\doibase 10.1016/S0370-2693(97)00370-5} {\bibfield
  {journal} {\bibinfo  {journal} {Phys. Lett. B}\ }\textbf {\bibinfo {volume}
  {400}},\ \bibinfo {pages} {379} (\bibinfo {year} {1997})}\BibitemShut
  {NoStop}%
\bibitem [{\citenamefont {Nogueira}(1993)}]{Nogueira:1991ex}%
  \BibitemOpen
  \bibfield  {author} {\bibinfo {author} {\bibfnamefont {P.}~\bibnamefont
  {Nogueira}},\ }\href {\doibase 10.1006/jcph.1993.1074} {\bibfield  {journal}
  {\bibinfo  {journal} {J. Comput. Phys.}\ }\textbf {\bibinfo {volume} {105}},\
  \bibinfo {pages} {279} (\bibinfo {year} {1993})}\BibitemShut {NoStop}%
\bibitem [{\citenamefont {Harlander}\ \emph {et~al.}(1998)\citenamefont
  {Harlander}, \citenamefont {Seidensticker},\ and\ \citenamefont
  {Steinhauser}}]{Harlander:1997zb}%
  \BibitemOpen
  \bibfield  {author} {\bibinfo {author} {\bibfnamefont {R.}~\bibnamefont
  {Harlander}}, \bibinfo {author} {\bibfnamefont {T.}~\bibnamefont
  {Seidensticker}}, \ and\ \bibinfo {author} {\bibfnamefont {M.}~\bibnamefont
  {Steinhauser}},\ }\href {\doibase 10.1016/S0370-2693(98)00220-2} {\bibfield
  {journal} {\bibinfo  {journal} {Phys. Lett. B}\ }\textbf {\bibinfo {volume}
  {426}},\ \bibinfo {pages} {125} (\bibinfo {year} {1998})}\BibitemShut
  {NoStop}%
\bibitem [{\citenamefont {Seidensticker}()}]{Seidensticker:1999bb}%
  \BibitemOpen
  \bibfield  {author} {\bibinfo {author} {\bibfnamefont {T.}~\bibnamefont
  {Seidensticker}},\ }in\ \href@noop {} {\emph {\bibinfo {booktitle}
  {{Proceedings of the 6th International Workshop on New Computing Techniques
  in Physics Research: Software Engineering, Artificial Intelligence Neural
  Nets, Genetic Algorithms, Symbolic Algebra, Automatic Calculation (AIHENP
  99), Heraklion, 1999}}}},\ \Eprint {http://arxiv.org/abs/hep-ph/9905298}
  {arXiv:hep-ph/9905298} \BibitemShut {NoStop}%
\bibitem [{\citenamefont {Vermaseren}()}]{Vermaseren:2000nd}%
  \BibitemOpen
  \bibfield  {author} {\bibinfo {author} {\bibfnamefont {J.~A.~M.}\
  \bibnamefont {Vermaseren}},\ }\href@noop {} {\ }\Eprint
  {http://arxiv.org/abs/math-ph/0010025} {arXiv:math-ph/0010025} \BibitemShut
  {NoStop}%
\bibitem [{\citenamefont {Kuipers}\ \emph {et~al.}(2013)\citenamefont
  {Kuipers}, \citenamefont {Ueda}, \citenamefont {Vermaseren},\ and\
  \citenamefont {Vollinga}}]{Kuipers:2012rf}%
  \BibitemOpen
  \bibfield  {author} {\bibinfo {author} {\bibfnamefont {J.}~\bibnamefont
  {Kuipers}}, \bibinfo {author} {\bibfnamefont {T.}~\bibnamefont {Ueda}},
  \bibinfo {author} {\bibfnamefont {J.~A.~M.}\ \bibnamefont {Vermaseren}}, \
  and\ \bibinfo {author} {\bibfnamefont {J.}~\bibnamefont {Vollinga}},\ }\href
  {\doibase 10.1016/j.cpc.2012.12.028} {\bibfield  {journal} {\bibinfo
  {journal} {Comput. Phys. Commun.}\ }\textbf {\bibinfo {volume} {184}},\
  \bibinfo {pages} {1453} (\bibinfo {year} {2013})}\BibitemShut {NoStop}%
\bibitem [{\citenamefont {Misiak}\ and\ \citenamefont
  {Munz}(1995)}]{Misiak:1994zw}%
  \BibitemOpen
  \bibfield  {author} {\bibinfo {author} {\bibfnamefont {M.}~\bibnamefont
  {Misiak}}\ and\ \bibinfo {author} {\bibfnamefont {M.}~\bibnamefont {Munz}},\
  }\href {\doibase 10.1016/0370-2693(94)01553-O} {\bibfield  {journal}
  {\bibinfo  {journal} {Phys. Lett. B}\ }\textbf {\bibinfo {volume} {344}},\
  \bibinfo {pages} {308} (\bibinfo {year} {1995})}\BibitemShut {NoStop}%
\bibitem [{\citenamefont {Chetyrkin}\ \emph {et~al.}(1998)\citenamefont
  {Chetyrkin}, \citenamefont {Misiak},\ and\ \citenamefont
  {Munz}}]{Chetyrkin:1997fm}%
  \BibitemOpen
  \bibfield  {author} {\bibinfo {author} {\bibfnamefont {K.~G.}\ \bibnamefont
  {Chetyrkin}}, \bibinfo {author} {\bibfnamefont {M.}~\bibnamefont {Misiak}}, \
  and\ \bibinfo {author} {\bibfnamefont {M.}~\bibnamefont {Munz}},\ }\href
  {\doibase 10.1016/S0550-3213(98)00122-9} {\bibfield  {journal} {\bibinfo
  {journal} {Nucl. Phys. B}\ }\textbf {\bibinfo {volume} {518}},\ \bibinfo
  {pages} {473} (\bibinfo {year} {1998})}\BibitemShut {NoStop}%
\bibitem [{\citenamefont {Marquard}\ and\ \citenamefont {Seidel}()}]{crusher}%
  \BibitemOpen
  \bibfield  {author} {\bibinfo {author} {\bibfnamefont {P.}~\bibnamefont
  {Marquard}}\ and\ \bibinfo {author} {\bibfnamefont {D.}~\bibnamefont
  {Seidel}},\ }\href@noop {} {\ }\Eprint {http://arxiv.org/abs/unpublished}
  {unpublished} \BibitemShut {NoStop}%
\bibitem [{\citenamefont {Laporta}(2000)}]{Laporta:2001dd}%
  \BibitemOpen
  \bibfield  {author} {\bibinfo {author} {\bibfnamefont {S.}~\bibnamefont
  {Laporta}},\ }\href {\doibase 10.1016/S0217-751X(00)00215-7,
  10.1142/S0217751X00002157} {\bibfield  {journal} {\bibinfo  {journal} {Int.
  J. Mod. Phys. A}\ }\textbf {\bibinfo {volume} {15}},\ \bibinfo {pages} {5087}
  (\bibinfo {year} {2000})}\BibitemShut {NoStop}%
\bibitem [{\citenamefont {Ihrig}\ \emph
  {et~al.}(2018{\natexlab{a}})\citenamefont {Ihrig}, \citenamefont {Mihaila},\
  and\ \citenamefont {Scherer}}]{ihrig2018}%
  \BibitemOpen
  \bibfield  {author} {\bibinfo {author} {\bibfnamefont {B.}~\bibnamefont
  {Ihrig}}, \bibinfo {author} {\bibfnamefont {L.~N.}\ \bibnamefont {Mihaila}},
  \ and\ \bibinfo {author} {\bibfnamefont {M.~M.}\ \bibnamefont {Scherer}},\
  }\href {\doibase 10.1103/PhysRevB.98.125109} {\bibfield  {journal} {\bibinfo
  {journal} {Phys. Rev. B}\ }\textbf {\bibinfo {volume} {98}},\ \bibinfo
  {pages} {125109} (\bibinfo {year} {2018}{\natexlab{a}})}\BibitemShut
  {NoStop}%
\bibitem [{\citenamefont {Ihrig}\ \emph
  {et~al.}(2018{\natexlab{b}})\citenamefont {Ihrig}, \citenamefont {Janssen},
  \citenamefont {Mihaila},\ and\ \citenamefont {Scherer}}]{ihrig2018b}%
  \BibitemOpen
  \bibfield  {author} {\bibinfo {author} {\bibfnamefont {B.}~\bibnamefont
  {Ihrig}}, \bibinfo {author} {\bibfnamefont {L.}~\bibnamefont {Janssen}},
  \bibinfo {author} {\bibfnamefont {L.~N.}\ \bibnamefont {Mihaila}}, \ and\
  \bibinfo {author} {\bibfnamefont {M.~M.}\ \bibnamefont {Scherer}},\ }\href
  {\doibase 10.1103/PhysRevB.98.115163} {\bibfield  {journal} {\bibinfo
  {journal} {Phys. Rev. B}\ }\textbf {\bibinfo {volume} {98}},\ \bibinfo
  {pages} {115163} (\bibinfo {year} {2018}{\natexlab{b}})}\BibitemShut
  {NoStop}%
\bibitem [{\citenamefont {Gracey}(1993{\natexlab{b}})}]{gracey1993b}%
  \BibitemOpen
  \bibfield  {author} {\bibinfo {author} {\bibfnamefont {J.~A.}\ \bibnamefont
  {Gracey}},\ }\href {\doibase 10.1088/0305-4470/26/6/024} {\bibfield
  {journal} {\bibinfo  {journal} {J. Phys. A}\ }\textbf {\bibinfo {volume}
  {26}},\ \bibinfo {pages} {1431} (\bibinfo {year}
  {1993}{\natexlab{b}})}\BibitemShut {NoStop}%
\end{thebibliography}%

\end{document}